\documentclass[aps,twocolumn,showpacs,preprintnumbers,amsmath,amssymb,nofootinbib,superscriptaddress,showkeys,prd]{revtex4-1}
\usepackage{graphicx,graphics,epsfig}
\usepackage{amsmath,amssymb,slashed,dcolumn,amsfonts,slashed,tabulary}
\usepackage{booktabs}
\usepackage{comment}
\usepackage{multirow}
\usepackage{mathptmx}      

\newcommand{\beq}{\begin{equation}}
\newcommand{\eeq}{\end{equation}}
\newcommand{\beqa}{\begin{eqnarray}}
\newcommand{\eeqa}{\end{eqnarray}}
\newcommand{\bal}{\begin{align}}

\newcommand{\diff}{\text{d}}
\newcommand{\dr}{\text{dr}}

\renewcommand{\Im}{\text{Im}\,}

\newcommand{\smax}{s_{\rm max}}

\allowdisplaybreaks[1]

\providecommand{\diff}{\!\!\tn{d}}

\begin{document}

\title{Coarse graining $\pi\pi$ scattering} \author{Jacobo Ruiz de
  Elvira}\email{elvira@itp.unibe.ch} \affiliation{ Albert Einstein
  Center for Fundamental Physics, Institute for Theoretical Physics,
  \\ University of Bern, Sidlerstrasse 5, CH--3012 Bern, Switzerland.}
\author{Enrique Ruiz Arriola}\email{earriola@ugr.es}
\affiliation{Departamento de F\'{\i}sica At\'omica, Molecular y
  Nuclear and Instituto Carlos I de F{\'\i}sica Te\'orica y
  Computacional \\ Universidad de Granada, E-18071 Granada, Spain.}

\date{\today}

\begin{abstract} 
\rule{0ex}{3ex} 
We carry out an analysis of $\pi\pi$ scattering in the
$IJ=00$, $11$ and $20$ channels in configuration space up to a maximal
center-of-mass energy $\sqrt{s}=1.4$ GeV.  We separate the interaction
into two regions marked by an elementarity radius of the system;
namely, a long distance region above which pions can be assumed to
interact as elementary particles and a short distance region where
many physical effects cannot be disentangled. The long distance
interaction is described by chiral dynamics, where a two-pion-exchange
potential is identified, computed and compared to lattice calculations. 
The short distance piece corresponds to a coarse grained description 
exemplified by a superposition of delta-shell potentials
sampling the interaction with the minimal wavelength. 
We show how the so constructed non-perturbative scattering
amplitude complies with the proper analytic structure,
allowing for an explicit N/D type decomposition in terms of the corresponding
Jost functions and fulfilling dispersion relations without subtractions.
We also address renormalization issues in coordinate space and
investigate the role of crossing when fitting the scattering
amplitudes above and below threshold to Roy-equation results.
At higher energies, we show how inelasticities can be described
by one single complex and energy dependent parameter. A successful
description of the data can be achieved with a minimal number of
fitting parameters, suggesting that coarse graining is a viable approach
to analyze hadronic processes. 
\end{abstract}

\pacs{12.38.Gc, 12.39.Fe, 14.20.Dh} \keywords{$\pi\pi$ interaction,
  Partial Wave Analysis, Chiral symmetry, Optical potential, Analytical properties}

\maketitle

\section{Introduction}

Hadronic interactions at low and intermediate energies are typically
characterized by a combination of elementary and composite particle
features. While at long distances hadrons behave as elementary
particles and their interactions can be described in terms of purely
color singlet degrees of freedom, at short distances their composite
character becomes manifest in terms of quark and gluon fields in the
fundamental and adjoint representations of the color group,
respectively.  The relevant scale separating between this dual
description marks the onset of a confinement scale and we expect it to
be of the order of the hadron size, which generally is found to be
about 1 fm. While the hadronic dynamics can be organized quite often
as a {\it long distance} perturbative hierarchy with an increasing
number of exchanged particles, it is by itself incomplete; some
further either {\it ab initio} or phenomenological information
reflecting the underlying quark-gluon structure is needed to provide a
full description of the scattering process.

The way how this separation is visualized in the complex energy plane
is not completeley straightforward. Traditionally, and within a
genuinely hadronic picture, one appeals to Mandelstam
analyticity~\cite{Mandelstam:1958xc}, i.e. the assumption that a
scattering amplitude can be expressed by double dispersive integrals
in terms of double-spectral density functions, where the integration
ranges extend over those regions in the Mandelstam plane where the
corresponding double-spectral functions have non-vanishing
support~\cite{Hoferichter:2015hva}.  This viewpoint is ultimately
grounded in the Mandelstam conjecture, which holds in lowest order in
the coupling constant in quantum field
theory~\cite{Mandelstam:1958xc,Mandelstam:1959bc} or to all orders
within a non-relativistic context in potential
scattering~\cite{Blankenbecler:1960zz}, and, which, in the $\pi\pi$
scattering case, has been rigorously proved in a finite
domain~\cite{Martin:1965jj,Martin:1966}.  It is noteworthy that under
this same assumption an equivalent local and energy dependent optical
potential of non-relativistic form was derived many years ago by
Cornwall and Ruderman~\cite{cornwall1962mandelstam,omnes1965optical}.
For a balanced review on these issues at the textbook level see, for
e.g.,~\cite{nussenzveig1972causality,goldberger2004collision}.  The
existence of a finite analyticity domain suggests in turn the very
existence of a finite cut-off on a purely hadronic basis but without
an explicit reference to the underlying quark-gluon dynamics and in
particular to the confinement scale, so that the cut-off may be
determined phenomenologically from data.

Pion-pion scattering is the simplest reaction in QCD mediated by
strong interactions involving the lightest hadrons.  Tight theoretical
constraints based on analyticity, crossing, unitarity, chiral symmetry
and Regge behavior can be imposed (see for e.g.~\cite{martin1976pion}
for an early review).  The machinery of effective field theories
(EFT)~\cite{Weinberg:1978kz} and in particular its implementation in
Chiral Perturbation Theory ($\chi$PT)~\cite{Gasser:1983yg} has enabled
as a consequence, the most precise theoretical extraction of the
$\pi\pi$ S-wave scattering lengths to date with about an order of
magnitude more precision than the
experiment~\cite{Colangelo:2000jc,Ananthanarayan:2000ht,Colangelo:2001df,Caprini:2003ta,Pelaez:2004vs,Kaminski:2006yv,Kaminski:2006qe,GarciaMartin:2011cn},
an unprecedented case in strong interactions, where invariably just
the opposite situation happens.  A historic overview is given
in~\cite{Gasser:2009zz}.  Along these lines, the most precise
$\pi\pi$-scattering analyses to date have been obtained
in~\cite{Colangelo:2001df,GarciaMartin:2011cn}.  The latter
corresponds to a $\pi\pi$ description up to $\sqrt{s}=1.42$ GeV,
obtained by fitting the available experimental data from $\pi N \to
\pi \pi N$ and $K_{e4}$ decays while imposing as further constraints
Roy and Roy-like equations, and with statistical uncertainties
satisfying the necessary normality requirements of the residual
distributions~\cite{Perez:2015pea}, (see for
e.g.~\cite{RuizdeElvira:2012mbw,Pelaez:2015qba} for reviews).  We
stress that despite all these tight mathematical constraints, most of
its non-perturbative setup rests upon the validity of the Mandelstam
conjecture~\cite{Mandelstam:1958xc,Mandelstam:1959bc}, a result which,
as already mentioned, has not yet been rigorously proven since it was
first proposed in 1958.  This tacit assumption will also be made
throughout our work.

In the present paper we invoke the equivalent local and energy
dependent optical potential approach suggested long
ago in~\cite{cornwall1962mandelstam,omnes1965optical} to describe
$\pi\pi$ scattering in coordinate space. In order to do so, we
consider a relativistic Schr\"odinger equation and define a potential
to describe the $\pi\pi$ interaction by matching the field theoretical
result to an equivalent quantum mechanical problem in perturbation
theory. Phenomenological precursors of $\pi\pi$ scattering analyses in
coordinate space were prompted in~\cite{au1963pion,au1964pi} within the
boundary condition model of strong interactions~\cite{feshbach1964boundary}. 
Equivalent coordinate space potentials using the Mandelstam representation or the Bethe-Salpeter
equation as a starting point were also proposed to all
orders in~\cite{balazs1965low,balazs1965lowII,Balazs:1969bt}.  
As it will become clear below, it is remarkable if not surprising that so little
work on $\pi\pi$ scattering has been conducted within this approach as
compared to more popular momentum space methods. Our work fills this
gap by implementing Wilsonian ideas inspired by recent developments in
the NN case~\cite{NavarroPerez:2011fm,Perez:2013jpa,Perez:2013oba}.
These NN investigations had as a consequence a selection of the
largest np and pp database up to energies about pion production
threshold of $3 \sigma$ mutually consistent data.  
Our present investigation within $\pi\pi$ is in a sense of exploratory character
and it pretends also to provide some training playground with an eye put
on the more compelling $\pi N$ case, where the selection of the
currently existing database is largely needed (see for
e.g.~\cite{Matsinos:2016fcd,RuizdeElvira:2017stg,RuizArriola:2017kqs}
and references therein).

At short distances, where the interaction is non perturbative, we will
assume a complete ignorance of the strong interaction behavior and
consider a coarse graining of the interaction instead, very much in
the spirit of the work done
in~\cite{NavarroPerez:2011fm,Perez:2013jpa,Perez:2013oba} for the NN
case.  The basic idea is to separate the $\pi\pi$ interaction into an
inner and outer region at a given separation distance, $r_c$, located
at about some {\it elementarity radius}.  This radius is defined so
that at larger distances pions behave effectively as point-like
particles.  We will assume that in this long distance regime their
interactions are ruled by chiral symmetry, and hence they become
calculable within $\chi$PT.  Thus, for $r>r_c$, we will construct a
chiral potential with the correct low-energy analytic properties by
matching both quantum mechanical and field theoretical scattering
amplitudes in perturbation theory~\footnote{ This is similar to the
  unitarization method based on the Bethe-Salpeter
  equation~\cite{Nieves:1998hp,Nieves:1999bx}.}.  On the contrary, the
inner region, $r<r_c$, is regarded as unknown and sampled with the
minimal de Broglie wave length determined by the maximum energy we
want to describe.  This corresponds to a coarse graining of the short
range piece and, in its simplest realization, the inner potential will
be written as a superposition of equidistant delta-shell interactions.
A key issue is to confidently determine the numerical value of the
separation scale $r_c$, since, as noted
in~\cite{Perez:2013cza,Fernandez-Soler:2017kfu} and we will see below,
the combination $p r_c $ will fix the total number of {\it
  independent} fitting parameters.  The longest range interaction
corresponds to a $2\pi$-exchange which is ${\cal O} ( e^{-2 m_\pi r})
$, so that a naive estimate suggests $r_c \sim 2/(2\,m_\pi)\sim 1.4 $
fm~\footnote{Details here are important. The extra factor 2 is to
  ensure that $e^{-2 m_\pi r_c} = 1/e^2 \sim 0.13$ is {\it really}
  negligible. This is confirmed by our analysis below.}, a number
which will be corroborated by our numerical analysis.

While the potential approach has been explained in great detail in
previous works within the NN context (see for
instance~\cite{Fernandez-Soler:2017kfu}), it is unconventional within
the $\pi\pi$ scattering folklore.  Thus, we will assume no previous
knowledge from the side of the reader and for the sake of completeness
we will briefly go through all the important issues along the paper.
Moreover, $\pi\pi$ scattering is characterized because at resonance
energies relativistic effects cannot be ignored.  For instance, for
the prominent case of the $\rho$-meson $\sqrt{s}=m_\rho \gg 2 m_\pi$.
Unlike the NN case, a new important aspect in the discussion is
related to crossing symmetry, which actually intertwines the $s$, $t$
and $u$ channels~\footnote{Crossing for NN relates the two-pion
  exchange interaction with the $N \bar N \to 2 \pi$ production
  channel. This implies an exponentially suppressed effect in the $NN$
  potential$\sim e^{-2 M_N r}$ and hence having little practical
  relevance.}.  In addition, the current extraordinary precision
achieved theoretically in extracting the S-waves scattering lengths or
the lightest $\pi\pi$ resonance pole
parameters~\cite{Caprini:2005zr,GarciaMartin:2011jx,Masjuan:2014psa,Caprini:2016uxy}
provides a great confidence on the theoretical ideas supporting these
benchmarking extractions.  The fact that the coarse graining approach
works for NN scattering in a regime where relativistic and inelastic
effects become important, such as pp scattering up to $
\sqrt{s}~\sim2$ GeV~\cite{Fernandez-Soler:2017kfu}, suggests extending
the method to other hadronic reactions under similar operating
conditions~\footnote{We remind that within such a context the methods
  based in analyticity, dispersion relations and crossing are
  currently considered to be, besides QCD, the most rigorous
  framework. We stress again that such an approach is based on the
  validity of the double spectral representation of the four-point
  function conjectured by Mandelstam.}.

Finally, for the sake of completeness let us mention that lattice
calculations are naturally formulated in coordinate space. These
calculations attack the problem on the finite lattice spacing and the
finite volume in two different fashions: either an (energy dependent)
potential is determined and the Schr\"odinger equation is solved
subsequently in the continuum or, alternatively, the energy level
shifts are determined on the lattice and converted into phase-shifts
by means of Luscher's formula~\cite{Luscher:1990ux}. Actually, in a
pioneering work~\cite{Beane:2011sc} , the $I=2$ $\pi\pi$ S-wave
scattering phase shifts from Lattice QCD have been determined.  Later
on, $\pi\pi$ scattering has been studied from $N_{\mathrm{f}} = 2+1$
and $N_{\mathrm{f}} = 2+1+1$ flavors in~\cite{Bulava:2016mks}
and~\cite{Helmes:2015gla}, respectively.  Connected and Disconnected
Contractions have also been analyzed in~\cite{Acharya:2017zje}.  In
addition, the $\pi\pi$ $I=2$ channel has also been studied within the
potential approach~\cite{Yamazaki:2015nka}.  A comparison between
potential and Luscher's approaches has been undertaken
in~\cite{Kurth:2013tua} for the $I=2$ case, with rather similar
results. We remind that both methods have potential drawbacks.  On the
one hand, the potential method uses interpolating fields which may
distort the physics at short distances, and we will explicitly show
that in a chiral expansion such potential presents a short distance
singularity, which evades the conventional solutions of the
Schr\"odinger equation. On the other hand, the current applicability
of this Luscher's method~\cite{Luscher:1990ux} requires the
interaction to sharply vanish at the edges of the volume (in the
relative coordinate), a fact that has been often ignored in momentum
space treatments (see for e.g.~\cite{Doring:2011vk,Doring:2012eu}) but
needs to be established for the $\pi\pi$ case.  Our analysis below
supports this assumption.

The paper is organized as follows. In Section~\ref{sec:general}, we
provide a general and brief field theoretical overview of $\pi\pi$
scattering to fix our notation in a way that our problem can be easily
formulated. In Section~\ref{sec:QM}, we show our choice for a quantum
mechanical description in terms of a complex local and energy
dependent optical potential. We analyze the long-range contributions
within $\chi$PT in Section~\ref{sec:chpt}, where an expression for the
potential is obtained from the discontinuities of the $\pi\pi$
scattering amplitude in the $t$-channel. This requires introducing a
short distance cut-off to handle the strong short distance power
divergences of the chiral potential, an issue which we discuss at
length in Section~\ref{sec:renor}. In Section~\ref{sec:elem}, we
analyze the concept of effective elementarity in order to display in
two examples how the elementarity radius depends on the particular
process. In Section~\ref{sec:coarse}, we address the problem of coarse
graining $\pi\pi$ interactions with and without the long-range contributions. 
The analytical properties of the
scattering amplitude and the relation of our approach with the N/D
method is discussed in Section~\ref{sec:analy}. The implementation of
inelasticities within a coarse grained perspective is explained in
Section~\ref{sec:inel}. We also analyze some aspects concerning low
energy constants and the number of parameters in
Section~\ref{sec:LECs}.  Finally, in Section~\ref{sec:conl} we
summarize our main results and provide some outlook for future work.

\section{Formalism for $\pi\pi$ scattering}
\label{sec:general}

We start by summarizing the relevant formulae for $\pi\pi$ scattering
to fix our notation and to provide a proper perspective of our
subsequent analysis. A comprehensive presentation at the textbook
level can be seen in~\cite{martin1976pion} and also in the
lecture~\cite{Yndurain:2002ud}. More recent upgrades can be consulted
in~\cite{RuizdeElvira:2012mbw,Pelaez:2015qba}

\subsection{Kinematics}

For a pion state $\varphi_\alpha$ with $\alpha=\{\pm,0\}$, the $\pi_
\alpha (p_1)+ \pi_\beta (p_2) \to \pi_ \gamma (p_1')+ \pi_ \delta
(p_2') $ relativistically invariant scattering amplitude can be
written as
\begin{eqnarray}\nonumber
  T_{\alpha \beta; \gamma \delta} &=& (\varphi_\gamma^* \cdot \varphi_\delta^*)
  (\varphi_\alpha \cdot \varphi_\beta ) A(s,t,u) \nonumber \\  &+& (\varphi_\gamma^* \cdot
  \varphi_\alpha ) (\varphi_\delta^* \cdot \varphi_\beta ) B(s,t,u) \nonumber\\ &+& (\varphi_\delta^*
  \cdot \varphi_\alpha ) (\varphi_\beta \cdot \varphi_\gamma^* ) C(s,t,u) \, ,
\end{eqnarray}
with $s=(p_1+p_2)^2$, $t=(p_1-p_1')^2$ and $u=(p_1-p_2')^2$ the
standard choice of Mandelstam variables. If we take
$\varphi_{\pm} = ( \phi_1 \pm i \phi_2) /\sqrt{2} $ and
$\varphi_{0}= \phi_3$, with $\phi_a \cdot \phi_b = \delta_{ab}$,
in the Cartesian basis we obtain
\begin{eqnarray}\nonumber
T_{ab;cd} = A(s,t,u) \delta_{ab} \delta_{cd}
+ B(s,t,u) \delta_{ac} \delta_{bd}
+ C(s,t,u) \delta_{ad} \delta_{bc}\, ,
\end{eqnarray}
where $A(s,t,u)$ stands for the $\pi^+\pi^-\to \pi^0 \pi^0$ amplitude. 
This amplitude is the the only independent one thanks to isospin, crossing and Bose-Einstein symmetries, 
$B(s,t,u)=A(t,s,u)$  and $C(s,t,u)=A(u,t,s)$. 
Denoting $T_{I}(s,t,u)$ as the isospin combination with
well defined isospin $I$ in the $s$-channel, one has
\begin{eqnarray}
T_{I=0}(s,t,u) &=& 3A(s,t,u) + A(t,s,u)+
A(u,t,s) \, ,\nonumber \\
T_{I=1}(s,t,u) &=& A(t,s,u)-
A(u,t,s) \, ,\nonumber\\
T_{I=2}(s,t,u) &=& A(t,s,u)+
A(u,t,s) \, .
\end{eqnarray}
For the normalization, we will use here the conventions
in~\cite{GomezNicola:2010tb,RuizdeElvira:2010cs,GomezNicola:2012uc}. The partial-wave decomposition in the
$s$-channel becomes
\begin{eqnarray}
T_{I}(s,t,u) = 16\pi \sum_{J=0}^\infty \left[1+(-1)^{J+I} \right](2J+1) t_{IJ}(s) P_J (z)\, ,
\label{eq:pwe}
\end{eqnarray}
where $z = 1 + 2 t/(4m_\pi^2 -s)$ is the $s$-channel scattering angle, $m_\pi=139.57$ MeV the
pion mass, $P_J( \cos \theta)$ the Legendre polynomials and
$t_{IJ}(s)$ is the partial-wave projection of the $\pi\pi$ scattering amplitude with isospin $I$ and total angular momentum $J$. 
Thus,  for waves fulfilling the relation $(-1)^{J+I}=1$ one has  
\begin{eqnarray}\label{eq:pw-def}
t_{IJ}(s) &=& \frac{1}{64\pi} \int^{+1}_{-1}\, \diff z;
P_J(z)\, T_I\left(s,t(s,z), u(s,z)\right)
\nonumber \\
&=&
\left (\frac{\eta_{IJ} (s) e^{2{\rm i}\delta_{IJ}(s)}-1}{2{\rm i}\,\sigma(s)}\right)\, ,
\end{eqnarray}
with
\begin{equation}\label{eq:sigma}
\sigma(s) = \sqrt{1-\frac{4 m_\pi^2}{s}}\, ,
\end{equation}
the $\pi\pi$ phase factor and $\delta_{IJ}$ the scattering phase shift. 
The in-elasticity $\eta_{IJ} (s)= 1 $ for $s< 16 m_\pi^2$ and 
the unitarity condition for the partial wave amplitude reads in the
elastic region
\begin{eqnarray}
{\rm Im}\,t_{IJ} (s) = \sigma(s)|t_{IJ}(s)|^2\quad{\rm for} \quad 4 m_\pi^2 \le s\le 16 m_\pi^2.
\label{eq:unit}
\end{eqnarray}
Of course, for $s> 16 m_\pi^2$ one has absorption $\eta_{IJ} (s) < 1 $
and inelastic processes such as $2\pi \to n \pi $ take place at
$\sqrt{s}=0.56,\,0.84,\,1.12$ and $1.40$ GeV for $n=4,6,8$ and $10$,
respectively, as well as $K \bar K$ and $\eta \eta$ at $\sqrt{s} \sim
1\,{\rm GeV}$, etc.


In our discussion we will also use the quantum mechanical amplitude $f_{IJ}(p)$ defined by 
\begin{eqnarray}\label{eq:rel-no-rel-eq}
f_{IJ}(p)= \frac{2}{\sqrt{s}} t_{IJ}(s),\,\qquad s = 4 (p^2+m_\pi^2) \, , 
\end{eqnarray}
with $p$ the CM momentum.  For elastic scattering one has
$f_{IJ}(s)^{-1}= p \cot \delta_{IJ} - i p$, so that at low energies
one has the threshold expansion
\begin{eqnarray}
{\rm Re} f_{IJ}(s) = p^{2J}  \left[ a_{IJ} + b_{IJ} p^2 + \dots \right]  \, , 
\label{eq:thres}
\end{eqnarray}
with $a_{IJ}$ and $b_{IJ} $ the lowest threshold parameters. An
equivalent way of representing the low energy behavior is 
\begin{eqnarray}
\frac{\tan \delta_{IJ}(s)}{p^{2J+1}}= a_{IJ} + b_{IJ} p^2 + \dots  \, , 
\label{eq:thres3}
\end{eqnarray}
or by an effective range expansion
\begin{eqnarray}\label{eq:effective-range-exp}
  p^{2J+1} \cot \delta_{IJ} (s)= - \frac1{\alpha_{IJ}} + \frac12 r_{IJ} p^2 + \dots \, , 
\label{eq:thres2}
\end{eqnarray}
where $\alpha_{IJ}=-a_{IJ}$ and $r_{IJ}/2=-b_{IJ}/a_{IJ}^2 $ is the
effective range, which is generally positive (see below). Usually, the
expansion~\eqref{eq:thres3} works for {\it small} scattering
lengths, such as $\pi\pi$ whereas~\eqref{eq:thres2}
works for {\it large} scattering lengths, such as $NN$ (see,  e.g. ,   
\cite{Adhikari:1983jb,Adhikari:2018ukk} for a discussion)

\subsection{Anatomy of the $\pi\pi$ interaction}

The purpose of the present paper is to coarse grain the {\it unknown}
pieces of the $\pi\pi$ interaction in configuration space. It is thus
important to gather some features emerging from comprehensive studies
over the last
decades~\cite{Colangelo:2000jc,Ananthanarayan:2000ht,Colangelo:2001df,Caprini:2003ta,Pelaez:2004vs,Kaminski:2006yv,Kaminski:2006qe,GarciaMartin:2011cn}. According
to these findings the partial wave expansion in~\eqref{eq:pwe} is
decomposed into two contributions: the low energy contribution
described by means of a partial pave (PW) expansion to finite order
and the high energy contribution assumed to be given by the leading
Regge trajectories,
\begin{eqnarray}
T_I = T_I|_{\rm PW} + T_I|_{\rm Regge} \, , 
\end{eqnarray}
which accounts for the long and short distance behavior of the
scattering amplitude respectively. 

A standard quantum mechanical argument based on the impact parameter
provides in the semi-classical limit and for an interaction of finite
range $r_c$, the number of necessary partial waves~\footnote{These
  arguments provide in addition a justification for
  analyticity~\cite{Omnes:1966pp,Kugler:1966zz}.}.  The impact
parameter is defined as $b = L/p$ with $p$ the CM momentum and $L$ the
orbital angular momentum, which in our case equals the total angular
momentum $J$.  The quantization condition for the angular moment
yields $L \approx \sqrt{J(J+1)} \sim (J+1/2)$ for $J \gg 1$.  For a
finite range, the maximal impact parameter where scattering happens is
$b_{\rm max} \sim r_c$.  Thus, for a maximum CM momentum $p_{\rm
  max}$, the maximum angular momentum $J_{\rm max}$ for which the
phase shift is compatible with zero within uncertainties is
\begin{equation}\label{bmax}
J_{\rm max} + 1/2 \sim p_{\rm max} r_c,\quad {\rm with} \quad |\delta_{J_{\rm max}}|\lesssim \Delta \delta_{J_{\rm max}}. 
\end{equation}
For a maximum energy $s_{\rm max}=2$ GeV$^2$, corresponding to $p_{\rm max} \sim 0.7 $ GeV, 
it was found in~\cite{GarciaMartin:2011cn}  that waves beyond $J_{\rm max}=4$ are vanishingly small for $\pi\pi$ scattering.
Therefore, one obtains from~\eqref{bmax} a range $r_c \sim 1.3$ fm. 
This simple estimate will be explicitly exploited below as an educated guess.

Low energies close to threshold are encoded by the threshold
parameters, see~\eqref{eq:thres} and~\eqref{eq:thres2}. The
S-wave scattering lengths are $\alpha_{00}=-0.3$ fm and
$\alpha_{20}=0.03$ fm whereas for the P-wave we have $\alpha_{11}=
-(0.48\,{\rm fm})^3$~\cite{GarciaMartin:2011cn}. 
These are {\it unnaturally small} numbers
compared with our above estimate of the range of the interaction, $r_c
\sim 1.3\,{\rm fm}$ and the elementarity radius, $r_e \sim 1.2\,{\rm
  fm}$ (see the discussion in Section~\ref{sec:elem}).  While the
behavior of the isotensor S-wave resembles a repulsive core, with
positive effective range $r_{20}=131.4\,{\rm fm}$, the effective range
in the isoscalar S-wave and isovector P-wave are {\it negative},
$r_{00}=-8.08\,{\rm fm}$ and $r_{11}=-5.25\,{\rm fm}$, respectively. 
For $S$ waves the Wigner causality bound~\cite{Wigner:1955zz}
(see also \cite{Phillips:1996ae}) restricts the maximum value of the
effective range by the inequality
\begin{eqnarray}
r_{I0} \le 2 r_c \left( 1- \frac{r_c}{\alpha_{I0}}+\frac{r_c^2}{3 \alpha_{I0}^2}\right),
\end{eqnarray}
which for $I=2$ implies $r_c \ge 0.95\,{\rm fm}$.  The positivity of
the effective range is not implied by this condition, and is usually
violated in the presence of resonances. This requires some
unconventional shape for the S-wave potential as we will see.

\subsection{Chiral Perturbation theory}

The scattering amplitude can be computed perturbatively in Quantum
Field Theory and in particular in $\chi$PT as a sum of Feynmann
diagrams in an expansion in $1/f$, with $f \sim 86\,{\rm MeV}$ the pion
weak decay constant in the chiral limit. In the partial waves basis
the expansion can schematically be written as
\begin{eqnarray}
t_{IJ} (s)= t_{IJ}^{(2)} (s) + t_{IJ}^{(4)} (s) + \dots
\end{eqnarray}
where $t_{IJ}^{(n)}={\cal O} (f^{-2n})$.  To one loop order, they were first
computed in~\cite{Weinberg:1978kz,Gasser:1983yg} and the
relevant non-polynomial contributions are reproduced for completeness
in appendix~\ref{sec:pots-amps}. Explicit analytical expressions for
the corresponding partial wave amplitudes are displayed in~\cite{Nieves:1999bx}. 
They obey the perturbative unitarity relation 
\begin{eqnarray}
{\rm Im}\,t_{IJ}^{(4)}(s)=\sigma(s)|t_{IJ}^{(2)}(s)|^2 \, , \qquad 4 m_\pi^2 \le s\le 16
m_\pi^2 \, .
\end{eqnarray}
At lowest order (LO) in the chiral expansion the threshold parameters
are unnaturally small, a fact naturally accommodated by $\chi$PT with pions
coupled derivatively.

\subsection{Unitarization vs Crossing}

The requirement of crossing is a fundamental one which stems from the
local character of Quantum Field Theories. Chiral Perturbation Theory
implements this symmetry at any order in the chiral expansion. The
problems with perturbation theory, however, are on the one hand the
lack of exact unitarity given by~\eqref{eq:unit} and on the other hand
the impossibility of describing outstanding non-perturbative features
such as the generation of resonances, which emerge as poles of the
scattering amplitude on unphysical Riemann sheets.  Within a $\chi$PT
framework, many methods have been proposed (see for
instance~\cite{Nieves:1999bx,Guo:2012ym,Guo:2012yt} and references
therein) based on imposing exact unitarity while matching perturbation
theory at low energies.  Most of them are nothing but algebraic tricks
or functional solutions to a set of {\it a priori} conditions.  As
such, unitarization methods are not unique but strongly driven by
experimental information, which explains partly their success.  The
Bethe-Salpeter method discussed at length in~\cite{Nieves:1999bx}
preserves an identification of Feynman diagrams but it is not free
from field reparameterizations or off-shell ambiguities.  In addition,
they violate crossing symmetry, which, in general, is only fulfilled
order by order, although these violations can be statistically
not-significant~\cite{Nieves:2001de}.

In Sections~\ref{sec:chpt} and~\ref{sec:coarse}, we will propose yet a
new method based on first defining an equivalent quantum mechanical
problem and, more importantly, on coarse graining the interaction.  Of
course, above the inelastic threshold $s \ge 16 m_\pi^2$ one may
wonder what condition should be imposed instead of
just~\eqref{eq:unit}~\footnote{Usually the coupled channel unitarity
  condition is implemented instead. Typically analyses within such a
  setup leave out the ``small'' multiple production channels, $2\pi
  \to n \pi$, see e.g.~\cite{Ledwig:2014cla} and works cited
  therein.}. We will extend the coarse graining idea to the case with
inelasticities.

\section{Quantum mechanics Formalism}
\label{sec:QM}

\subsection{Relativistic equation}
\label{sec:el}

At the maximum CM energy we will be considering in this work $s_{\rm max}=2\;{\rm GeV}^2$, 
relativity and inelasticities are crucial physical ingredients
since firstly $\sqrt{\smax} \gg  m_\pi$ and secondly we can produce up to
$n=(\sqrt{\smax}-2 m_\pi)/m_\pi \sim 8$ pions as well as one $K \bar K
$ and $\eta \eta $ pair in the final state.  From a field
theoretical point of view, this could be solved by using a
multichannel Bethe-Salpeter equation for the several $2\pi$, $4\pi$,
$6\pi$, $8 \pi $, $K \bar K$ and $\eta \eta$ coupled channels, but it
would be an extremely difficult task, which has never been
accomplished to our knowledge. Even in the simplest elastic case the
off-shell ambiguities are present for calculations with a {\it
  truncated} kernel~\cite{Nieves:1998hp,Nieves:1999bx}. In order to
grasp the nature of the ambiguities, consider for instance the case
of $\pi^0(p_1) \pi^0(p_2) \to \pi^0(k_1)\pi^0(k_2)$ scattering, in the elastic
regime. The Bethe-Salpeter (BS)  equation reads,
\begin{align}
T_P (p,k) =& V_P (p,k) + \frac{i}{2} \int \frac{\diff^4 q}{(2\pi)^4}
V_P (p,q) \Delta(q_+) \Delta(q_-) T_P (q,k)  \nonumber \\ 
\label{eq:BS}
\end{align}
where $p=\frac{p_1-p_2}{2}$, $ k=\frac{k_1-k_2}{2}$, $P=p_1+p_2$ and $
q_\pm= P/2 \pm q$.  $\Delta (q_\pm) = 1/(q_\pm^2-m_\pi^2+i0^+)$ is the
free pion propagator and $V_P (p,k)$ and $T_P (p,k)$ stand for the
two-particle irreducible kernel or {\it potential} and the scattering
amplitude, respectively.  The factor $1/2$ comes from the scattering
of identical particles.

While the BS equation has been the subject of extensive research for
a given potential, the main point of~\cite{Nieves:1998hp,Nieves:1999bx} was the flexible
interpretation of the BS equation within $\chi$PT or more
generally within EFT. Indeed, while the potential $V_P (p,k)$ can
be organized as a power series $V_P (p,k)=V_P^{(2)} (p,k)+V_P^{(4)}(p,k)+ \dots $ 
with reference to the same expansion of
the scattering amplitude $T_P (p,k)=T_P^{(2)} (p,k)+T_P^{(4)} (p,k)+\dots$, 
it can be done only in at {\it on-shell mass} scheme, i.e. for
\begin{eqnarray}
T(s,t)=T_P (p,k),\quad p^2=k^2 = \frac{s}{4}-m_\pi^2, \quad P \cdot p = P \cdot k=0 .\nonumber \\  
\end{eqnarray}
Thus, there is an inherent ambiguity in the definition and form
of the potential, which has no consequences perturbatively but become
relevant in the solution of the BS equation~(\ref{eq:BS}) where
the off-shellness enters explicitly. 
This was mended in~\cite{Nieves:1998hp,Nieves:1999bx} by invoking an on-shell
scheme, namely considering {\it only} on shell intermediate states,
i.e.  $q^2 = s/4-m_\pi^2$ and $ P \cdot q=0 $, so that the on-shell
amplitude $T(s,t)$ depends only on the on-shell potential $V(s,t)$. 
Unfortunately, it also gives rise to pathologies in the
coupled channel case producing spurious singularities due to an
improper treatment of the crossed-channel exchanges~\cite{Ledwig:2014cla}. 
The present paper pretends to address
crossed-channel exchanges {\it without} invoking the on-shell scheme.

\subsection{Invariant mass and equivalent Schr\"odinger equation}

We will follow here the invariant mass
formulation~\cite{Allen:2000xy}~\footnote{These authors wondered if
  there was a way to promote non-relativistic fits of NN scattering to
  a relativistic formulation without refitting parameters. The answer
  is in the affirmative by just reinterpreting the CM momentum by its
  relativistic counterpart.}, already used for NN scattering with an
optical potential~\cite{Arriola:2016bxa,Fernandez-Soler:2017kfu}.
This is the simplest way of retaining relativity without solving a BS
equation but with a {\it phenomenological} optical potential that we
review here for completeness. The idea is to write the total squared
mass operator as
\begin{eqnarray}
  {\cal M}^2 = P^\mu P_\mu + W ,
\end{eqnarray}
where $W$ represents the (invariant) interaction, which can be
determined in the CM frame by matching in the non-relativistic limit
to a non-relativistic potential $V(\vec x)$. This yields for $\pi\pi$ scattering
after quantization ${\cal \hat M}^2 = 4(\hat p^2+m_\pi^2) + 4 m_\pi V $, with $\hat p = - i \nabla$.  
Thus, the relativistic equation can be written as ${\cal \hat M}^2 \Psi = 4 (p^2+m_\pi^2) \Psi $,
with $p$ the CM momentum,  i.e. as a non-relativistic Schr\"odinger equation
\begin{eqnarray}
(-\nabla^2 + m_\pi V ) \Psi = (s/4-m_\pi^2) \Psi \, . 
\label{eq:mass-sq}
\end{eqnarray}
This corresponds to the simple rule that one may effectively implement
relativity by just promoting the non-relativistic CM momentum to the
relativistic CM momentum. This {\it minimal relativity} ansatz is as
good as the more fundamental one based on the Bethe-Salpeter equation
as long as we use scattering data to determine the corresponding
potential rather than an {\it ab initio} determination (see
Ref.~\cite{Nieves:1999bx} for an in-depth discussion).

To take into account the inelasticity within the mass-squared
construction, we assume a local and energy-dependent phenomenological potential, 
$V(\vec r,s)= {\rm Re}\, V(\vec r,s) + i\,{\rm Im}\,V(\vec r,s)$,
 which could be obtained by fitting inelastic scattering data. 
Due to causality, the optical potential in the $s$-channel
satisfies a dispersion relation for each CM radial distance $r$  of the form
~\cite{cornwall1962mandelstam}
\begin{eqnarray}
{\rm Re}\,V(r,s)= V(r) + \frac1{\pi} \int\limits_{s_0}^\infty \diff s'\,\frac{{\rm
    Im}\,V(r,s')}{s'-s- i \epsilon}\,,
\label{eq:dr}
\end{eqnarray} 
where $\sqrt{s_0}= 4 m_\pi $ is the first $4\pi$ inelastic threshold
and $V(r)$ is an energy independent component. The complete potential
includes also the crossed $u$ channel component.
The simple looking equation~\eqref{eq:mass-sq}, together with the
fixed-$r$ dispersion relation~\eqref{eq:dr}, incorporates the
necessary physical ingredients present in any theoretical approach: 
relativity and inelasticity consistent with analyticity.

\subsection{Isospin and exchange potential}

The incorporation of isospin into the game is straightforward.
Rotational, isospin and particle exchange invariance requires the
representation of the potential to be given by
\begin{eqnarray}
  V(r) &=& \left[V_A (r) + V_B (r)\,\vec I_1 \cdot \vec I_2 + V_C (r)\,(\vec I_1 \cdot \vec I_2)^2 \right] (1+ {\cal P}_{12})\nonumber \\
  &=& V_{\rm D}(r) +V_{\rm X}(r),
\end{eqnarray}
where $V_{\rm D}$ and $V_{\rm X}$ stand for the direct and
exchange potential pieces, respectively.  
${\cal P}_{12}$ is the particle exchange operator,
which implements the Bose-Einstein symmetry 
and that can be factorized as ${\cal P}_{12} = {\cal P}_{x}{\cal P}_{I}$.
Moreover, for states with a well defined total isospin $\vec I = \vec I_1+ \vec I_2$, 
we can use the relation $\vec I_1 \cdot \vec I_2 = I(I+1)/2-2$ with $I=0,1,2$, 
so that $ {\cal P}_{I} = (-1)^I$.
In addition, for angular momentum eigenstates ${\cal P}_{x} = (-1)^J $, 
so that ${\cal P}_{12} = (-1)^{I+J}$. 
Therefore, in the isospin basis the potential can be decomposed as
\begin{eqnarray}
  V = \sum_{I=0,1,2} P_I V_I (1+ {\cal P}_{12}),
\end{eqnarray}
where we have introduced the  projection operators
\begin{eqnarray}
  P_0 &=& \frac{1}{3} (\vec I_1 \cdot \vec I_2-1) (\vec I_1 \cdot \vec I_2+1), \nonumber\\ 
  P_1 &=& -\frac{1}{2} (\vec I_1 \cdot \vec I_2-1) 
  (\vec I_1 \cdot \vec I_2+2),\nonumber\\ 
P_2 &=& \frac{1}{6} (\vec I_1 \cdot \vec I_2+1) (\vec I_1 \cdot \vec I_2+2),
\end{eqnarray}
fulfilling the orthogonality relations $P_I P_{I'} = \delta_{II'}P_I$.

In the partial wave representation, the exchange symmetry of the
potential is preserved by just solving the Schr\"odinger equation for
the direct potential for the allowed $IJ$ channels with
$(-1)^{J+I}=1$. In addition, for a spherically symmetric potential we
have the usual factorization of the wave
function~\cite{galindo2012quantum}
\begin{eqnarray}
  \Psi(\vec x) = \frac{u_l(r)}{r} Y_{l m_l} (\hat x),
\end{eqnarray}
where $Y_{l m_l} (\hat x)$ are the spherical harmonics and $u_l(r)$ is
the reduced wave function, fulfilling the radial Schr\"odinger equation
\begin{equation}\label{radial-schr}
  -u_l'' (r) + \left[ U(r) + \frac{l(l+1)}{r^2} \right] u_l(r) = p^2 u_l (r),
\end{equation}
where $U_I (r)=U_I (\vec  r)$ is the central potential with isospin $I$.
This equation is indeed regular at the origin~\footnote{We are assuming that at short distances the centrifugal barrier dominates, i.e. $r^2 U(r) \to 0 $. 
Nevertheless, chiral potentials diverge as $1/r^7$, as it is discussed below, and require special treatment if extended to the origin.}
\begin{equation} \label{eq:bc-short}
  u_l (r)  \to  r^{l+1}
\end{equation}
and it satisfies the asymptotic scattering condition at infinity 
\begin{eqnarray}
  u_l(r) \to \sin \left( p r -  \frac{l \pi}{2}+ \delta_l \right).
\end{eqnarray}
Thus, the partial wave expansion for the quantum mechanical scattering
amplitude with isospin I in the CM system is defined by:
\begin{equation}
f_I(p,\cos\theta)=\sum_{J=0}^\infty(2J+1)P_J(\cos{\theta}) \frac{\eta_{IJ}(s)e^{2i\delta_{IJ}(s)}-1}{2ip}.
\end{equation}

\subsection{Inverse scattering problem}

Although we will be determining the potentials from fits to phase
shifts, it is worth reminding that the inverse scattering problem
allows one to determine a local and continuous potential directly from
scattering data by solving for each partial wave either the
Gelfand-Levitan or Marchenko equations (see for
e.g. \cite{Chadan:1977pq} for a review).  It can be shown that for
holomorphic S-matrix functions both methods yield the same {\it local}
potential. While usually the discussion is conducted within a
non-relativistic setup, according to our discussion above, the
analysis can directly be overtaken and interpreted at the relativistic
level.

This inverse scattering approach was adopted in~\cite{Sander:1997br},
where a holomorphic S-matrix was used to parameterize the scattering
data. In that work it was found that the S and P-wave potentials have
a range around $0.25$ fm with strengths between $100-200$ GeV. Quite
remarkably they also found a barrier in the isoscalar S-wave and a
repulsive core in the isotensor S-wave. While this is a very
insightful and mathematically rigorous approach, this method requires
{\it exact} knowledge of the phase shifts at {\it all energies}.  In
practice, a meromorphic function is fitted up to a maximum energy
corresponding to a maximum momentum $p_{\rm max}$.  As we will see
below, this puts in practice a limitation to the resolution $\Delta r
\sim 1/p_{\rm max}$ with which the potential $V(r)$ may be determined,
so that a suitable coarse graining makes sense.

\section{The Chiral $\pi\pi$ local potential}
\label{sec:chpt}

In this section we outline the perturbative matching procedure between
quantum mechanic (QM) and quantum field theory (QFT) calculations
in order to determine the local and energy dependent chiral potential. 
The connection between the QFT and
QM scattering amplitudes is given by
\begin{equation}
T_I(s,t)=16 \pi \sqrt{s} f_I(p,\cos\theta) \,. 
\end{equation}
The potential appearing in this equation will be determined in
perturbation theory. For the quantum mechanical problem 
we have the Born series 
\begin{eqnarray}\label{eq:Born-series}
f(p, \cos \theta) &=&  
- \frac1{4 \pi} \int{\diff^3\vec r\,\, U (r) e^{-i \vec{q}\cdot\vec{r}}} \nonumber \\
&-& \int \diff^3 \vec r_1 d^3 \vec r_2 e^{i (\vec p' \cdot \vec r_2 - \vec p \cdot \vec r_1 )}
\frac{e^{i p r_{12}}}{r_{12}} U (r_1)  U(r_2) + \dots,\nonumber\\
\end{eqnarray}
where $r_{12}=\vert \vec r_1-\vec r_2\vert$, $\vec p$ and $\vec p'$ are the initial and final CM momenta, respectively,  and $\vec q=\vec p'-\vec p=2 \vec p\sin{(\theta/2)}$ is the momentum transfer.  
The potential $U(r)$ is directly defined from the two-particle irreducible states included in
the scattering amplitude.  We will define the potential through the
$t$-channel exchanges of the amplitude, so that crossing symmetry will
be incorporated exactly when simmetryzing the partial wave expansion~\footnote{
We have checked that one can either work either in the particle
or the isospin basis, the resulting potential is the same.}.
Moreover, in a coordinate space description, contact terms are
irrelevant as long as the field theoretical potential is not extended
to the origin $r=0$ since
\begin{equation}\label{pol-part}
\int{\diff q\,\,q^2\hat P(q^2)\,\frac{\sin{q r}}{qr}}= 0 \, ,  \qquad r > r_c > 0 \,, 
\end{equation} 
with $\hat P(q^2)$ a generic polynomial in $q^2$. 
Thus, any polynomial part of the scattering amplitude gives a vanishing
contribution to the long-range piece of the potential.  
Therefore, we will analyze only the effect of pion loop contributions on the $t$-channel.

\subsection{Leading order }

We will first discuss the lowest non-trivial order since it provides just contact terms.  
In the Born approximation, i.e. just taking the first term in~\eqref{eq:Born-series},  
the scattering amplitude just becomes the Fourier transform of the potential~\cite{galindo2012quantum}
\begin{eqnarray} \label{bpw}
f^{B}(p,\cos \theta) &=& -\frac{1}{4\pi}\int{\diff^3\vec r\,\, U(r) e^{-i \vec{q}\cdot\vec{r}}} \nonumber \\ 
&=& -\int\limits_0^\infty{\diff r\,\, r^2\,\, U(r)\frac{\sin{q r}}{q r}}\,,
\end{eqnarray}
where $q=2p\sin{(\theta/2)}$ is the momentum transfer. This equation can be inverted to give 
\begin{eqnarray} 
U(r,s) = - 4 \pi \int \frac{\diff^3 q}{(2\pi)^3} e^{i \vec{q}\cdot\vec{r}} f_B (\vec q)\,,
\label{bpw}
\end{eqnarray}
so in the Born approximation, the scattering amplitudes can be related to the potential by:
\begin{equation}
T_I(s,t)\big\vert_{B}= -4 \sqrt{s} \int{\diff^3\vec r\,\, U_I(r) e^{-i \vec{q}\cdot\vec{r}}}\,,
\end{equation}
where $T_I(s,t)\big\vert_{B}$ denotes the disconnected part of the
amplitude, i.e. contact terms and $t$-channel exchange.  In the same
way, the potential (defined in spatial coordinates) is defined from
the disconnected part of the amplitude by:
\begin{eqnarray}\label{pot}
U_I(r,s)&=&\frac{-1}{4 \sqrt s} \int{\frac{\diff^3\vec q}{(2\pi)^3}\,\, e^{i \vec{q}\cdot\vec{r}}}  \, \, 
T_I (s,-\vec q^2)\big\vert_{B}  \nonumber \\ &=& \frac{-1}{8  \pi^2 \sqrt s}
\int\limits_0^\infty {\diff q\,\,q^2\, \, T_I (s,t)\big\vert_{t=-q^2} \frac{\sin{q r}}{qr} }\,.
\end{eqnarray}
Using the $\chi$PT lowest order amplitudes~\cite{Gasser:1983yg} we get 
\begin{eqnarray}\label{eq:LOpot}
U_0^{(2)} (r,s) &=& \frac{-1}{4 \sqrt s} \frac{m_\pi^2-2s}{2 f^2} \delta^{(3)} (\vec r)\,, \nonumber\\
U_1^{(2)} (r,s) &=& \frac{-1}{4 \sqrt s} \frac{4 m_\pi^2- 2 \nabla^2-s}{2 f^2} \delta^{(3)} (\vec r)\,,\nonumber\\
U_2^{(2)} (r,s) &=& \frac{-1}{4 \sqrt s} \frac{s-2 m_\pi^2}{2 f^2} \delta^{(3)} (\vec r)\,.
\end{eqnarray}
These algebraic manipulations are purely formal, and in fact is
unspecified what is the meaning of solving the wave equation with
these highly singular potentials. Already at this level, we can see the
need of introducing a regularization~\footnote{There is a conservation
of difficulty principle here, 
one could stay in momentum space in which case the potential is well defined, 
but the scattering equation is UV divergent.}.

\subsection{Next-to-leading order}

\begin{figure}
\centering
\includegraphics[scale=0.7]{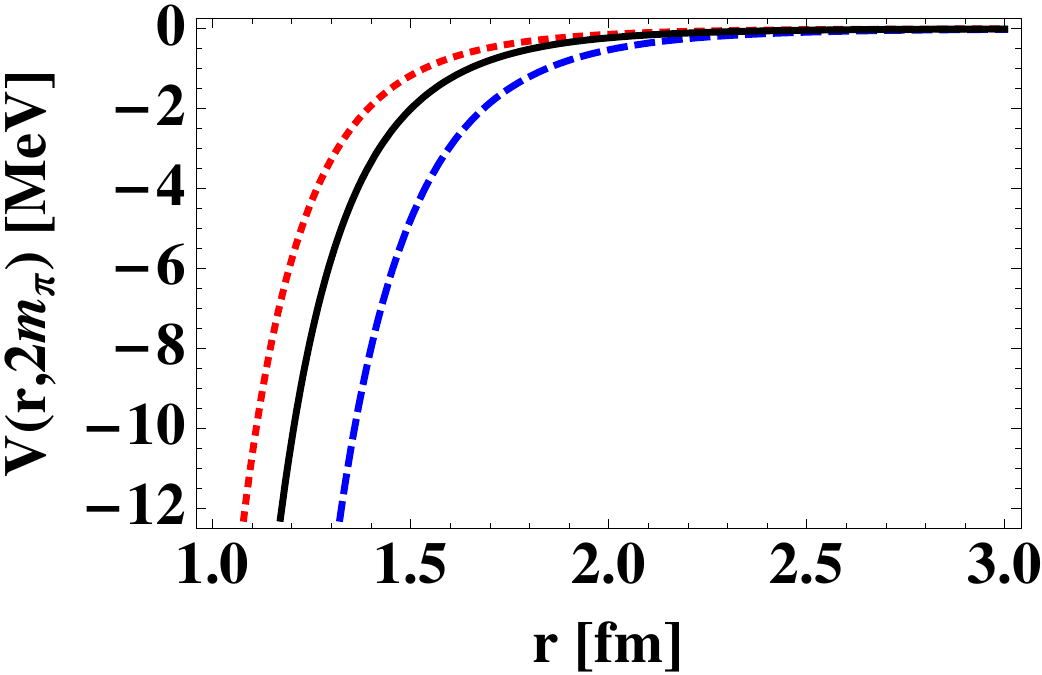} 
\caption{Color online: chiral $2\pi$ exchange potentials $V_{IJ}$ at threshold $\sqrt{s}= 2 m_\pi$ as a function of the distance for $IJ=00$ (dashed), $IJ=11$ (full) and $IJ=20$ (dotted) channels.}\label{Fig:Upipi}
\end{figure}

The NLO contribution becomes more cumbersome. 
Firstly, we take the potential to be expanded as 
\begin{eqnarray}
U_I(r,s)=U_I^{(2)} (r,s) + U_I^{(4)} (r,s) + \dots\,, 
\end{eqnarray}
so we get the matching condition
\begin{eqnarray} \label{eq:NLO}
-\frac{T^{(4)}_I (s,t)}{4\sqrt{s}} &=&  \int{\diff^3\vec r\,\, U_I^{(4)}(r,s) e^{-i \vec{q}\cdot\vec{r}}} \nonumber \\ 
&-& \int \diff^3\vec r_1 d^3 \vec r_2 e^{i (\vec p' \cdot \vec r_2 - \vec p \cdot \vec r_1 )}
\frac{e^{i p r_{12}}}{r_{12}} U^{(2)} (r_1,s)  U^{(2)} (r_2,s) \nonumber\\
&+& \dots .
\end{eqnarray}
Due to the Dirac delta functions in the LO potential~\eqref{eq:LOpot},
we have a divergence for the real part of $ e^{i p r_{12}}/r_{12} $,
albeit it can be absorbed in the real part of the NLO potential
$U^{(4)}(r,s)$.  Besides, the non-polynomial pieces in $T^{(4)}$
amplitude corresponding to the t-channel exchange can generally be
written as
\begin{equation}\label{2pi}
T_I(s,t)\big\vert_{\rm 2\pi}=P(s,t)J(t)\,,
\end{equation}
where $P(s,t)$ is a polynomial in both $t$ and $s$,
which analytical expression can be read from~\eqref{eq:a4chpt}, 
and $J(t)$ denotes the one-loop $2\pi$ function.
In order to integrate this amplitude, we will take advantage of the
analytic structure of the loop function $J(t)$, which is analytic in
the whole complex plane but for a cut above $4m_\pi^2$ with a discontinuity, 
${\rm disc}\, J(t)=2\,i\,{\rm Im}\, J(t)=2\pi\,i\sigma(t)/16\pi$, 
with $\sigma(t)$ the phase-space factor defined in~\eqref{eq:sigma}.
Thus, up to subtractions one finds the dispersion relation
\begin{equation}
\label{Jdisp}
J(t)= \frac{t-4m_\pi^2}{16\pi^2}
\int\limits_{4m_\pi^2}^{\infty}{\diff t^\prime\frac{\sigma(t^\prime)}{(t^\prime-t)(t^\prime-4m_\pi^2)}} + {\rm
    C.T.},
\end{equation}
where C.T. is a subtraction constant that can be fixed by setting the value of 
$J(4m_\pi^2) = 1/8\pi^2 $. Likewise we have
\begin{equation}\label{2pi}
P(s,t)J(t) =  \frac{t-4m_\pi^2}{16\pi^2}
\int\limits_{4m_\pi^2}^{\infty}{\diff t^\prime\frac{ P(s,t^\prime) \sigma(t^\prime)}{(t^\prime-t)(t^\prime-4m_\pi^2)}} + {\rm
    C.T.}\,.
\end{equation}
Thus, taking into account the Yukawa integral 
\begin{eqnarray}
\int \frac{\diff^3 q}{(2\pi)^3} \frac{e^{i \vec q \cdot \vec r }}{q^2 + \mu^2} = 
\frac1{4\pi} \frac{e^{-\mu r}}{r}
\end{eqnarray}
and the inversion of~\eqref{eq:NLO}, the NLO potential becomes
\begin{equation}
U_I^{(4)}(r,s)=\int\limits_{2 m_\pi}^\infty \diff\mu  \rho^I(\mu,s) \frac{e^{-\mu r}}{r}+ {\rm C.T.},
\label{eq:spec-pot}
\end{equation}
where $t=\mu^2$ and the spectral function $\rho_I(\mu,s)$ is defined
as
\begin{equation}\label{eq:rhoI}
\rho^I(\mu,s)=\frac{-1}{128\pi^3\sqrt s} P_I(s,\mu^2)(\mu^2-4m^2)^{1/2} \, , 
\end{equation}
with $P_I(s,\mu^2)$ polynomials in $s$ and $\mu^2$ of fourth degree
(see Appendix~\ref{sec:pots-amps}) and C.T. map into contact terms,
which are distributions at the origin and hence vanish elsewhere. All
necessary integrals can be obtained from the general integral valid
for $r>0$,
\begin{eqnarray}
\int\limits_{2m_\pi}^{\infty}{\diff\mu (\mu^2-4m_\pi^2)^{n/2} \frac{e^{-\mu r}}{r}} &=&
\frac{ n 2^n m_\pi^{\frac{n+1}{2}} }{\sqrt{\pi } r^{\frac{n+3}2} }
   \Gamma \left(\frac{n}{2}\right) K_{\frac{n+1}{2}}(2 m_\pi r)\,, \nonumber \\ 
\end{eqnarray}
with $K_n(x)$ the modified Bessel function of order $n$ and
$\Gamma(z)$ the Euler's Gamma.  Polynomials in $\mu$ can be generated
from derivation with respect to $r$.  The chiral potentials obtained
directly from the spectral representation~(\ref{eq:spec-pot}), read
then
\begin{eqnarray}\label{eq:chiral-pot}
U_{0}(r,s)  &=&\frac{\left(-23 m_\pi^5 r^2-200 m_\pi^3\right) K_1(2 m_\pi r) }{128 \pi ^3 f^4 r^4
   \sqrt{s}}  \nonumber \\ &+&\frac{\left(-24 m_\pi^4 r^2-m_\pi^2 r^2 s-100 m_\pi^2\right) K_2(2 m_\pi r)  }{32 \pi ^3 f^4 r^5
   \sqrt{s}}, \nonumber\\
U_{1}(r,s)  &=&\frac{\left(-13 m_\pi^5 r^2-40 m_\pi^3\right) K_1(2 m_\pi r)}{128 \pi ^3 f^4 r^4
   \sqrt{s}} \nonumber \\ &+&\frac{\left(-18 m_\pi^4 r^2-m_\pi^2 r^2 s-40 m_\pi^2\right) K_2(2 m_\pi r)}{64 \pi ^3 f^4 r^5
   \sqrt{s}},  \nonumber\\
U_{2}(r,s)  &=&
\frac{\left(-17 m_\pi^5 r^2-80 m_\pi^3\right) K_1(2 m_\pi r)}{128 \pi ^3 f^4 r^4
   \sqrt{s}} \nonumber \\ &+& \frac{\left(-30 m_\pi^4 r^2+m_\pi^2 r^2 s-80 m_\pi^2\right) K_2(2 m_\pi r)}{64 \pi ^3 f^4 r^5
   \sqrt{s}}.
\end{eqnarray}
From a more general point of view, these potentials play for the
$\pi\pi$-system the role of relativistic van der Waals interactions
(see~\cite{Feinberg:1989ps} for a review in the atomic case) and
hence display their characteristic features: they are attractive and 
diverge at short distances as $\sim 1/( r^7 f^4 \sqrt{s}) $, i.e.
\begin{eqnarray}\label{eq:pot-short-aprox}
U_0 (r,s) &=& -\frac{25}{16 \pi ^3 f^4 r^7 \sqrt{s}} + \dots\,, \nonumber\\ 
U_1 (r,s) &=& -\frac{5}{16 \pi ^3 f^4 r^7
   \sqrt{s}} 
 + \dots\,,  \nonumber\\
U_2 (r,s) &=& -\frac{5}{8 \pi ^3 f^4 r^7 \sqrt{s}}
 + \dots \, , 
\end{eqnarray}
and have the expected exponentially suppressed long distance behavior
$\sim e^{-2 m_\pi r}$, namely
\begin{eqnarray}
U_0 (r,s) &=& -\frac{23 m_\pi^{9/2} e^{-2 m_\pi r}}{256 \pi ^{5/2} f^4 r^{5/2}
  \sqrt{s}} + \dots\nonumber \\ 
U_1 (r,s) &=& -\frac{13 m_\pi^{9/2} e^{-2 m_\pi r}}{256 \pi ^{5/2} f^4 r^{5/2}
  \sqrt{s}} + \dots \nonumber\\
U_2 (r,s) &=&
-\frac{17 m_\pi^{9/2} e^{-2 m_\pi r}}{256 \pi ^{5/2} f^4 r^{5/2}
  \sqrt{s}} + \dots \, .  
\end{eqnarray}
In Fig.~\ref{Fig:Upipi} we show the threshold combination $V_{I}(r, 4
m_\pi^2) \equiv m_\pi U_{I}(r, 4 m_\pi^2)$ and, as we can see, they
are attractive at {\it all} distances. The energy dependence generates
a repulsive effect for increasing values of $s$, i.e.
\begin{eqnarray}
\frac{\partial U_I(r,s)}{\partial s} > 0 \qquad s  > 4 m_\pi^2. 
\end{eqnarray}

On the lattice, the energy dependence of the potential is generated
from the Nambu-Bethe wave function~\cite{Kurth:2013tua}.  As already
stated in the introduction, the $\pi\pi$ potential in the $I=2$
channel has been computed on the lattice by the HAL QCD
collaboration~\cite{Kurth:2013tua,Kawai:2017goq} for $a \approx 0.12$
fm on a $163 \times 32 $ lattice and with a pion mass $m_\pi \approx
870$ MeV.  For these pion masses the value the chiral potentials
in~\eqref{eq:chiral-pot} become smaller than for the physical case
depicted in Fig.~\ref{Fig:Upipi}.  In addition, the HAL QCD lattice
potential presents a repulsive core below $0.5$\,fm. This is a feature
one can not obtain using the chiral potentials
in~\eqref{eq:chiral-pot} which display strong short distance
singularities. This fact already suggests that they can not be used at
arbitrary short distances. At this point it is worth stressing that
both the lattice as well as the present approach based on chiral
perturbation theory {\it assume} point-like sources, a unrealistic
feature. In the next sections we analyze this topic in more detail.

\section{Renormalization}
\label{sec:renor}

The renormalization of non-perturbative amplitudes is a tricky matter,
particularly with the highly power-divergent kernels deduced from $\chi$PT
(see for e.g.~\cite{Nieves:1998hp,Nieves:1999bx} for a discussion within
the Bethe-Salpeter framework in momentum space).  The chiral potential
deduced in coordinate space by a perturbative matching procedure
presents an energy dependence. In this section we show how the
scattering amplitude stemming from the iteration of the two-pion exchange (TPE) chiral
potentials in~\eqref{eq:chiral-pot} can be renormalized from a coordinate space
point of view if the energy dependence is ignored by taking, say, the
threshold value $\sqrt{s}=2 m_\pi$. Hence, we will implement as
renormalization conditions the scattering amplitude at threshold. 
We will see that, while this is a mathematically viable approach, it fails
phenomenologically. Furthermore, the consideration of energy dependence
will prevent a sensible non-perturbative renormalization procedure.
For large values of the coordinate space cut-off $r_c \gtrsim 1.2
{\rm fm}$, the results will not be affected by taking either $U_I(s,r)$ or $U_I(4 m_\pi^2,r)$.

\subsection{Discussion}

One of the advantages of the energy dependent coordinate space
representation of the potential is that off-shell and field
reparameterization ambiguities manifest as contact interactions at the
origin. Thus, they reflect the cut-structure of the amplitude, which is hence
unambiguously defined. This is unlike their momentum space counterpart,
where both polynomial and cut contributions are treated on equal
footing~\cite{Nieves:1999bx}.

On the other hand, a difficulty with the chiral $\pi\pi$ potentials in
the previous section is that they become singular at short distances.
Hence, the solution of the Schr\"odinger equation is not well defined in
a conventional sense, since the short distance behavior is not
dominated by the centrifugal barrier and the regular solution given
in~\eqref{eq:bc-short} is not suitable.  An early review on the
subject can be found in~\cite{Frank:1971xx}.  Singular potentials are
commonplace within EFT and finite solutions exist in a renormalization
sense within well specified conditions, as discussed at length in the
NN scattering
case~\cite{PavonValderrama:2005gu,PavonValderrama:2005wv,PavonValderrama:2005uj}.
Applications for
$\alpha\alpha$-scattering~\cite{Arriola:2007de,RuizArriola:2008cy} and
atom-atom scattering~\cite{Cordon:2009wh,RuizArriola:2009wi} are well
documented by now (see for e.g.~\cite{RuizArriola:2007wm} for a sucint and
pedagogical presentation).  While these renormalized solutions
represent theoretically a viable solution to the problem, we will
consider here a more phenomenological interpretation by introducing a
short distance cut-off $r_c$, which value reflects short distance
effects not taken into account in the derivation of the chiral
$\pi\pi$ potential~\footnote{The renormalization procedure would
  correspond to take $r_c \to 0$ while keeping scattering lengths
  fixed.  This consistent choice assumes point-like hadrons.}.  This
leaves undefined the short distance dynamics.

The energy dependence of the potential takes into account retardation
effects.  This can be seen as follows; if the potential is given as a
function of the difference of two space-time causally related events
$K(x-x')$ then we have
\[
\int\limits_0^\infty e^{-\sqrt{s}\,t} \diff t = \frac1{\sqrt{s}}\,,
\]
explaining the $1/\sqrt{s}$ factor in~\eqref{eq:chiral-pot}.  In the
case of ``heavy pions'' or very low energies $ p \ll m_\pi$, we can
take $\sqrt{s} \sim 2 m_\pi$ and work within a non-relativistic
approximation. A compelling consequence of causality is the
verification of dispersion relations in the complex energy plane.  We
will dedicate Section~\ref{sec:analy} to prove that the right
analytical properties hold for the partial wave amplitudes.

The spectral representation~\eqref{Jdisp} suggests the use of a
spectral regularization consisting of introducing a cut-off $\Lambda$
at a given value of $\mu$, so that the potential at short distances
becomes regular.  We find that for $\Lambda > 5 m_\pi$ the
regularization quenches the potential for $r < 1/m_\pi$. We explore 
this issue in more detail in Sect.~\ref{sec:spec}.

\subsection{The short distance cut-off and boundary condition
regularization}

\begin{figure*}
\centering
\includegraphics[scale=0.5]{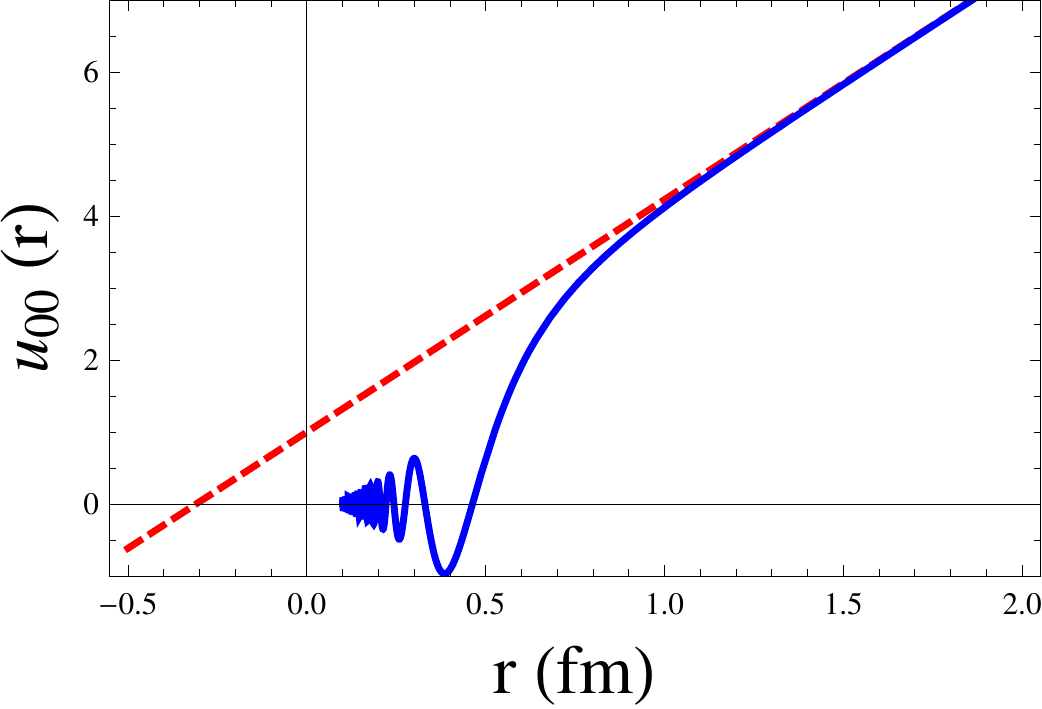}
\includegraphics[scale=0.5]{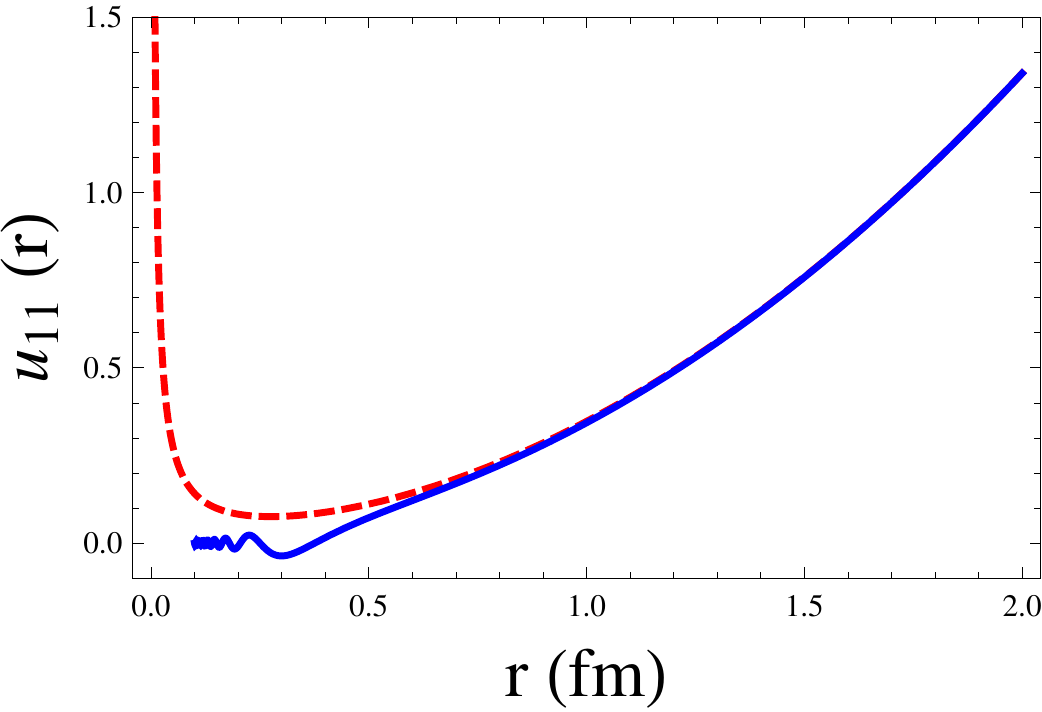}
\includegraphics[scale=0.5]{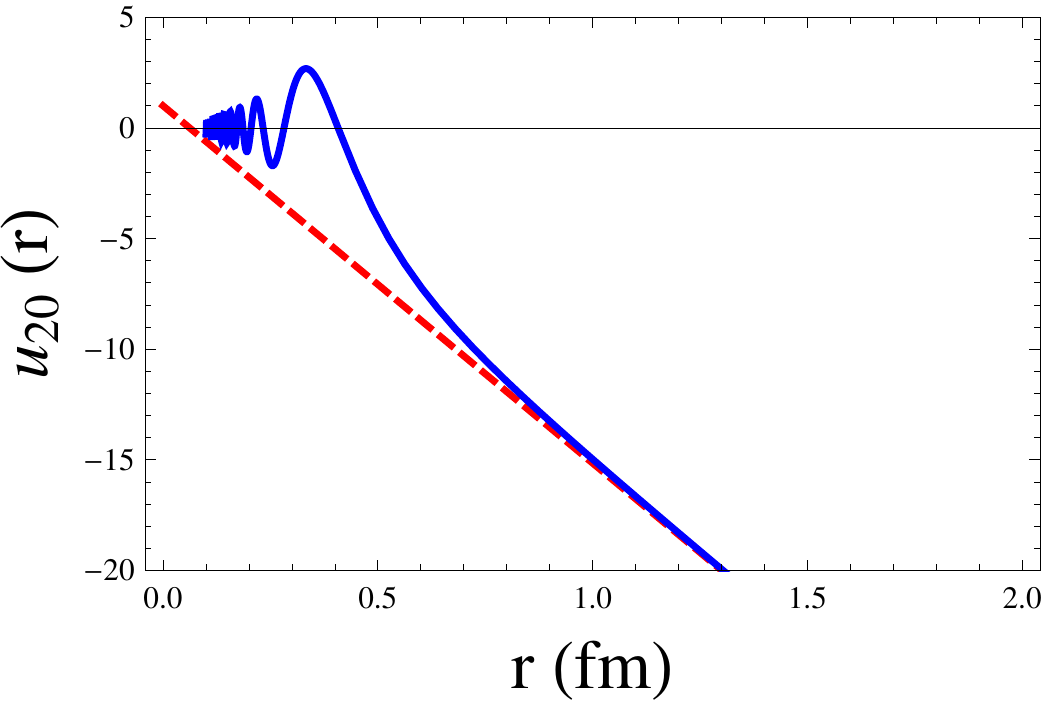}
\caption{Color online: zero momentum integrated-in wave functions
  (solid) from large distances using the scattering length as input
  and the chiral $2\pi$ exchange potentials $V_{\pi\pi}$ at threshold
  $\sqrt{s}= 2 m_\pi$ as a function of the distance for $I=0$ (left) ,
  $I=1$ (middle) and $I=2$ (right). We also draw the asymptotic zero
  energy wave function (dashed) for comparison. }\label{Fig:uzero}
\end{figure*}

One way to analyze the range of validity of the chiral potentials is
to discuss the zero momentum scattering, $p=0$. Thus, we solve the
Schr\"odinger equation using the asymptotic solutions at zero momentum
(we discard the isospin label here),
\begin{eqnarray}
u_l (r) = \frac{(2l-1)!!}{r^{l}} - \frac{r^{l+1} }{\alpha_l (2l+1)!!}\,, 
\end{eqnarray}
with $\alpha_l \equiv - \lim_{p \to 0} \delta_l(p) /p^{2 l+1}$ and
integrating inward. The result is illustrated in Fig.~\ref{Fig:uzero}
where we see that the zero momentum wave function presents oscillations
at short distances. This behavior can qualitatively be understood if
we write the potential at short distances in the form
\begin{eqnarray}
U(r) &=& - \frac1{R^2}\left(\frac{R}{r} \right)^7 \,, \label{eq:pot-short}
\end{eqnarray}
where $R \propto 1/f $ is the van der Waals scale, 
which in our case and from~\eqref{eq:pot-short-aprox} takes the values  
$R = 0.94,\,0.68,\,0.79$ fm for $I=0,1,2$, respectively. Actually, for $r \ll R $ the
centrifugal barrier,  $l(l+1)/r^2$ , can be neglected and a semi-classical
approximation holds since $\lambda'(r) \ll 1$, where $\lambda(r)$ is the local
wavelength, $\lambda(r) \equiv 1/k(r)=1/\sqrt{-U(r)}$, with $k(r)$ the
local wavenumber. Thus, one finds $\lambda'(r) = 7/2\,(r/R)^{5/2} \ll 1$
for~\eqref{eq:pot-short}.  In this case, the WKB wave function
reads~\cite{galindo2012quantum}
\begin{eqnarray}
u(r) &\approx&  \frac{1}{\left[k(r) \right]^\frac14} \sin \left[ \int
  k(r) \diff r \right] \nonumber \\  
&\to& r^{7/4} \sin \left[ (R/r)^{5/2} + \varphi
  \right] \ , 
\end{eqnarray}
where $\varphi $ is an arbitrary phase, 
which is fixed by the long distance solution.  

\begin{figure*}
\centering
\includegraphics[scale=0.7]{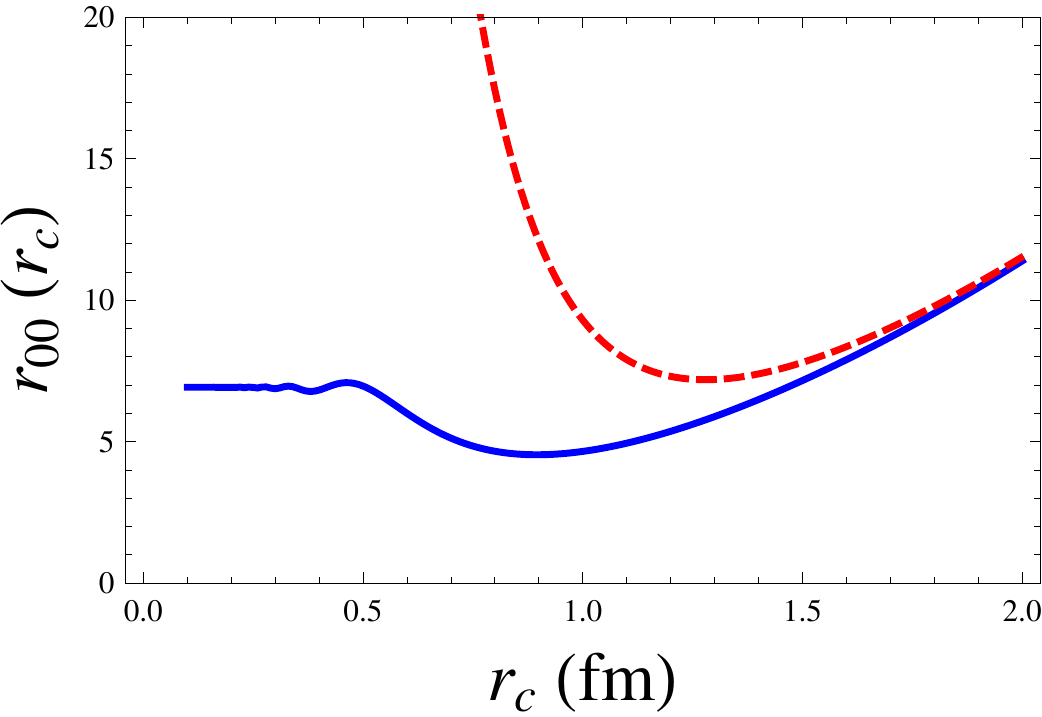} \hskip.3cm 
\includegraphics[scale=0.7]{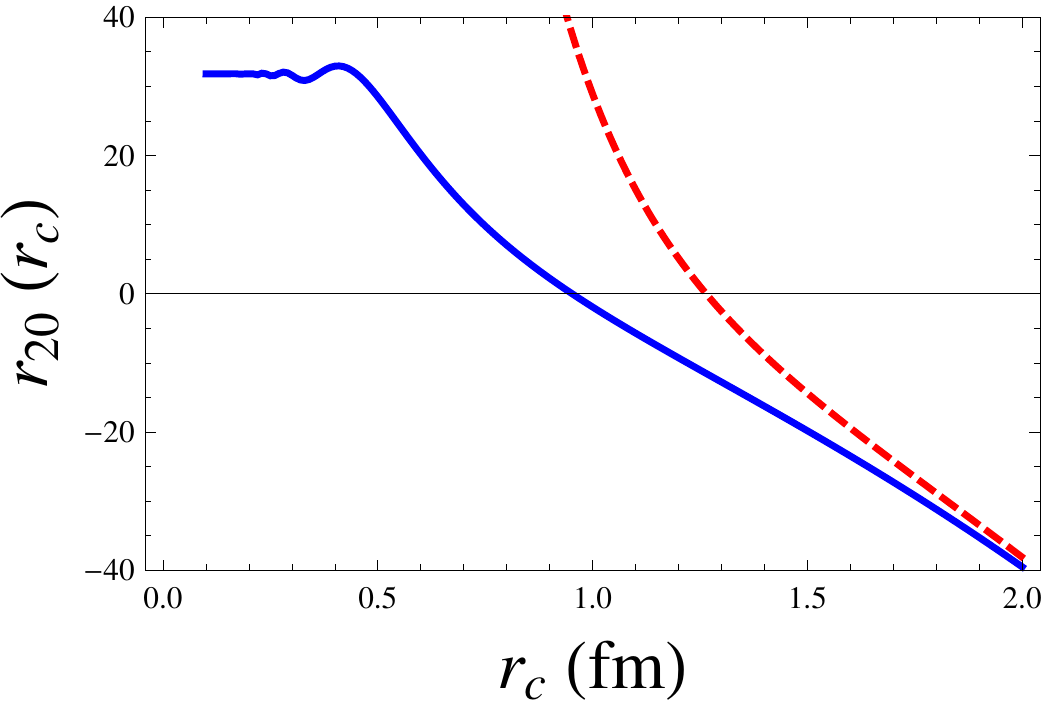}
\caption{Color online: renormalized effective range for the
  integrated-in large-distance wave functions (solid),
  as a function of the  distance for $I=0$ (left) and $I=2$ (right) 
  It is obtained using the scattering length as input and the chiral $2\pi$ exchange potentials
  $V_{\pi\pi}$ at threshold $\sqrt{s}= 2 m_\pi$. 
  In addition, we also plot the energy dependence case (blue-dashed), 
  which leads to a divergent value at $r \to 0$.}\label{Fig:reff}
\end{figure*}

According to the oscillation theorem~\cite{galindo2012quantum}, the
number of nodes of the wave function at zero momentum corresponds to the
number of bound states. However, as we know, there are no bound states in the
$\pi\pi$ scattering case.  Thus, if we want to avoid these spurious solutions, 
we cannot remove the cut-off completely, 
but it should be larger than the outmost right node of the wave functions 
depicted in Fig.~\ref{Fig:uzero}, which  are located at $r\sim0.5,\,0.4,\,0.4$ fm for $I=0,1,2$, respectively.
In practice, we will take a cut-off slightly above the van der Waals scale. 
 Note that there is {\it a priori} no other reason to discard the chiral potential down to
these scales. In fact, phase shifts in any partial wave $\delta_l(p)$
are convergent when the short distance cut-off goes to zero, {\it
  provided} the scattering length $\alpha_l $ is fixed. 
This corresponds to a renormalization program already developed in previous
studies~~\cite{PavonValderrama:2005gu,PavonValderrama:2005wv,PavonValderrama:2005uj,Arriola:2007de,RuizArriola:2008cy,Cordon:2009wh,RuizArriola:2009wi,RuizArriola:2007wm}
that will not pursued any further here. 
The upshot of these considerations in the $\pi\pi$ case is that generally one needs 
a cut-off $r_c > 0.5$ fm to prevent the appearance of unphysical bound states generated
by the TPE potential. However, for these kind of short distance 
attractive singularity $U(r) \propto -1/r^7$ finite $r_c$
effects are minor in physical observables if the energy dependence of the potential is neglected.  
Take for instance the effective range defined in~\eqref{eq:effective-range-exp} 
and depicted in Fig.~\ref{Fig:reff}. 
As one can see, the chiral potential and the
scattering length lead to a finite result at short distances, $r < R$.
Up to minor oscillations, it provides values which {\it differ} from the
experimental ones. In Section~\ref{sec:elem}, we will see that this turns
out to be {\it much smaller} than the elementarity radius, $r_e \sim 1.2$ fm.  
Moreover, there is no finite cut-off $r_c$ which reproduces
the experimental values $r_{00}=-8.08\,{\rm fm}$ and $r_{20}=131.4\,{\rm fm}$.

In the previous discussion the energy dependence of the potential
was neglected. On the one hand, if we take the energy dependence into account, 
we see in Fig.~\ref{Fig:reff} a quite different trend at short distances,  
namely the effective range is {\it not}
convergent when the cut-off is removed, i.e. for $r_c \to 0$. On the other
hand, we also see that for $r_c \gtrsim 1.2 {\rm fm}$ this energy
dependence in the chiral potential becomes irrelevant.

\section{Effective elementarity}
\label{sec:elem}

As we have seen, even at the perturbative level we must introduce a
short distance cut-off, $r_c$. In this section we elaborate on
sensible choices of this cut-off on the light of the onset of
effective elementarity and its corresponding radius $r_e$. 
This is the scale above which particles interact as if they were pointlike.
Thus, they can be taken as elementary, so that a hadronic field can be attached to the particle.

\subsection{Hadron sizes and form factors}

Hadrons have a finite size, which is usually characterized by their form
factors and corresponding radii. Roughly speaking, we expect that
for two hadrons of size $r_1$ and $r_2$ they will not overlap at
relative separations above the mean average distance $(r_1+r_2)/2$,
and their interaction will correspond to that of two elementary particles. 
However, hadronic sizes obtained from, say, electroweak
form factors are specific on that particular process, as we will
illustrate below. In fact, in order to describe $\pi \pi$ interactions
we are interested on the corresponding effective elementary size,
$r_e$, as {\it seen} by the interaction. Unfortunately, without a
microscopic calculation, there is no way to know this size in the
case of strong interactions.  Nonetheless, for our discussion on the
relevant scales it is important to estimate {\it a priori} the
separation distance $r_c$ between the inner and outer pieces of the
potential, as this determination has an impact on the minimal number of
fitting parameters. Quite generally, this separation distance should be
larger than the elementarity radius, $r_e \le r_c$, and the optimal
choice would be to take both distances equal.
\begin{figure*}
\centering
\includegraphics[scale=0.7]{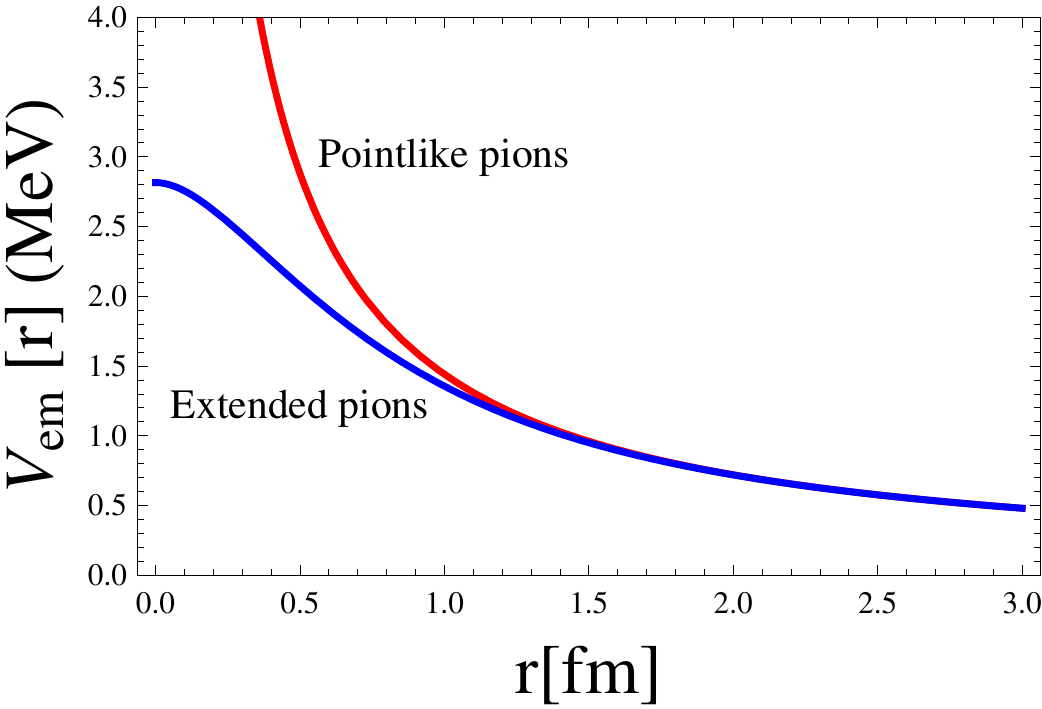} \hskip.5cm 
\includegraphics[scale=0.7]{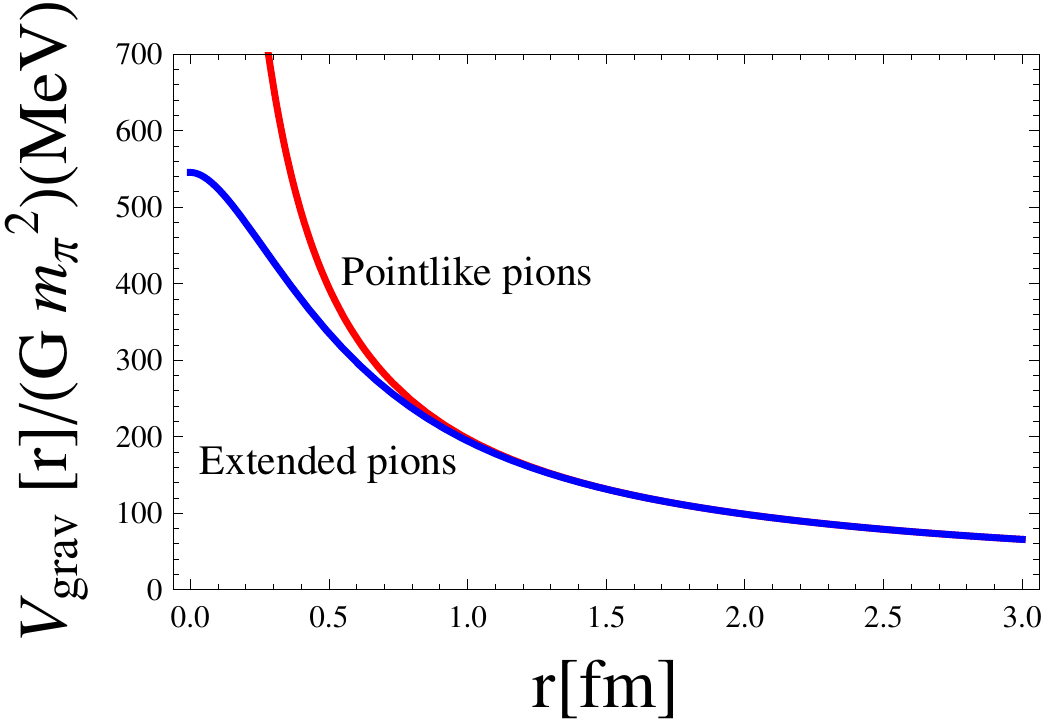}
\caption{Color online: effective elementarity of the pion. We show
  the Electrostatic (left panel) and gravitational (right panel)
  potentials for $\pi\pi$ interactions and compare the point-like
  limit (red line) with the finite size case (blue
  line).}\label{Fig:VMD}
\end{figure*}

Since the pion size is around 1 fm, it is natural to expect that the
inelastic non-perturbative description of the interaction will take
place in the region below $2$ fm.

In order to illustrate our point, let us consider the implications of
elementarity for electric and gravitational interactions between
pions. The electromagnetic pion form factor reads
\begin{eqnarray}
\langle \pi^+ (p') | J^\mu(0) | \pi^+ (p) \rangle = (p'^\mu + p^\mu) F_{\rm em} (q)\,,
\end{eqnarray}
whereas the gravitational form factor is given by
\begin{eqnarray}
\langle \pi^b(p') \mid \Theta^{\mu \nu}(0) \mid \pi^a(p) \rangle &=& \frac{1}{2}{\delta^{ab}}  \nonumber \\ 
\times \left[ (g^{\mu \nu} q^2- q^\mu q^\nu) 
\Theta_1(q^2) \right. &+& \left. 4 P^\mu P^\nu \Theta_2(q^2) \right]\,,
\end{eqnarray}
where $q=p'-p$, $P=p'+p$ and $\Theta_2(q^2)$ stands for the Heaviside step function.
In the Breit frame, where there is no energy transfer, 
form factors can be interpreted as the Fourier transform of a
density~\cite{Ernst:1960zza}
\begin{equation}
F(q)=\int \diff^3r\, e^{i q\cdot r}\rho(r)\,,
\end{equation}
so that the charge form factor determines the charge density and the
gravitational form factor the mass density of the pion. 

Quite generally, form factors are matrix elements of local operators between hadronic states. 
For the case of operators with well-defined $J^{PC}$ quantum numbers, 
a generalized meson dominance is expected to
work in harmony with the high energy behavior deduced from QCD
counting rules (see for e.g.~\cite{Masjuan:2012sk} for a thorough
discussion and comparison with lattice QCD data). For example, in the
electromagnetic case, it can be parameterized according to vector
meson dominance as,
\begin{equation}
F_{\rm em}(q)=\int \diff^3 r \, e^{i q\cdot r} \, \rho(r) = \frac{m_\rho^2}{m_\rho^2+q^2}\,,
\end{equation}
where we have kept only the $\rho$ meson with $m_\rho = 0.77\,{\rm GeV}$, 
as we are only interested in the long distance properties.

\subsection{Point-like vs extended particles interactions}

According to the previous discussion, the electrostatic potential can be written as
\begin{eqnarray}
V^{el}_{\pi\pi}(r) &=&\int\diff^3r_ 1d^3r_2 \frac{\rho(r_1)\rho(r_2)}{\vert \vec r_1-\vec r_2 -\vec r\vert}\nonumber \\
&=& \int \frac{\diff^3 q}{(2\pi)^3} \frac{4\pi}{q^2}| F_{\rm em} (q) |^2
e^{i \vec q \cdot \vec r} \nonumber\\ &=& \frac{1}{r} - e^{-m_\rho r} \left[ \frac12
  m_\rho + \frac1{r}\right]\nonumber\\
&\sim& \frac{1}{r} \quad {\rm for}\quad r> r_e\,, 
\end{eqnarray}
which is depicted in the left panel of Fig.~\ref{Fig:VMD} and reflects
that for $r>1.2\sim1.5$~fm, pions start to interact as expected from
point-like particles~\footnote{The value at $r=0$ corresponds to the
  electromagnetic mass difference between charged and neutral pions, $
  V_{\pi\pi}^{\rm el}(0)=\Delta m_\pi|_{\rm EM} = m_{\pi^+}-m_{\pi^0}
  |_{\rm EM} = e^2 m_\rho/2= 2.8\,{\rm MeV}$, which provides a reasonable value.}. 
In the gravitational case, a good description is found with the tensor
$f_2(1270)$ meson~\cite{Masjuan:2012sk} with the mass scale given by
$m_f = 1.2\,{\rm GeV}$, so that
\begin{eqnarray}
V^{g}_{\pi\pi}(r) &=&\frac{1}{r} - e^{-m_f r} \left[ \frac12
  m_f + \frac1{r}\right]\nonumber\\ &\sim& \frac{1}{r} \quad {\rm for}\quad r> r_e\,. 
\end{eqnarray}
In this case, right panel of Fig.~\ref{Fig:VMD}, the interaction
becomes elementary at $r_e \sim 1\,{\rm fm}$, a shorter scale.
The previous discussion illustrates our point, namely effective
elementarity depends on the particular process.

Before proceeding further, it should be noted that within a more
microscopic point of view, for instance a cluster quark model, these
are not the only possible contributions to the interactions between
pions (see e.g.
\cite{Barnes:1991em,Swanson:1992ec,Barnes:2000hu}). Actually, in a
Hartree-Fock approximation they would correspond to the direct
interaction (Hartree) term. In addition, one also has the exchange
(Fock) term where the quarks inside different pions are interchanged.
These terms are genuinely non-local at short distances, so that they
are exponentially suppressed at long distances. Therefore, they are
expected to contribute {\it below} the elementarity radius, $r_e$, and
hence can be regarded as finite size effects.
\begin{figure*}[tb]
\includegraphics[scale=0.55]{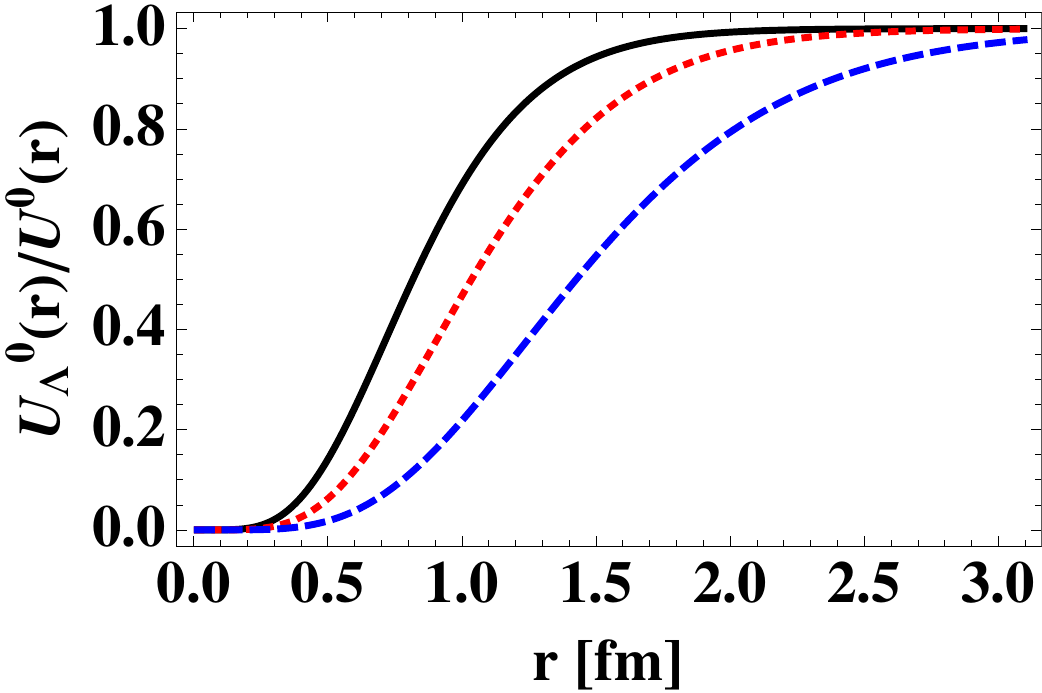}
\includegraphics[scale=0.55]{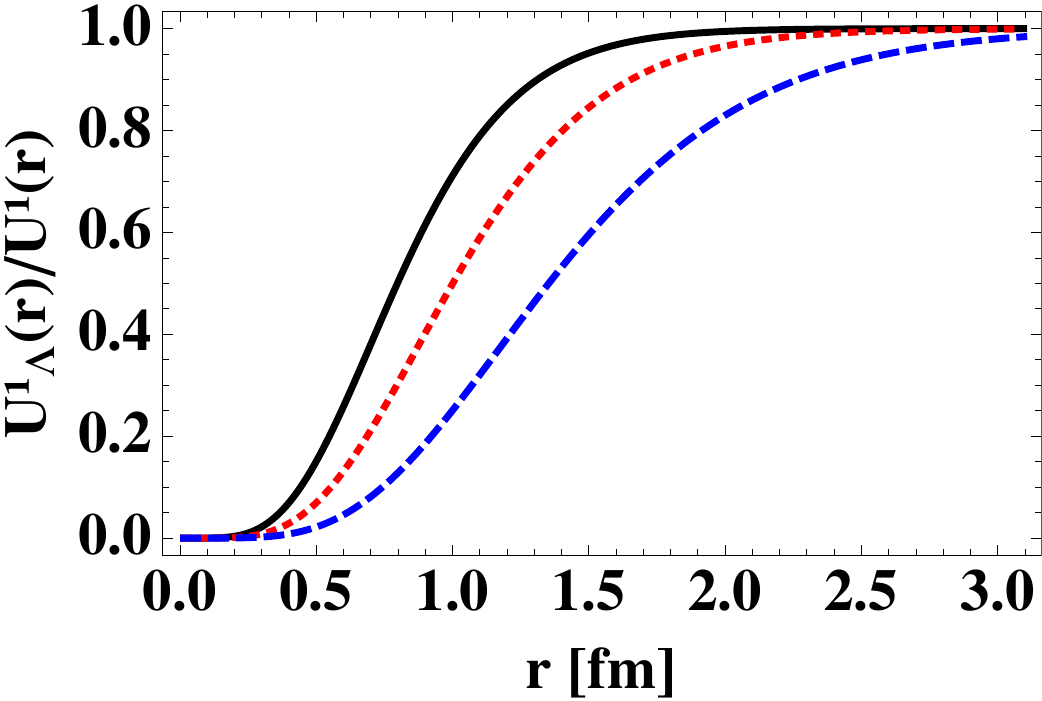}
\includegraphics[scale=0.55]{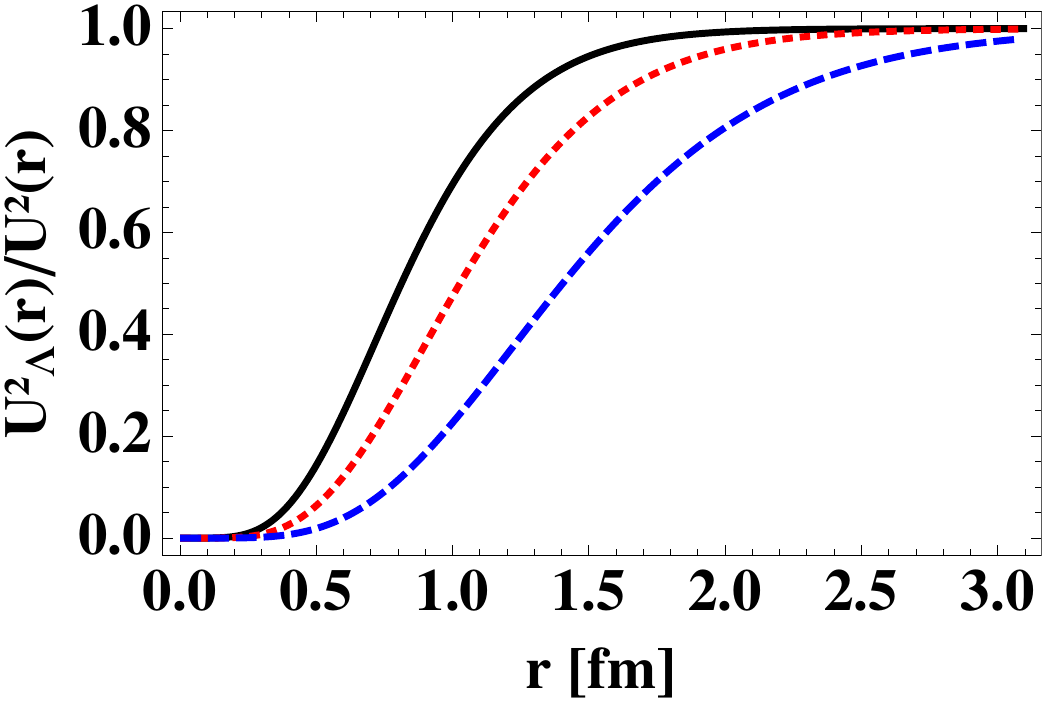}
\caption{Color online: ratio of spectral regularized and
  unregularized TPE chiral potentials $U_\Lambda^I(r)/U^I(r) $ for
  spectral cut-offs of $\Lambda=1.4$ GeV (solid, black),
  $\Lambda=1.12$ GeV (dotted, red) $\Lambda=0.84$ GeV (dashed, blue) at
  threshold $\sqrt{s}= 2 m_\pi$ as a function of the distance for
  $I=0$ (left) , $I=1$ (middle) and $I=2$ (right) channels. 
}\label{Fig:spectral}
\end{figure*}

\subsection{Spectral regularization}
\label{sec:spec}

While we might estimate the elementarity radius for the chiral
potentials from the corresponding folding of ``strong'' densities, 
we prefer to analyze instead the effect of introducing a cut-off
$\Lambda$ in the spectral function in~\eqref{eq:spec-pot}, i.e. 
\begin{eqnarray}
U_I^\Lambda(r,s) = \int\limits_{2 m_\pi}^\Lambda \diff\mu  \rho_I(\mu,s) \frac{e^{-\mu r}}{r},
 \end{eqnarray}
which corresponds to the two-pion invariant mass spectrum. 

In Fig.~\ref{Fig:spectral} we show the ratio $U_\Lambda^I(r)/U^I(r) $ for
spectral cut-offs of $\Lambda=1.4,\,1.12,\,0.84$ GeV at threshold
$\sqrt{s}= 2 m_\pi$ as a function of the distance. This estimate
yields a larger elementarity radius, which becomes $r_e = 1.2$ fm for
the largest spectral cut-off and a quenching of about a half for
$\Lambda \sim m_\rho$. This numerical exercise shows that the naive
estimate $r_e \sim 1/\Lambda$ is numerically rather inaccurate. Of
course, one could consider $\Lambda$ as a fitting parameter in the
analysis of $\pi\pi$ scattering. Nevertheless, in our view this has the disadvantage
that a model dependence is introduced by the cut-off procedure above
the elementarity radius.

Note also that the larger $r_e$ the smaller the TPE potential, since
$U(r_e) = {\cal O} (e^{-2 m_\pi r_e})$, so that one may end up a the
situation where a model independent treatment becomes only possible
when the chiral contribution is actually vanishingly small. In the
next section we will present a different strategy. Overall, our
numerical results will confirm this pessimistic expectation.

\subsection{Potential separation}

The previous discussion suggests that we should decompose the potential for
each isospin channel as
\begin{eqnarray}
U^I (r) = U_{\rm Short}^I (r) \theta (r_c - r ) + U_{\rm Long}^I (r)
\theta (r - r_c)\,,
\label{eq:full-pot}
\end{eqnarray}
where $U_{\rm Short} (r)$ is a short distance and 
$U_{\rm Long} (r)$ stands for the long distance
contribution.  The natural choice is to take for the long-range part
the {\it unregularized} chiral TPE potential
\begin{eqnarray}
U_{\rm Long}^I (r) = U_{\chi}^I (r)\,,
\end{eqnarray}
i.e. potential computed in $\chi$PT to a given order in the chiral
expansion.  As we have seen in the previous section, the lowest order
effect in the chiral potential comes from the two pion exchange, which
gives a contribution that at long distances is $U_{\chi} (r) = {\cal
  O} (e^{- 2 m_\pi r} /f^4)$.  For $r_c \sim 1/m_\pi$ this
contribution will in principle play a role.  Of course, these most
peripheral contributions to the interaction will contain perturbative
corrections to {\it all} orders in $m_\pi/f$, which are expected to
modify slightly the tail of the potential. Therefore the question is
what is the relative importance of the {\it unknown} short distance
piece and the {\it known} long distance contribution. Finally, let us
mension that while we will keep the energy dependence already present
in the chiral potential we will assume for simplicity that the short
distance potential is energy independent, as long as the inelasticity
is negligible (see also Section~\ref{sec:inel}).

\begin{figure}
\epsfig{figure=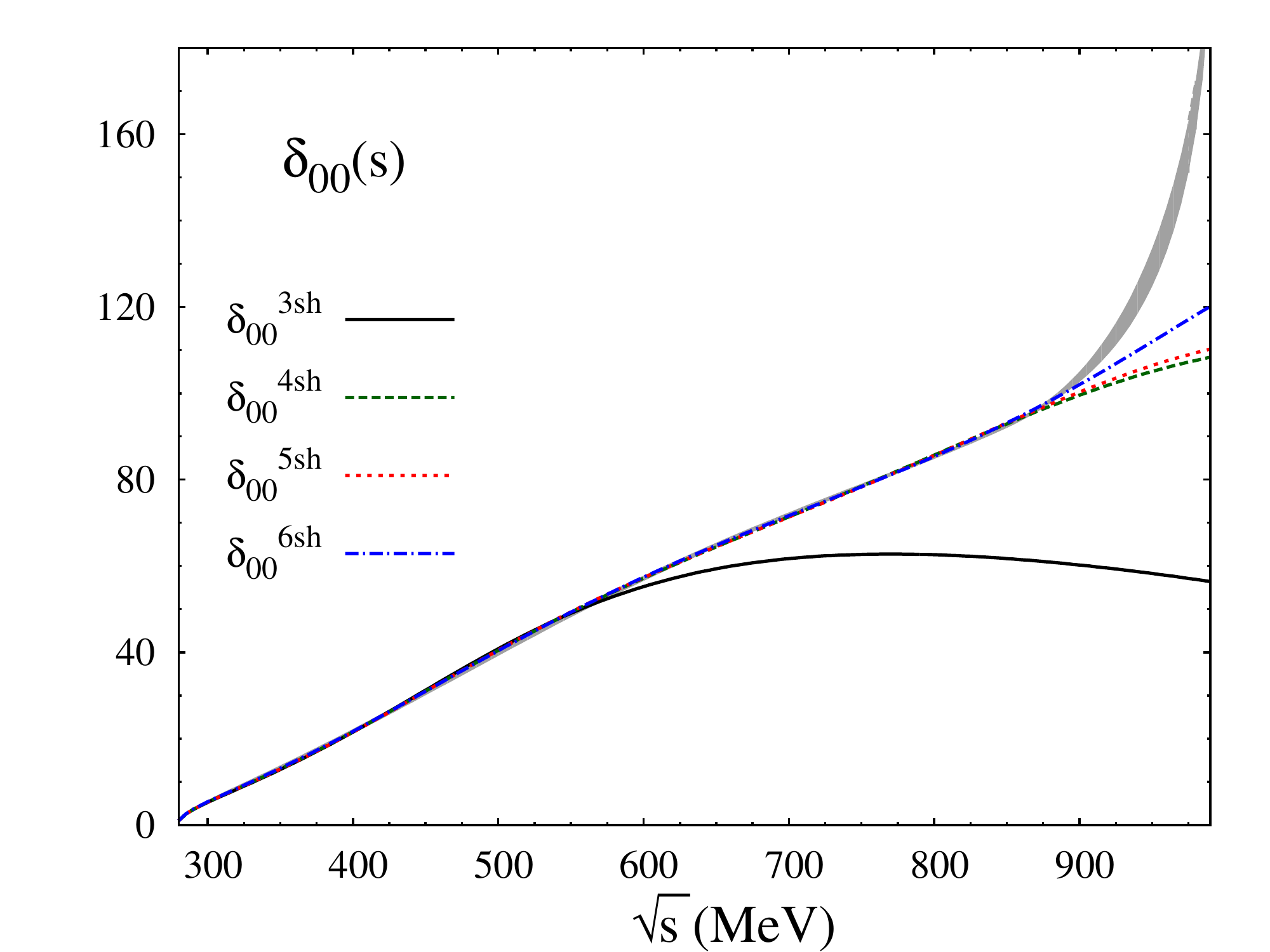,width=5.5cm,angle=-90}
\epsfig{figure=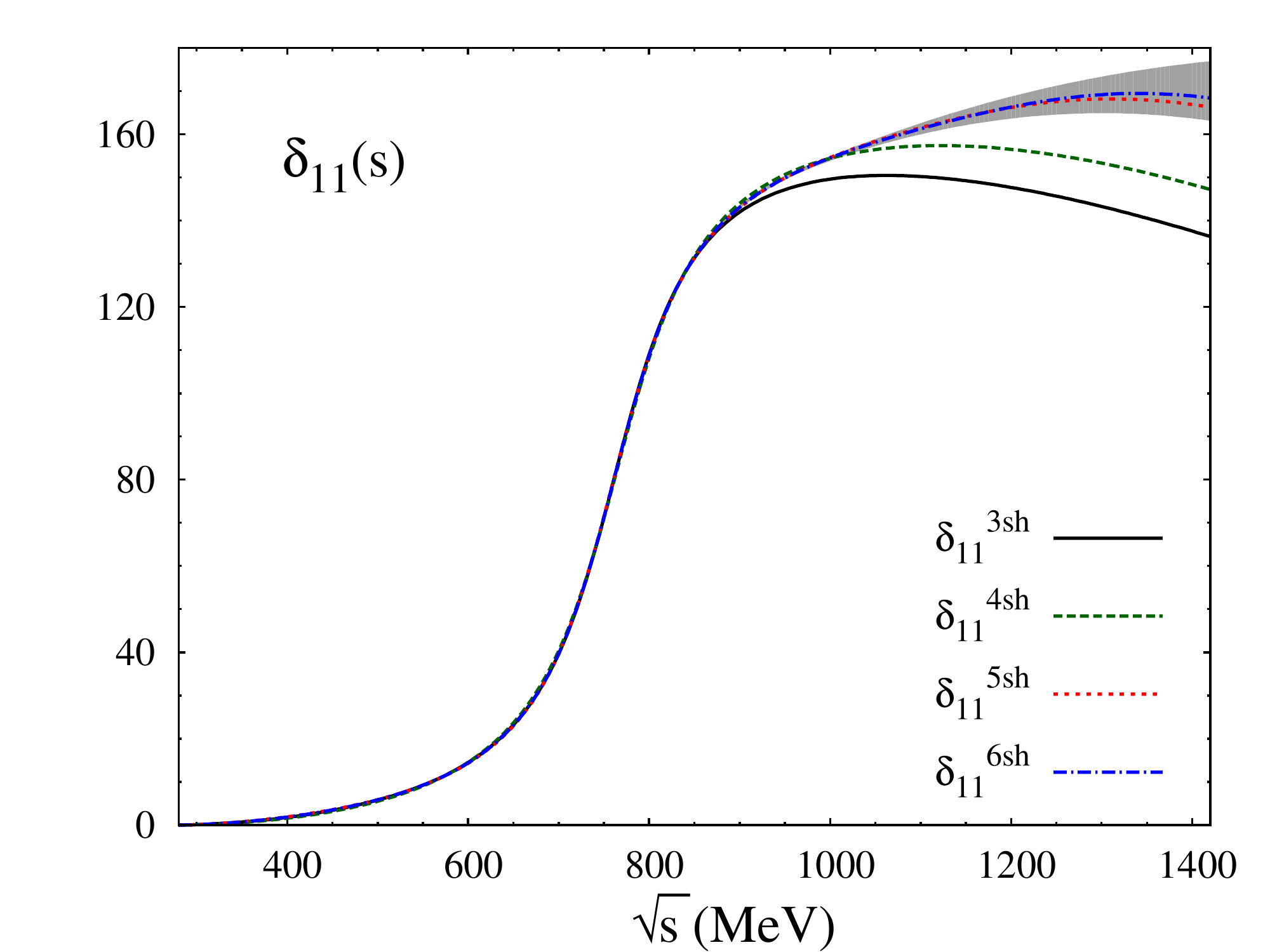,width=5.5cm,angle=-90}
\epsfig{figure=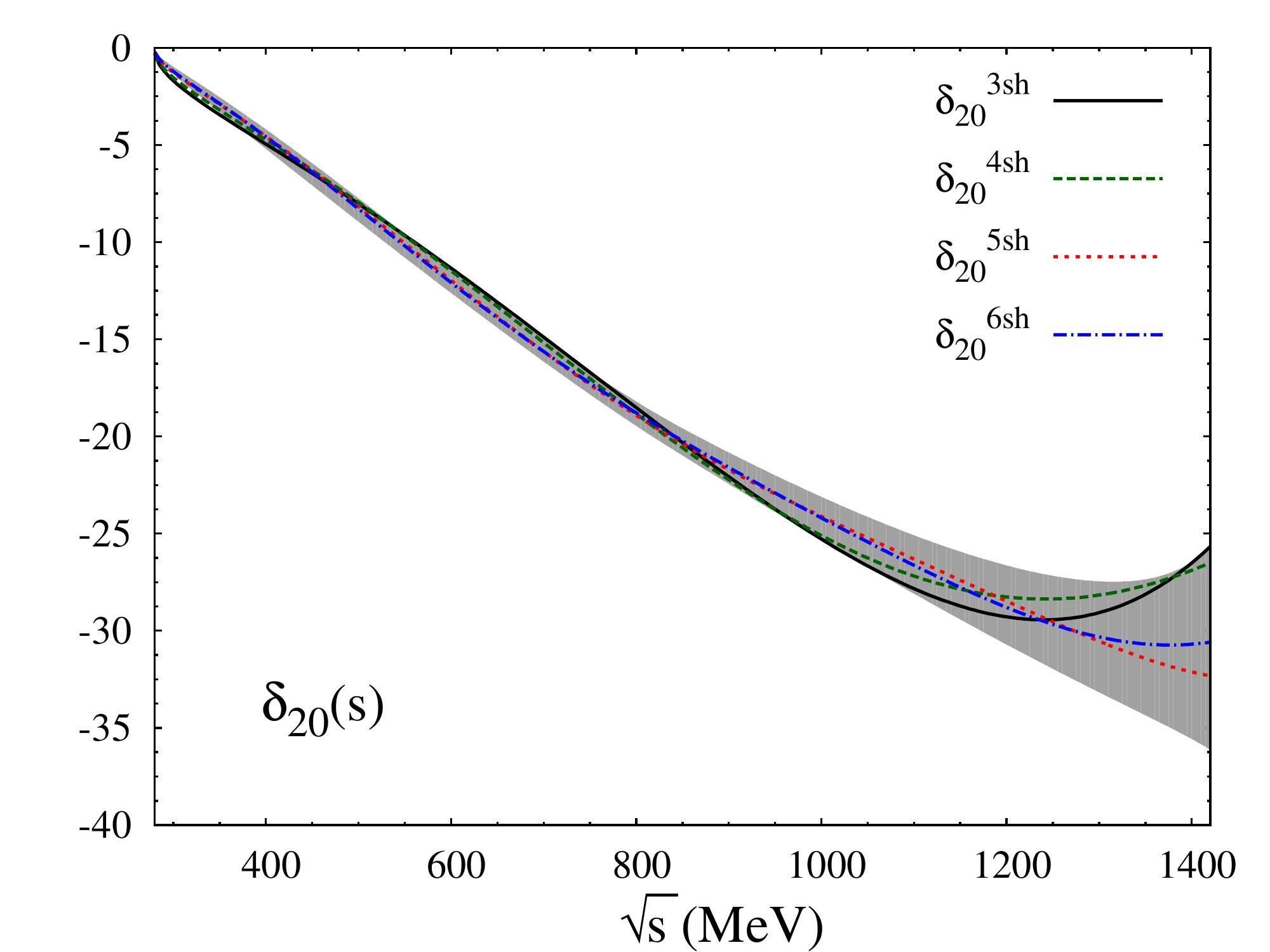,width=5.5cm,angle=-90}
\caption{Energy independent fits with the delta-shells potential given
  in~\eqref{eq:deltash} using $\Delta r=0.3$ fm for the $\pi\pi$
  $S0$-, $P$- and $S2$-wave phase shifts.  The uncertainties are those
  quoted in~\cite{GarciaMartin:2011cn}, whereas solid-black,
  green-dashed, red-dotted and blue dot-dashed lines stand for for the
  central results for $r_c=\,0.9,\, 1.2,\, 1.5,\, 1.8$ fm,
  respectively.
}
\label{Fig:phasessh}
\end{figure}

\section{Coarse graining}
\label{sec:coarse}

The basic notion of coarse graining in the scattering problem was
outlined by Avil\'es as early as 1972 in an insightful but forgotten
paper~\cite{Aviles:1973ee} within the context of NN interactions.  In
this article the potential was effectively represented as a sum of
delta-shell potentials. This form has important simplifications and a
recent comprehensive mathematical study of this specific case has been
carried out~\cite{albeverio2013spherical}. Here we extend the approach
to account for the $\chi$PT potential tail at long distances (see also
Refs.~\cite{NavarroPerez:2011fm,Perez:2013jpa,Perez:2013oba} for a
paralell treatment of the NN case). 

\subsection{Short distance potential}

The basic idea is as follows. If we want to describe the two-particle
CM wave-functions limited to the range $\Delta p$, only gross
information can be determined in an interval $\Delta r$, with $\Delta
r\Delta p\sim 1$. Thus, for a maximal $s_{\rm{\max}}=2$ GeV$^2$, we
have $\Delta p \sim \sqrt{s_{\rm{max}}/4-m_\pi^2}\sim 0.70$ GeV and
one obtains $\Delta r \sim 0.3$~fm.  This uncertainty suggests that for a limited
energy range the potential only needs to be known in a limited number
of points.  With this in mind, we consider for $r< r_c$ the $\pi\pi$
potential as a sum of a subsequent number of $\delta$ functions
separated by about $\Delta r$.  Thus, up to $r_c$, it requires to
introduce $N_\delta$ delta shells~\footnote{The specific form of the
  potential is not essential, see for e.g.~\cite{Fernandez-Soler:2017kfu}
  for a detailed comparison. In our case, we take delta-shells for simplicity.}
\begin{equation}\label{eq:deltash}
U_{\rm Short} (r)= U_{\Delta r}(r) \equiv \sum_{n}^{N_\delta}U (r_n) \Delta r \delta{(r-r_n)}\,.
\end{equation} 
Thus, we have two relevant scales in our setup: the separation distance
$r_c$ and the coarse graining scale $\Delta r$. While one expects the
results and main properties of the potential to be independent of the
particular choices of $r_c$ and $\Delta r$, this can only happen when
accurate information on the interaction is available at {\it all}
energies. In our case $\Delta r$ acts as an UV regulator whereas $r_c$
works as an IR regulator.

We will analyze this problem for four different values of $r_c$.  
For convenience, we will set $r_c=0.9$, $1.2$, $1.5$ and $1.8$~fm. 
The solution of~\eqref{radial-schr} for the delta-shell potential in~\eqref{eq:deltash} is straightforward. 
One has
\begin{eqnarray}
u_{l,n} (r) = {\hat j}_l (pr)  - \tan \delta_{l,n}\,{\hat y}_l (pr)\, ,  \qquad r_n < r < r_{n+1}\,,
\end{eqnarray}
where $\hat j_l(x) = x\,j_l(x)$ and $\hat y_l(x) = x\,y_l(x)$ are the reduced
Bessel functions of first and second kind, respectively, and $\delta_{l,n}$ is the accumulated phase shift.  
The discontinuity in the logarithmic derivative at $r=r_n$
\begin{eqnarray}
\frac{u_l' (r_{n}^+)}{u_l(r_{n}^+)} -\frac{u_l' (r_n^-)}{u_l(r_n^-)} =U(r_n)\,\Delta r\,,
\end{eqnarray}
with $r_n^+ \equiv r_n +0^+$ and $r_n^- \equiv r_n +0^-$
leads after using the unit Wronskian condition $\hat j_l'(x) \hat y_l (x)- \hat y_l'(x) \hat j_l (x)=-1 $ 
to the bilinear recursion relation for $\tan \delta_{l,n}$
\begin{eqnarray}
  \tan \delta_{l,n+1}= \frac{A_{l,n} (p) + B_{l,n} (p) \tan \delta_{l,n}  }{ C_{l,n} (p) +D_{l,n} (p) \tan \delta_{l,n}}\,,
  \label{eq:rec-1}
\end{eqnarray}
with
\begin{eqnarray}
  A_{l,n} (p)&=& \Delta r  \, U (r_n) j_l (p r_n)^2\,, \nonumber\\
  B_{l,n} (p)&=& \Delta r  \, U (r_n) j_l (p r_n) y_l (p r_n)+k\,,\nonumber  \\ 
  C_{l,n} (p)&=&-\Delta r  \, U (r_n) j_l (p r_n) y_l (p r_n)+k\,, \nonumber \\  
  D_{l,n} (p)&=& \Delta r \,U (r_n) y_l (p r_n)^2\,,
   \label{eq:rec-2}
\end{eqnarray}
with the initial and final conditions
\begin{eqnarray}
  \delta_{l,0}(k)=0\,, \qquad \delta_{l} (k) = \delta_{l,N} (k)\,.
\end{eqnarray}
As shown in~\cite{Entem:2007jg}, these equations can be interpreted as
the discrete version of the variable phase equation of
Calogero~\cite{calogero1967variable}, corresponding to the limit
$\Delta r \to 0$.  Of course, the previous equations define an
integration method in this limit.  We stress that the idea of coarse
graining is that if $U(r)$ is determined from data with a maximum CM
momentum $p_{\rm max}$, the natural resolution of the problem is
$\Delta r \sim 1/p_{\rm max}$, and $U(r_n)$ are the natural fitting
parameters~\footnote{One may argue that the coarse grained potential
  is generally nonlocal. While this is certainly expected, our
  assumption is compatible with allocating the non-locality for scales
  smaller than the resolution scale $\Delta r$.}.

\subsection{Fitting procedures}

In this section we present our numerical results based on standard
$\chi^2$-fits.  Precise $\pi\pi$-scattering phase shifts have been
obtained in~\cite{GarciaMartin:2011cn,Perez:2015pea} using Roy
equations up to $\sqrt{s_{\rm{max}}}=1.42$ GeV and we will use their
tabulated values to determine our fitting parameters. While the
standard strategy in $\pi\pi$ scattering studies has been to use the
physical region $\sqrt{s}> 2 m_\pi$, for reasons to be justified in
Section~\ref{sec:cross-roy} we also include in the S0 fit the
subthreshold region, $0 < \sqrt{s} \le 2m_\pi$ , also deduced in those
Roy-equation analyses~\cite{GarciaMartin:2011cn,Perez:2015pea}.

In order to exemplify this formalism, we will focus just on the lowest
$\pi\pi$ partial-waves, namely the isoscalar and isotensor $S$ waves
$S0$ and $S2$, respectively, and the $P$ wave.  For any isospin
channel we will take the potential
\begin{eqnarray}\label{eq:U-lambda}
  U (r) &=&\left[ \sum_{n=0}^{N} \lambda_n \delta{(r-r_n)} \right]
  \theta( r_c- r)  +  U^{(4)}_\chi (r)\theta(r - r_c)\,,\nonumber\\ 
\end{eqnarray}
where $\lambda_n = U(r_n) \Delta r $. For such a maximal energy this
means $\Delta r =0.3 $ fm.  Thus, up to $r_c=\{0.9,1.2,1.5,1.8\}$ fm,
it requires to introduce $N_\delta=\{3,4,5,6\}$ delta shells.

Anticipating the result, we will carry out the study, first {\it
  without} including the chiral tail, since as we will see in
Sect.~\ref{sec:full}, it plays a minor role in the resulting fitting
parameters $\lambda_n$.

\subsection{Energy independent coarse graining}

To start with, we will assume on purpose sufficiently large separation
distances so that the long-distance field theoretical contribution
${\cal O} (e^{-2 m_\pi r})$ can be safely neglected. 

In order to constrain as much as possible the delta-shell coefficient
values $\lambda_j$, we will introduce each delta-shell adiabatically,
one by one.  In addition, we will impose at threshold the
corresponding $\pi\pi$ scattering lengths and slope parameters, 
so that the value of the most internal delta-shell parameters, $\lambda_0$ and
$\lambda_1$, are completely fixed.  The $\pi\pi$ $S0$-, $S2$- and
$P$-wave phase shift obtained in this way for the different values of
$r_c$ previously chosen are plotted in Fig.~\ref{Fig:phasessh},
whereas the value of the delta-shell parameters are given in
Table~\ref{Tab:phasessh}.
The maximum fitted energy $\sqrt{s}$ for each partial waves is chosen as the maximum one for which one can find a 
$\chi^2/d.o.f$ around 1. 
In addition, as we will discuss in Section~\ref{sec:cross-roy}, we also include for the S0 fit the subthreshold region $m_\pi <s<2m_\pi$.
\begin{table}
\begin{center}
\begin{tabular}{cc|llllll|ccc}\hline
$IJ$& $r_c$ (fm) &$\lambda_0$&$\lambda_1$&$\lambda_2$&$\lambda_3$&$\lambda_4$&$\lambda_5$&$\sqrt{s_{\rm{max}}}$ &$N_p$&$\chi^2/N_p$\\\hline
00&$0.9$ &-1.49&-2.34&-0.19&--&---&---  &0.6&65 &0.99\\
&$1.2$ &-1.31&-4.92&-0.73&0.21&---&---  &0.9&125&1.03\\
&$1.5$ &-1.31&-5.06&-0.76&0.26&-0.01&---  &0.92&130&0.92\\
&$1.8$ &-1.28&-5.93&-0.95&0.87.7&-0.20&0.04&0.95&135&1.01\\\hline
11&$0.9$ &-2.24&-4.52&-0.45&---&---&---&0.79&103&0.13\\
&$1.2$ &-2.22&-5.07&-0.57&0.06&---&---   &1.1&164&1.05\\
&$1.5$ &-2.21&-5.76&-0.76&0.32&-0.06&--- &1.42&230&0.06\\
&$1.8$ &-2.21&-5.81&-0.78&0.38&-0.09&0.08 &1.42&230&0.02\\\hline
20&$1.2$ &-0.04&0.42&-0.10&---&---&---   &1.42&230&0.98\\
&$1.2$ &0.21&0.20&-0.02&-0.02&---&---   &1.42&230&0.57\\
&$1.5$ &0.11&0.39&-0.15&0.06&-0.03&---  &1.42&230&0.05\\
&$1.8$ &0.25&0.23&-0.06&0.00&0.01&-0.01  &1.42&230&0.02\\
\end{tabular}
\end{center}
\caption{Energy independent fit values of the $\lambda_i \equiv
  U(r_i) \Delta r$ coefficients in GeV units defined
  in~\eqref{eq:U-lambda} for each partial wave and value of $r_c$.
  For $r_c=\,0.9,\,1.2,\,1.5$ and $1.8$ fm, the corresponding number of
  delta-shells is 3, 4, 5 and 6, respectively. The maximum fitted
  energy $\sqrt{s}$ (GeV) is chosen as the maximum one for which the
  $\chi^2/N_p$ is around 1, with $N_p$ the number of data
  points.} \label{Tab:phasessh}
\end{table}

We can see from these results that with a few free parameters, from 3
to 6 depending on the particular choice of $r_c$, this formalism
already allows one to describe the $\pi\pi$ scattering in the elastic
region.  In the case of the $S2$ wave, where the inelasticities are
very small at low energies, it is possible to obtain a perfect
description up to the maximum energy at $s_{\rm{max}}=2$~GeV$^2$.  
For the $P$ wave one obtains a good description, up to energies around the
$\bar KK$ threshold at $\sqrt s=1$ GeV, whereas for the S0-wave the
description is limited to the region around 0.9 GeV, where the phase
presents a huge rise due to the effect of the inelastic $f_0(980)$ resonance.

Our results reproduce qualitatively features found 
in~\cite{Sander:1997br} using inverse scattering methods, where by
construction the potential is a local and continuous function. In
particular, for the $S$ waves we find a short distance barrier in the isoscalar
and a repulsive core in the isotensor.  Of course, we only get a
coarse grained version of those potentials, which befits our idea of a
finite resolution sampling.

\subsection{Threshold parameters}
\label{sec:thpar}

Our results for the threshold parameters defined by~\eqref{eq:thres} are presented in Table~\ref{Tab:th}.  
As expected from the quality of the fits, they agree within uncertainties with the
results obtained in previous
analyses~\cite{Colangelo:2000jc,Ananthanarayan:2000ht,Colangelo:2001df,Caprini:2003ta,Pelaez:2004vs,Kaminski:2006yv,Kaminski:2006qe,GarciaMartin:2011cn}. 

\begin{table}
\begin{center}
\begin{tabular}{cc|cc}\hline
$IJ$&$r_c$ (fm) & $a_0$ $(m_\pi)^{-1}$ & $b_0$ $(m_\pi)^{-3}$\\\toprule\hline
00&$0.9$ &$0.212$&$0.262$\\
&$1.2$   &$0.214$&$0.264$\\
&$1.5$   &$0.215$&$0.272$\\
&$1.8$   &$0.218$&$0.274$\\\hline
20&$0.9$ &$-0.073$&$-0.048$\\
&$1.2$   &$-0.065$&$-0.049$\\
&$1.5$   &$-0.052$&$-0.056$\\
&$1.8$   &$-0.048$&$-0.059$\\\hline
11&$1.2$ &$31.2\cdot10^{-3}$&$5.9\cdot10^{-3}$\\
&$1.2$   &$33.3\cdot10^{-3}$&$5.6\cdot 10^{-3}$\\
&$1.5$   &$34.6\cdot10^{-3}$&$5.2\cdot10^{-3}$\\
&$1.8$   &$37.2\cdot10^{-3}$&$5.1\cdot10^{-3}$\\
\end{tabular}
\end{center}
\caption{Scattering lengths and slope parameters for the
different fits without TPE with an increasing number of delta shells separated
  by $\Delta r=0.3$ fm} \label{Tab:th}
\end{table}

\subsection{Resonance poles}
\label{sec:res}

Resonances are determined by looking for poles in the
second Riemann sheet, which is defined by the complex wavenumber 
$k_R= k_r + i k_i $ with $k_r \equiv {\rm Re}\,k_R > 0$ and $k_i \equiv {\rm
  Im}\,k_R <0$. This can be achieved directly by substituting $k \to
k_R$ in the Schr\"odinger equation~\eqref{eq:mass-sq}, or its
coarse-grained delta-shell implementation~\eqref{eq:rec-1} and~\eqref{eq:rec-2}.

Our numerical results are presented in Table~\ref{Tab:res}. The
numerical values are slightly different from those quoted in
benchmarking studies based on Roy equations and forward dispersion
relations~\cite{GarciaMartin:2011jx}. This is not fully surprising
since the analytical properties of our scattering amplitude are not
the same as in the analyses based on the Roy equations, but only to
leading order in the chiral expansion and hence the extension to the
complex plane is not determined from the phase-shift analysis only.
Nonetheless, we find that the result is encouraging and we expect to
return in the future in order to implement higher orders to analyze
the effect.

\begin{table}
\begin{center}
\begin{tabular}{c|cc}\hline
$r_c$ (fm) &$\sqrt{s_{f_0(500)}}$ (MeV)  &$\sqrt{s_{\rho(770)}}$ (MeV)\\\toprule\hline
$0.9$ &$453- i\, 212$&$762-i\, 70$\\
$1.2$ &$438-i\, 233$& $764-i\, 74$\\
$1.5$ &$441-i\, 247$& $762-i\, 72$\\
$1.8$ &$446-i\, 250$& $762-i\, 72$\\\hline
\end{tabular}
\end{center}
\caption{$f_0(500)$ and $\rho(770)$ resonance poles obtained from the
  different fits without TPE with an increasing number of delta shells separated
  by $\Delta r=0.3$ fm} \label{Tab:res}
\end{table}

\subsection{Crossing and comparison with Roy equations}
\label{sec:cross-roy}

\begin{figure}
\centering
\includegraphics[scale=0.39,angle=-90]{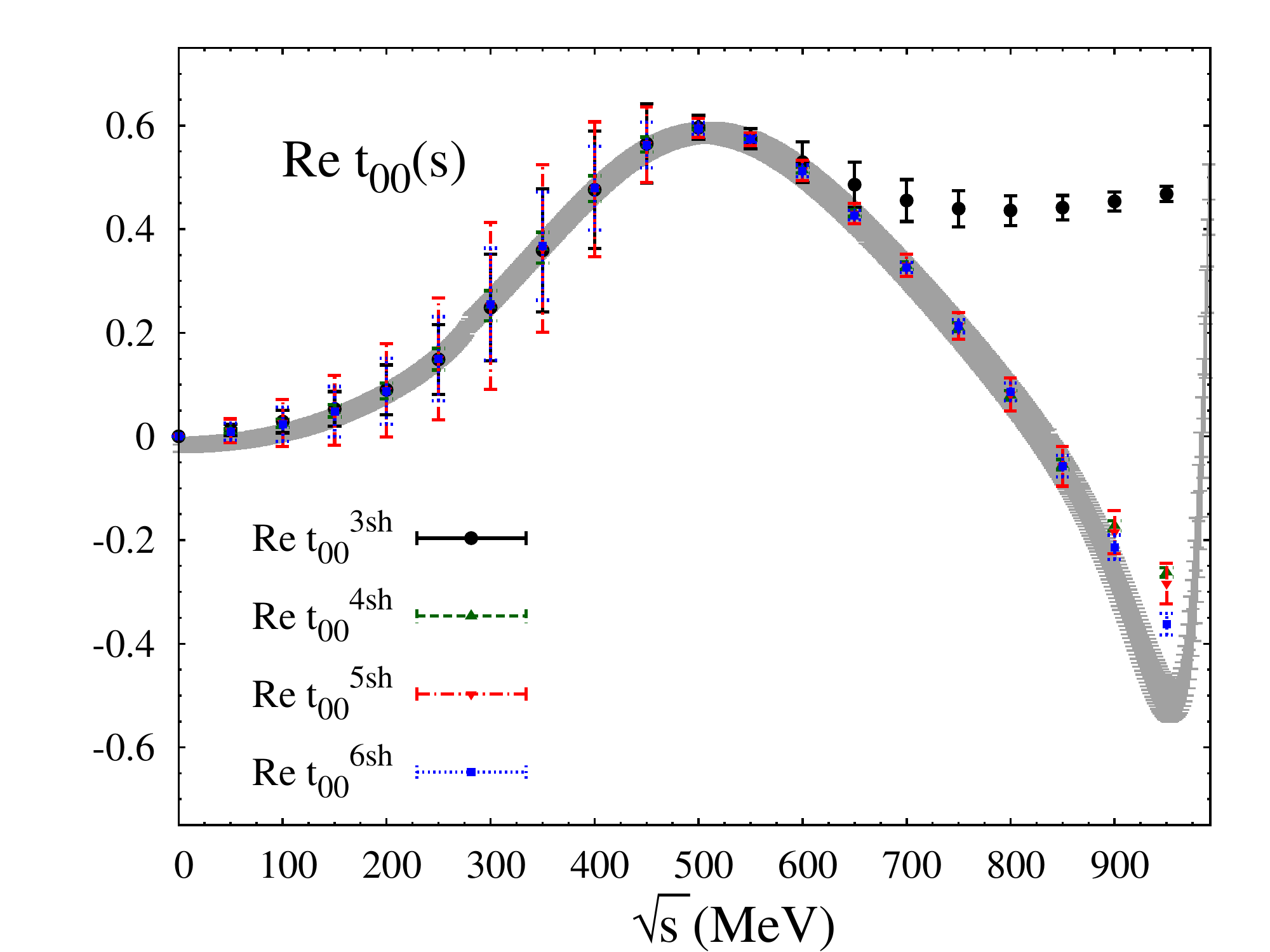}\\
\includegraphics[scale=0.39,angle=-90]{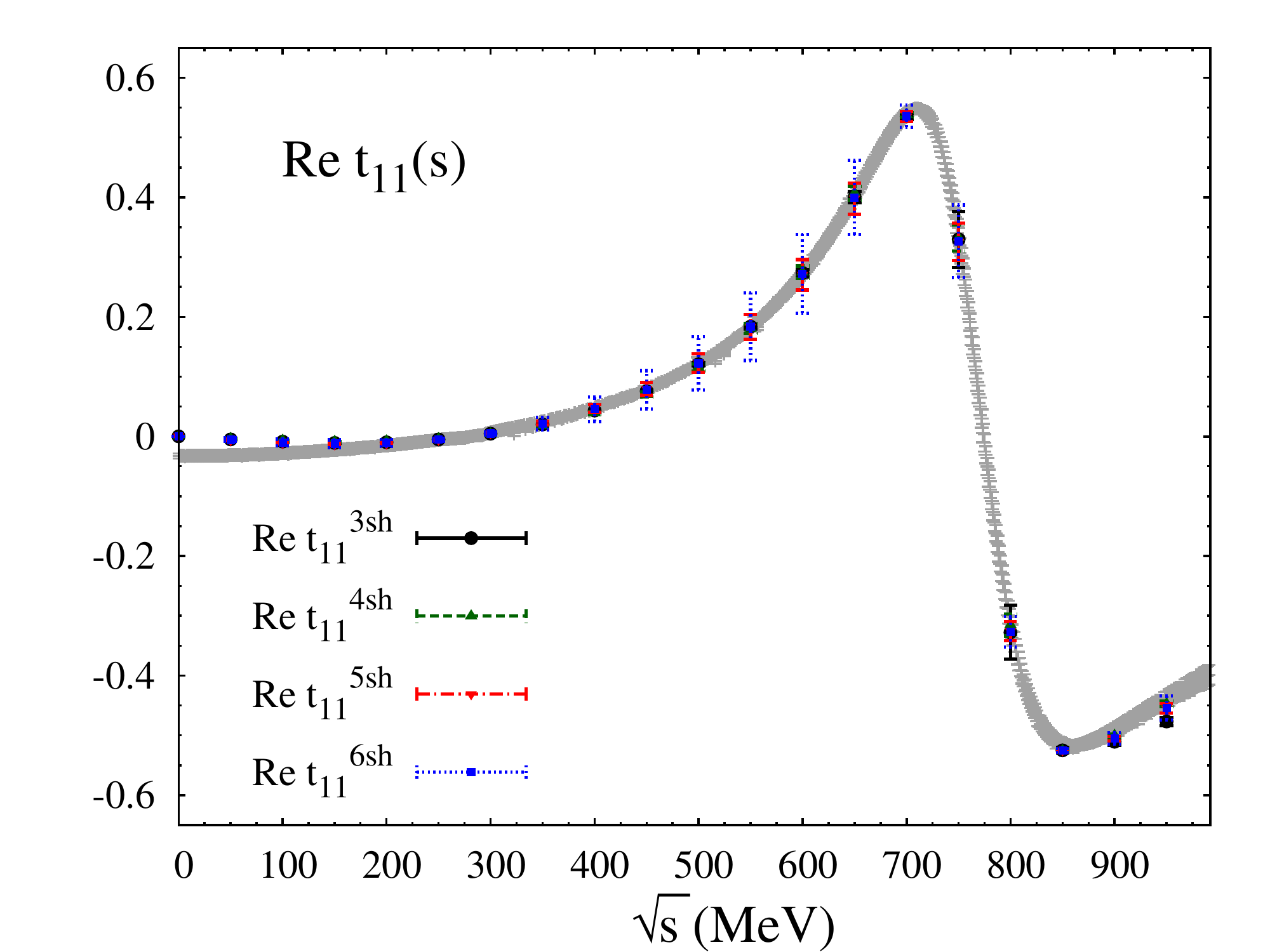}\\
\includegraphics[scale=0.39,angle=-90]{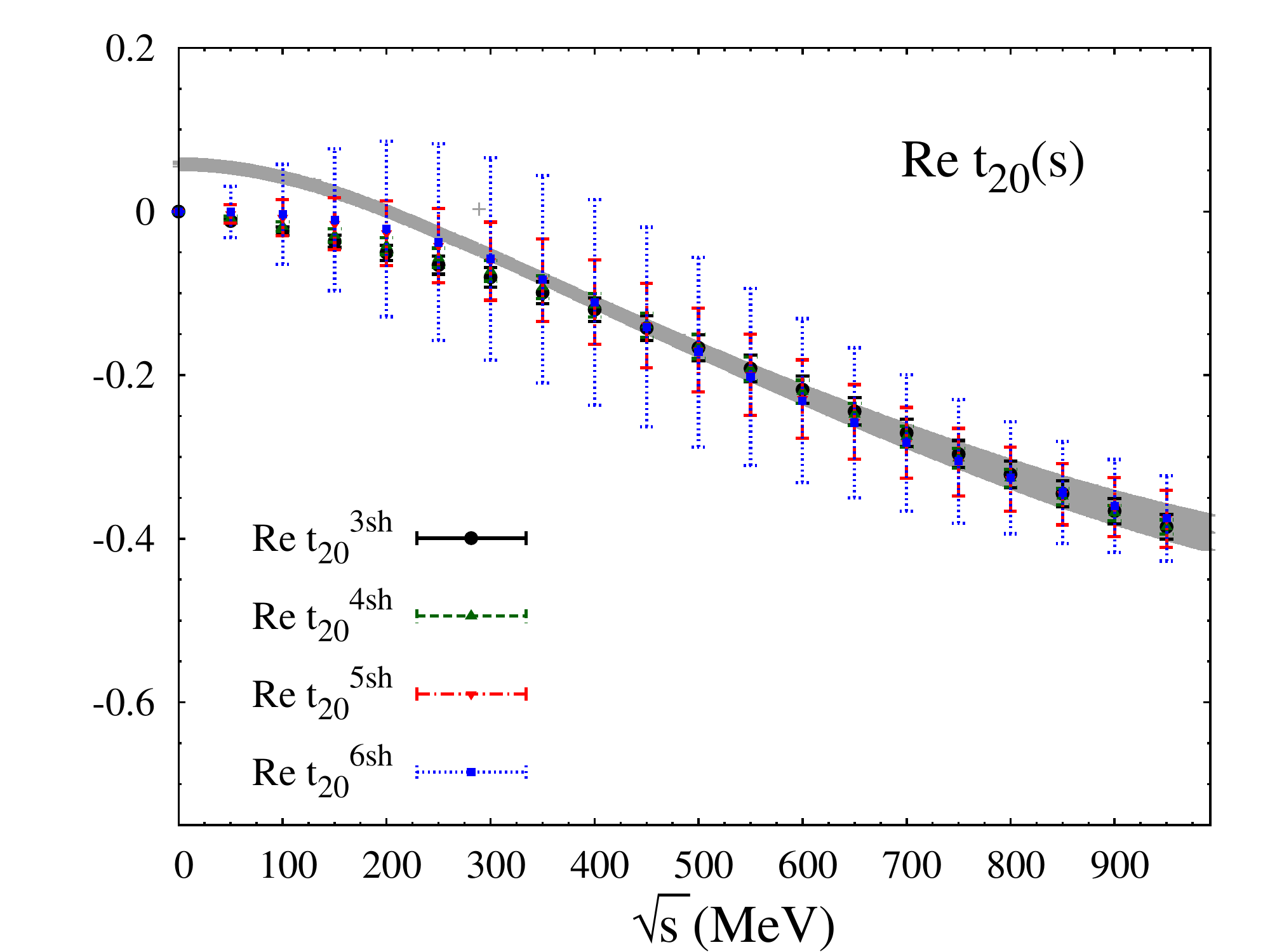} 
\caption{Color online: real $\pi\pi$ partial-wave amplitudes  as a function
  of the CM energy, starting from the subthreshold value $\sqrt{s}=0$ for $I=0$ (top), $I=1$ (middle) and $I=2$
  (bottom). We compare the different delta-shell results with the results of the Roy-equation analysis (gray band). 
}\label{Fig:cros-roy}
\end{figure}

As we have mentioned in the introduction, there are no {\it direct}
$\pi\pi$ scattering data. The closest thing to it are possibly the
outcome of Roy equations, which incorporate by construction crossing,
analyticity, unitarity and Regge
behavior~\cite{Colangelo:2001df,Caprini:2003ta,GarciaMartin:2011cn}. Moreover, a proper identification
of the LECs requires by definition the satisfaction of crossing,
particularly in the sub-threshold region. For definiteness, we show in
Fig.~\ref{Fig:cros-roy} the real part of the amplitude starting at the
edge of the left hand cut $s=0$, covering the subthreshold region $0 <
\sqrt{s} \le 2 m_\pi$ (where the amplitude is purely real) and the
physical elastic region, $2 m_\pi<\sqrt{s}<2 m_K$, where the imaginary
part of the amplitude is determined from unitarity and the
corresponding real part.  Note that we are neglecting possible
inelasticities coming from multi-pion channels.  In particular, one
can clearly see the Adler zeros of the S0 and S2-wave amplitudes, a
characteristic features of the subthreshold region and a distinct
fingerprint of chiral symmetry~\cite{martin1976pion}.

The comparison in Fig.~\ref{Fig:cros-roy} supports the view that fits
to phase shifts in the physical region, even to relatively high
energies, do not constrain the subthreshold region. This has a large effect
on the location of the $\sigma$-resonance and the implications will be discussed in
more detail elsewhere. In addition, our fits have not imposed crossing correlations
and the consideration or not of the subthreshold region in the S0
channel can be grasped by comparing the location of the corresponding
resonance. 
If the subthreshold region $m_\pi <s<2m_\pi$ is not imposed in the fit, 
one gets substantially smaller values for the real
part. For instance, for $r_c=0.9$ fm and $N_\delta=3$ delta-shells one gets $\sqrt{s_\sigma}=
(324 - i\,194 )$ MeV, whereas the pole moves to $\sqrt{s_\sigma}= (453 - i\,212 )$ MeV when 
the subthreshold region is also fitted. 
This large influence of the subthreshold region in the S0 wave is not surprising,
since the pole is rather far from the real axis.  In contrast, the
location of the $\rho$-resonance pole proves rather insensitive to the
subthreshold region in the P-wave.
Although we are not fitting the subthreshold region in this channel, 
one gets for $r_c=0.9$ fm and $N_\delta=3$ delta-shells
$\sqrt{s_\rho}= (762 - i\,70)$ MeV, 
even when the subthreshold extrapolation of the fit, 
middle panel of Fig.~\ref{Fig:cros-roy}, shows a significant
discrepancy with the outcome of the Roy equations.  Finally, we also
report a discrepancy in the isotensor channel, where the
phase shifts are compatible up to $\smax$. Actually, in all cases
the amplitude has a zero at $s=0$, due to the factor $\sqrt{s}$ in the
definition of $t_{IJ}(s)$. Thus, the failure to satisfy the subthreshold
behavior in the S2 wave is intrinsic~\footnote{This is in common with
  other unitarization methods, such as IAM where the unitarized
  amplitude $t_{IJ}(s)= t_{IJ}^{(2)}(s) + t_{IJ}^{(4)}(s) + \cdots =
  \frac{[t_{IJ}^{(2)}(s)]^2 }{t_{IJ}^{(2)}(s) - t_{IJ}^{(4)}(s)} $
  develops a double Adler zero instead of the expected single one (see
  e.g. Ref.~\cite{Nieves:2001de}) and Fig.~\ref{Fig:cros-roy}.}. We
have checked that this does not improve by incorporating
the TPE tail of the chiral potential (see also next subsection). The
fact that the subthreshold region is quantitatively as relevant as the
physical region, is another manifestation of the relevance of crossing
in $\pi\pi$ scattering, and calls for improvement when all relevant
partial waves are considered.

\begin{table}
\begin{center}
\begin{tabular}{cc|llllll|ccc}\hline
$IJ$& $r_c$ (fm) &$\lambda_0^\chi$&$\lambda_1^\chi$&$\lambda_2^\chi$&$\lambda_3^\chi$&$\lambda_4^\chi$&$\lambda_5^\chi$&$\sqrt{s_{\rm{max}}}$ &$N_p$&$\chi^2/N_p$\\\hline
00&$0.9$ &-1.58&-2.07&0.02 &--   &---  &---  &0.41&27 &1.04\\
&$1.2$   &-1.28&-7.47&-0.85&0.37 &---  &---  &0.65&75 &1.05\\
&$1.5$   &-1.32&-4.71&-0.68&0.13 & 0.02&---  &0.90&124&1.04\\
&$1.8$   &-1.28&-5.91&-0.95&0.88 &-0.21&0.05 &0.95&135&1.12\\\hline
11&$0.9$ &-2.24&-4.32&-0.44&---  &---  &---  &0.76&100&0.34\\
&$1.2$   &-2.22&-5.08&-0.57&0.06 &---  &---  &1.10&164&0.98\\
&$1.5$   &-2.21&-5.76&-0.76&0.32 &-0.06&---  &1.42&230&0.06\\
&$1.8$   &-2.21&-5.81&-0.78&0.38 &-0.09&0.01 &1.42&230&0.02\\\hline
20&$1.2$ &0.22 &0.22 &-0.05&---  &---  &---  &1.42&230&0.01\\
&$1.2$   &0.21 &0.20 &-0.04&-0.01&---  &---  &1.42&230&0.27\\
&$1.5$   &0.11 &0.38 &-0.14&0.06 &-0.02&---  &1.42&230&0.02\\
&$1.8$   &0.25 &0.23 &-0.06&0.01 & 0.00&-0.01&1.42&230&0.01\\
\end{tabular}
\end{center}
\caption{Energy independent fit including the chiral tail for $r>r_c$.
  Values of the $\lambda_i$ coefficients in GeV units defined
  in~\eqref{eq:U-lambda} for each partial wave and value of $r_c$.
  For $r_c=\,0.9,\,1.2\,,1.5$ and $1.8$ fm, the corresponding number of
  delta-shells is 3, 4, 5 and 6, respectively. The maximum fitted
  energy $\sqrt{s}$ (GeV) is chosen as the maximum one for which the
  $\chi^2/N_p$ is around 1, with $N_p$ the number of data
  points.} \label{Tab:phasessh-ch}
\end{table}

\begin{table}
\begin{center}
\begin{tabular}{cc|cc}\hline
$IJ$&$r_c$ (fm) & $a_0$ $(m_\pi)^{-1}$ & $b_0$ $(m_\pi)^{-3}$\\\toprule\hline
00&$0.9$ &$0.214$&$0.221$\\
&$1.2$   &$0.208$&$0.268$\\
&$1.5$   &$0.211$&$0.274$\\
&$1.8$   &$0.218$&$0.274$\\\hline
20&$0.9$ &$-0.044$&$-0.082$\\
&$1.2$   &$-0.056$&$-0.063$\\
&$1.5$   &$-0.049$&$-0.062$\\
&$1.8$   &$-0.048$&$-0.064$\\\hline
11&$1.2$ &$32.0\cdot10^{-3}$&$5.6\cdot10^{-3}$\\
&$1.2$   &$34.1\cdot10^{-3}$&$5.5\cdot 10^{-3}$\\
&$1.5$   &$35.0\cdot10^{-3}$&$5.2\cdot10^{-3}$\\
&$1.8$   &$37.5\cdot10^{-3}$&$5.1\cdot10^{-3}$\\
\end{tabular}
\end{center}
\caption{Scattering lengths and slope parameters for the different
fits including the TPE chiral tail with an increasing number of delta shells separated
by ∆r = 0.3 fm}
\label{Tab:th-ch}
\end{table}

\begin{table}
\begin{center}
\begin{tabular}{c|cc}\hline
$r_c$ (fm) &$\sqrt{s_{f_0(500)}}$ (MeV)  &$\sqrt{s_{\rho(770)}}$ (MeV)\\\toprule\hline
$0.9$ &$451- i\, 178$&$762-i\, 74$\\
$1.2$ &$420-i\, 210$& $764-i\, 74$\\
$1.5$ &$440-i\, 241$& $762-i\, 72$\\
$1.8$ &$443-i\, 246$& $762-i\, 72$\\\hline
\end{tabular}
\end{center}
\caption{$f_0(500)$ and $\rho(770)$ resonance poles obtained from the
  different fits including the TPE chiral tail with an increasing number of delta shells separated
  by $\Delta r=0.3$ fm} \label{Tab:res-ch}
\end{table}

\subsection{Full potential: Inclusion of Two Pion Exchange}
\label{sec:full}

Once the chiral tail has been defined, we can finally construct a full
potential. At short distances ($r< r_c$), the non-perturbative regime
of QCD is encoded by a sum of delta-shells separated by
the minimum De Broglie wave-length considered in the analysis. At
large energies ($r< r_c$), the chiral potential constructed
in~\eqref{eq:chiral-pot} with the correct analytical properties and
left-hand cut contribution is included.  
Higher order corrections in the chiral expansion are ${\cal O} (f^{-6})$
and the determination of the corresponding potential requires going 
in the quantum mechanical picture to second order in the Born approximation. 
Their analysis is left for future research. 

The resulting fits are reported in Table~\ref{Tab:phasessh-ch}. As one can
see, and with the exception of the S2 wave with three delta shells, we do
not find a substantial improvement due to the explicit incorporation of
the TPE potential.  A similar trend is found for the corresponding low
energy threshold and resonance parameters, see Tables~\ref{Tab:th-ch}
and \ref{Tab:res-ch}. In any case, chiral effects due to explicit TPE
are generally found to play a minor role.

In our analysis we have restricted to the lowest $S$- and $P$-waves. The
implementation of higher partial waves is straightforward and requires
introducing further short range contributions. Due
to the centrifugal barrier term, we expect that the number of
grid points will be reduced as the angular momentum increases.

\section{Analytic properties and the  N/D method}
\label{sec:analy}

Traditionally, much of the discussion of $\pi\pi$ scattering has been
marked by analyticity and  dispersion relations,
which at the partial-wave level and in the unsubtracted case read 
\begin{eqnarray}
t_{IJ}(s) &=& \frac1{\pi} \int\limits_{-\infty}^0 \diff s' \frac{{\rm Im}\,t_{IJ}(s')}{s'-s-i 0^+}+ \sum_{n} \frac{ g_n}{s_n-s}  
\nonumber \\ &+&
\frac1{\pi} \int\limits_{4 m_\pi^2}^\infty \diff s' \frac{{\rm Im}\,t_{IJ}(s')}{s'-s-i 0^+},
\end{eqnarray}
where $0< s_n < 4 m_\pi^2$ are the possible bound states, which we keep
for generality, and ${\rm Im}\,t_{IJ}(s) = \rho(s) |t_{IJ}(s)|^2$ for
elastic scattering. In this section, we will discuss this issue within the
context of our coordinate space framework and for the specific chiral
potential derived in Section~\ref{sec:chpt}, but the results are general. 
Actually, we will see that the subtractions are not explicitly needed, 
although they are somewhat encoded into the short distance component of the potential.  
While many of the issues discussed here have been known for potential
scattering since many years~\cite{fivel1960analytic,newton1960analytic} (see also
\cite{nussenzveig1972causality,goldberger2004collision}), 
we feel it is necessary to review the main aspects for completeness.

The analytical properties of the scattering amplitude in the complex
energy plane are determined by the long-distance behavior of the
interaction. This is the basis for dispersion relations, 
which own their popularity to their link to axiomatic field theory and its
straightforward implementation in terms of leading singularities. A
frequent method, which has been used in these regard to implement known
analytical properties, is the so-called N/D approach.
In the N/D approach the partial
wave amplitude is written in the form~\cite{Chew:1960iv} 
\begin{eqnarray}
t_{IJ}(s) = \frac{N_{IJ}(s)}{D_{IJ}(s)},
\end{eqnarray}
with $N_{IJ}(s)$ having only left-hand cut singularities and $D_{IJ}(s)$ having
only right-hand  cut singularities. This method has often been invoked 
in $\pi\pi$ scattering and implemented in several approximations
(see for e.g.~\cite{Oller:1998zr}).

\subsection{Jost functions}

The way of realizing the N/D representation in potential scattering is
well-known. We will describe here the formalism within the coarse grained
approach for completeness as well as to provide the $N_l(k)$ and $D_l(k)$
functions explicitly, (we drop the isospin index for simplicity). The
discussion is naturally carried in terms of the quantum mechanical
scattering amplitude defined in~\eqref{eq:rel-no-rel-eq},
$f_l(k)= 2 t_{l} (s)/\sqrt{s}$, as a function of
the CM momentum, $k$, with $s=4 (k^2+m_\pi^2)$. 
Note that $s$ is invariant under $k \to -k $ and hence two-valued. 
The discussion below entitles
to take ${\rm Im} k >0$ for the first Riemann sheet and ${\rm Im} k <0$ for the second.

For a regular potential, the Jost functions, $F_l(k)$, are
defined by the regular solutions of the wave equation at short
distances, i.e. $u_l(r) \to \hat j_l ( kr)$ and subjected to the asymptotic
condition at $r \to \infty$~\cite{galindo2012quantum}
\begin{eqnarray}
u_{k,l}(r) \to \frac12 \left[  F_l(k)\,\hat h_l^{(2)} (kr) 
+ F_l(-k)\,\hat h_l^{(1)} (kr)
 \right],
\label{eq:jost}
\end{eqnarray}
where $\hat h_l^{(1,2)} (x) = \hat j_l(x) \pm i \hat y_l(x) $ are the
reduced Hankel functions fulfilling $\hat h_l^{(1)} (x)^* = \hat
h_l^{(2)} (x) $ and $\hat h_l^{(1)} (-x) = (-1)^{l+1} \hat h_l^{(2)}(x) $.  
Thus, the S-matrix is then defined as
\begin{eqnarray}
S_l(k) = \frac{F_l(-k)}{F_l(k)} = e^{2 i \delta_l(k)},
\end{eqnarray}
so that the scattering amplitude becomes 
\begin{eqnarray}
f_l(k) = \frac{F_l(-k)-F_l(k)}{2 i k F_l(k)}.
\end{eqnarray}
The Jost functions fulfill the reflection conditions 
\begin{eqnarray}
  F_{l}(-k^*) = F_{l}(k)^*
  \label{eq:refl}
\end{eqnarray}
for complex $k$. Furthermore, due to~\eqref{eq:refl}, it can be shown
\cite{goldberger2004collision,nussenzveig1972causality} that the functions
\begin{eqnarray}\label{eq:n-and-d}
n_l(k) = \frac{F_l(-k)-F_l(k)}{2 i k}\quad{\rm with}\quad d_l(k) = F_l(k) 
\end{eqnarray}
fulfill the relations
\begin{eqnarray}
  n_l(k)^*  = n_l(-k^*), \qquad   n_l(k)  = n_l(-k),
\end{eqnarray}
which means that $n_l(k)$ is purely real for $\Im\,k=0$.
Moreover, ${\rm Re}\, n_l(k)={\rm Re}\,n_l(-k^*)$ and $ {\rm Im}\, n_l(k)  = -{\rm Im}\, n_l(-k^*)$,
so that $n_l(k)$ is purely imaginary for ${\rm Re}\,k=0$. In addition, 
\begin{eqnarray}
\lim_{k \to \infty} n_l (k) = 0,\qquad   \lim_{k \to \infty} d_l (k) = 1.
\end{eqnarray}
Thus, they have the desired properties for a potential constructed as a
superposition of Yukawa potentials.
A straightforward consequence of these properties is Levinson's
theorem.  While the proofs of these statements have been known for a
long time~\cite{fivel1960analytic,newton1960analytic}, to our
knowledge they have not been considered within the present context.
We will review them adapted to our complete potential, 
which can be decomposed into two pieces: a cut-off potential and a Yukawa superposition.  
While they are discussed separately in the literature,
our case at hand involves both cases at the same time. 
We will show next the pertinent steps.

\subsection{Analytic properties}

The scattering amplitude obtained perturbatively from $\chi$PT enjoys
analytical properties deduced from the corresponding Feynman diagrams,
i.e. particle exchange generated at the partial-wave level a left-hand cut,
which discontinuity has been used to reconstruct the chiral potential.
A relevant question is whether our resulting full amplitudes obtained
by solving the Schr\"odinger equation, share these properties beyond first
order in perturbation theory, i.e. whether they satisfy, as expected,
dispersion relations.

In order to check the analytical properties of the quantum mechanical
amplitude, we note that our chiral potential is indeed a superposition
of Yukawa potentials with the exception of the explicit $s$-dependence
in the spectral function. 
It is convenient to write the potential 
as a superposition of exponentials in the form
\begin{eqnarray}\label{eq:yukawa-sum}
U_I(r,s) = \int\limits_{2 m_\pi}^\infty \diff\mu \sigma_I (s,\mu) e^{-\mu r},
\end{eqnarray}
where
\begin{eqnarray}
\sigma_I(s,\mu) = \int\limits_{2 m_\pi}^\mu \diff\mu' \rho_I (s,\mu')
\end{eqnarray}
and $\rho_I$ was defined in~\eqref{eq:rhoI}. 
This is indeed our case for the TPE potential in $\pi\pi$ scattering, 
as one can see comparing~\eqref{eq:spec-pot} and~\eqref{eq:yukawa-sum}.  
Being a Laplace transformation, it corresponds to the so-called
{\it analytical potential} in the complex-$r$ plane for ${\rm Re}\, r
>0$~\cite{taylor1972scattering}, which fulfills the relation
$\lim_{\rho \to 0}\,U_I( \rho e^{i \theta},s)=0$ for $-\pi/2 < \theta < \pi/2$,
with $r=\rho e^{i \theta}$.

As we will see below, while the exponential falloff of the potential at
long distances suffices to prove the analyticity in the strip $ | {\rm
  Im}\,k | < m_\pi $, with $k=\sqrt{s/4-m_\pi^2}$, the spectral
decomposition is indeed needed to establish the cut along the line
${\rm Re}\,k=0$ and $ m_\pi < {\rm Im}\,k $ of the S-matrix. There are
several versions of the proof.  On the one hand, it can be proved by
estimating a bound for the Jost function using directly the spectral
representation.  On the other hand, by profiting from the analytical
character of the potential and deforming the integration in $r$ into
the complex plane so that $k r >0 $.  For completeness we will review
next and in a sketchy fashion the second method, as it does not
require a bounded spectral function or the use of an spectral
regularization (see Section~\ref{sec:spec}).

The determination of the analyticity domain of the quantum mechanical
problem is based on the equivalent integral equation for the Jost
Functions as follows
\begin{eqnarray}\label{eq:fixed-r}
u_{k,l} (r) = \hat j_l(kr) + \int\limits_0^r \diff r'  {\cal K}_{k,l}(r,r') U(r',s) u_{k,l} (r'),
\end{eqnarray}
which is a Volterra type integral equation and the kernel is given by
\begin{eqnarray}
{\cal K}_{k,l}(r,r') = \frac{i}{2k}
\left[ h_l^{(1)} (kr') h_l^{(2)} (kr) - h_l^{(1)} (kr) h_l^{(2)} (kr') 
\right]\,.\nonumber\\ 
\end{eqnarray}
Taking the large-$r$ limit and comparing with~\eqref{eq:jost}, we get 
\begin{eqnarray}
F_l(k) = 1 + \frac{i}{k} \int\limits_0^\infty \diff r\,\hat h^{(1)}_l (kr) U(r,s) u_{k,l} (r).
\end{eqnarray}
This equation is the basis for the analytical continuation to the
complex-$k$ plane. Actually, in the limit $r \to \infty$ one has
from~\eqref{eq:jost} that $ |h^{(1)}_l (kr) | = {\cal O} (e^{-{\rm
    Im}\,k\,r})$ for ${\rm Im}\,k> 0$.  Thus, even when $|u_{k,l} (r)|
={\cal O} (e^{{\rm Im }\,k\,r})$, one has a finite integral provided
that the potential goes to zero.  As a consequence, $F_l(k)$ is
analytical for ${\rm Im }\,k> 0$.  On the contrary, for $- m_\pi <
{\rm Im}\,k< 0$ one has $|h^{(1)}_l (k\,r)\,U(r,s)\,u_{k,l} (r) | =
{\cal O} (e^{-2{\rm Im}\,k\, r- 2 m_\pi r} )$ which is convergent for
${\rm Im k}> -m_\pi$. Therefore, $F_l(k)$ is analytical for ${\rm Im
  k}> -m_\pi$ and hence $F_l(-k)$ is analytical for ${\rm Im }\,k <
m_\pi$.  In conclusion, $S_l(k)= F_l(-k)/F_l(k)$ is analytical in the
strip $ | {\rm Im}\,k | < m_\pi $.

For $k=|k| e^{i\theta}$, due to the analytical
character of $U(r,s)$, we can deform the contour to $r=\rho e^{-i\theta}$, 
so that, for {\it any} term in the spectral integral over
$\mu$, one has a pole at $k = -i \mu/2$.
In the case of $F_l(k) $ and $F_l(-k) $, 
this pole becomes a cut after $\mu$-integration along the lines
${\rm Re}\,k=0$ and $-\infty < {\rm Im}\,k < -m_\pi$ and $ m_\pi < {\rm
  Im}\,k < \infty$, respectively. The fact that $F_l(k)$ is analytical
for ${\rm Im}\,k>0$ and $\lim_{k\to\infty} F_l(k) \to 1$ allows one to write a dispersion
relation in the upper half-circle
\begin{eqnarray}
F_l(k)=1 + \frac1{\pi} \int\limits_0^\infty \diff k' \frac{k'}{k'^2-k^2} {\rm Im} F_l(k'),
\end{eqnarray}
where the antisymmetry of ${\rm Im} F_l(-k)=-{\rm Im} F_l(-k)$ from~\eqref{eq:refl} has been used. 
Thus, passing to the variable
$s=4(k^2+m_\pi^2)$, which is one-valued in ${\rm Im} k >0$, one can define the function~\footnote{We use capital letters for the function $n_l$ and $d_l$ in~\eqref{eq:n-and-d} when referring to the $t_{l}(s)$ amplitudes.} 
\begin{eqnarray}
t_{l} (s) = \frac{N_l(s)}{D_l(s)},
\end{eqnarray}
so that we can identify $D_l(s)=F_l(k)$ for ${\rm Im} k > 0$
\begin{eqnarray}
D_l(s) \equiv 1 + \frac1{\pi} \int\limits_{4 m_\pi^2}^\infty \diff s'  \frac{{\rm Im} D_l(s')}{s'-s-i 0^+},
\end{eqnarray}
where $ {\rm Im}\, D_l(s') = {\rm Im}\,F_l(s') $ has a right-hand cut in
$4 m_\pi^2<s<\infty$ and is real for $s<4 m_\pi^2$.  
Furthermore, in the elastic approximation, using~\eqref{eq:unit} we get
\begin{eqnarray}
{\rm Im} D_l(s)= - \sigma(s) N_l(s) \, , \qquad s > 4 m_\pi^2 \, ,  
\end{eqnarray}
where 
\begin{eqnarray}
N_{l} (s) = \frac{F_l(-k)-F_l(k)}{2 i \sigma(s)} \, ,
\end{eqnarray}
is real for $s> 4 m_\pi^2$.  For real $s <0$ we get that $s+i0^+
\leftrightarrow 0^+ + i \kappa$ with $\kappa> m_\pi$ and hence 
\begin{eqnarray}
N_{l} (s+i0^+)  &=& \frac{F_l(-0^+ - i \kappa )-F_l(0^+ + i \kappa)}{2 i \sigma(s)} \, ,\\ 
N_l(s-i0^+) &=&  \frac{F_l(-0^+ + i \kappa)-F_l(0^+ - i \kappa)}{2 i \sigma(s)} \, . 
\end{eqnarray}
Thus, the discontinuity is 
\begin{eqnarray}
{\rm Disc}\,N_{l} (s) = 2 i\,{\rm Im}\,N_l(s) = \frac{{\rm Im}\,F_l(-0^+ - i \kappa)}{i\,\sigma(s)},
\end{eqnarray}
where we have used that for ${\rm Im} k>0$, $F_l(k)$ is analytical
and hence $F_l(0^+ + i \kappa)=F_l(-0^+ + i \kappa)$. 
Finally, note that the $s$-dependence appearing in the
spectral function does not spoil these analytic properties. This
completes the proof that the scattering amplitude obtained by solving
the Schr\"odinger equation for the potential in~\eqref{eq:full-pot} has the correct analytical properties.

\subsection{Coarse graining}

\begin{figure}
\centering
\includegraphics[scale=0.4,angle=-90]{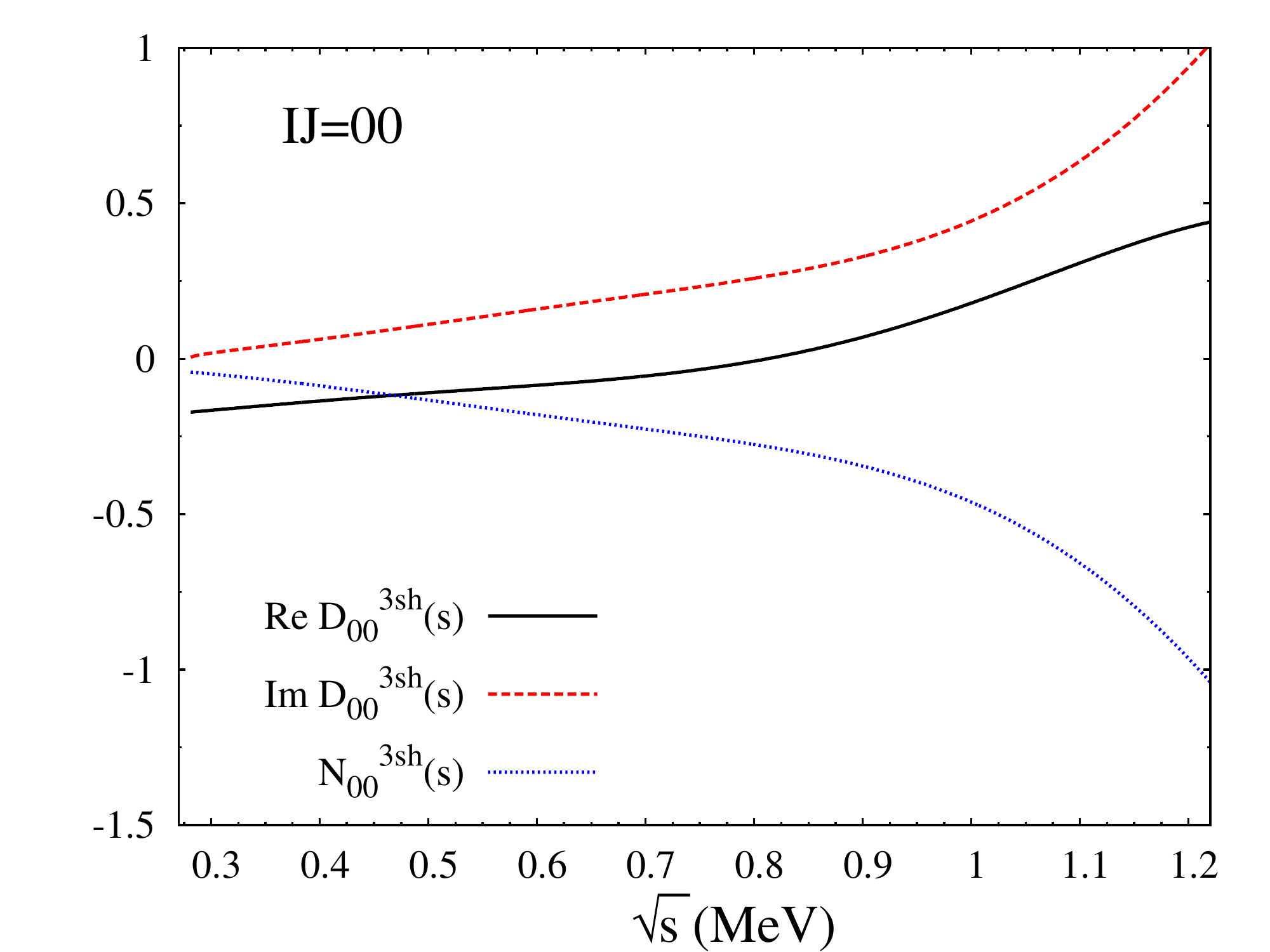}\\
\hspace{-0.14cm}\includegraphics[scale=0.411,angle=-90]{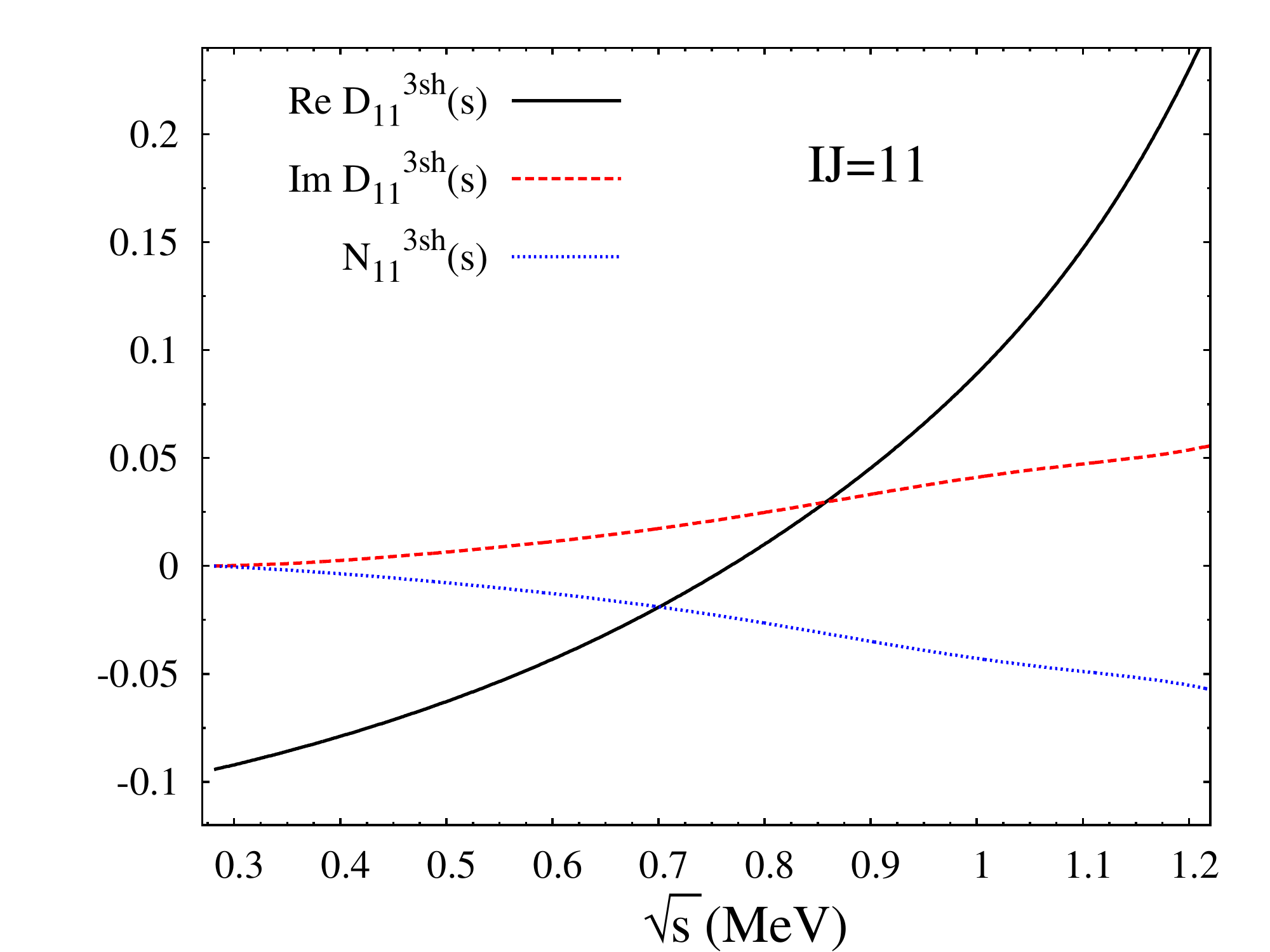}\\
\includegraphics[scale=0.4,angle=-90]{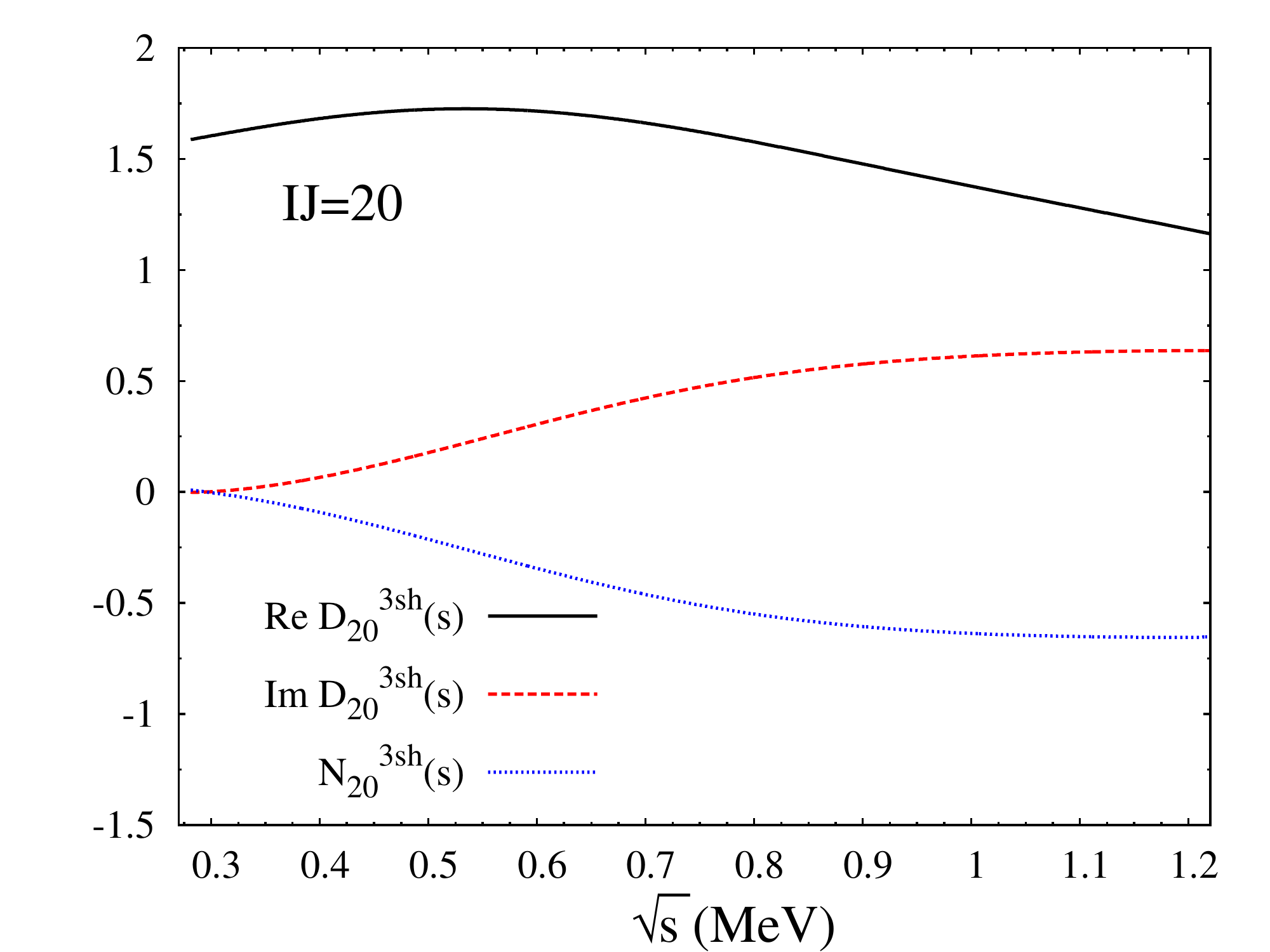} 
\caption{Color online: $N_{JI}$ and $D_{JI}$ (dimensionless)
  functions appearing in the N/D method (see main text) as a function
  of the CM energy for $I=0$ (top) , $I=1$ (middle) and $I=2$
  (bottom).}\label{Fig:nd}
\end{figure}

In order to calculate the Jost functions, in practice we coarse grain
the interaction using the delta-shells and we proceed as before. In
the discretized version we define the accumulated Jost functions
$F_{l,n}(k) $ and $F_{l,n}(-k)$ as
\begin{eqnarray}
u_{l,n}(r) &=& \frac12 \left[  F_{l,n}(k) \hat h_l^{(2)} (kr) 
+ F_{l,n}(-k) \hat h_l^{(1)} (kr) \right] \,, \nonumber \\ 
&& r_n < r < r_{n+1} \, . 
\end{eqnarray}
However, we keep track of both the continuity of the wave functions
and the discontinuity of the derivative separately, 
\begin{eqnarray}
u_l (r_{n}^+) -u_l (r_{n}^-) &=& 0 \nonumber\\ 
u_l' (r_{n}^+)- u_l' (r_n^-)&=& U(r_n) u_l (r_{n}) \,\Delta r\,,
\end{eqnarray}
with $r_n^+ \equiv r_n +0^+$ and $r_n^- \equiv r_n +0^-$. Again we use 
the fact that the Wronskian $2i ( \hat h_l^{(1)} (x) \hat h_l^{(2)\,\prime}(x)- \hat h_l^{(1)\,\prime}  (x) \hat h_l^{(2)} (x))= 1$  so that 
\begin{eqnarray}
F_{l,n+1}(k) &=& A_{l,n}(k) F_{l,n} (k) + B_{l,n}(k) F_{l,n} (-k) \nonumber\\
F_{l,n+1}(-k) &=& C_{l,n}(k) F_{l,n} (k) + D_{l,n}(k) F_{l,n} (-k) 
\end{eqnarray}
where we have introduced the coefficients, 
\begin{eqnarray}
A_{l,n}(k) &=& 1+\frac{i \Delta r \, U(r_n) h_l^{(1)}(k r_n) h_l^{(2)}(k r_n)}{2 k}\,, \nonumber\\ 
B_{l,n}(k) &=& \frac{i \Delta r \,  U(r_n) h_l^{(1)}(k r_n){}^2}{2 k}\,, \nonumber\\ 
C_{l,n}(k) &=& -\frac{i \Delta r \,  U(r_n) h_l^{(2)}(k r_n){}^2}{2 k}\,, \nonumber\\ 
D_{l,n}(k) &=& 1-\frac{i \Delta r \, U(r_n) h_l^{(1)}(k r_n) h_l^{(2)}(k r_n)}{2 k}
\end{eqnarray}
and the initial conditions are 
\begin{eqnarray}
F_{l,0}(k) = 1,  \quad F_{l,0}(-k) = 1, 
\end{eqnarray}
which corresponds to take the normalization $u_{l,0}(r) = \hat j_l(kº,r)$. 
The final values are 
\begin{eqnarray}
F_{l}(k) \equiv F_{l,N}(k) \, , 
  \qquad F_{l}(-k) \equiv F_{l,N}(-k).
\end{eqnarray}
For illustration purposes, we plot in Fig.~\ref{Fig:nd} the functions
$N_{IJ}(s)$ (which is real) and $D_{IJ}(s)$ (which is complex)
separately above the threshold for the five delta-shell case. 
Of course, these functions reproduce
the phase shifts presented before. As expected, ${\rm Im}\,D_{IJ} (4m_\pi^2)=0$. In this representation the Breit-Wigner position of the
resonance is given by computing the zeros of ${\rm Re}\,D_{IJ}(s_R)=0$.  
While it is not shown in the pictures, the Jost functions display some
oscillatory behavior at higher energies due to the explicit delta-shells. 
Nevertheless, they still go to the expected values $D(\infty)=1$ and $N(\infty)=0$.

\section{Coarse graining inelasticities}
\label{sec:inel}

\subsection{Energy dependent coarse graining}

\begin{figure}
\centering
\includegraphics[scale=0.4,angle=-90]{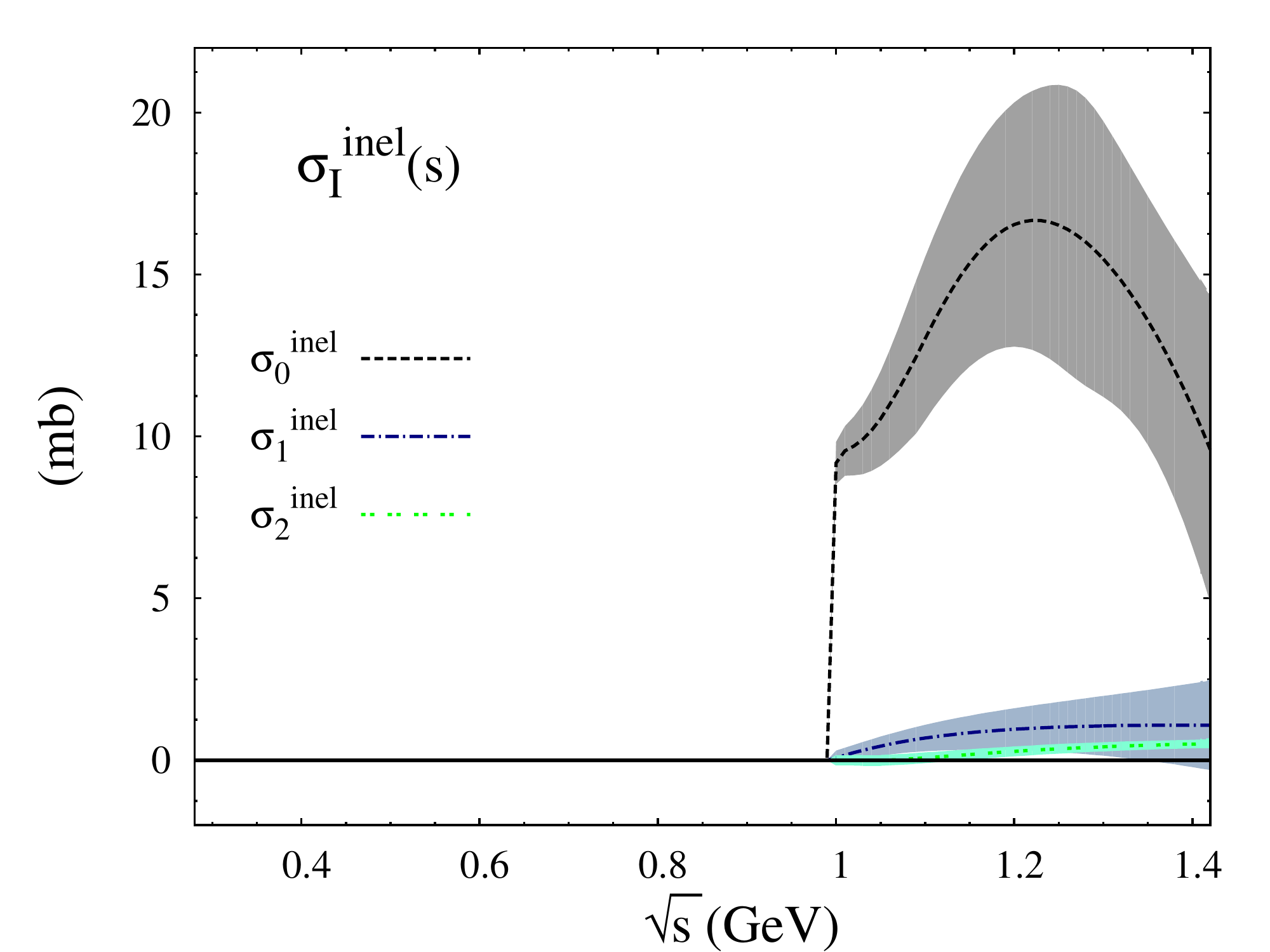} 
\caption{Color online: inelastic cross-section
in the different isospin channels as a function of the CM energy 
below   $\sqrt{s}=1.42$~GeV.}\label{Fig:sig-inel}
\end{figure}

As mentioned in the introduction, 
at sufficiently high energies particles are produced
and elastic scattering happens in the presence of absorption, which one
may view as a leak or hole in the probability.  The inelastic region
is characterized by the recombination of the two-pion internal
configuration, which, of course, demands a complex energy-dependent
potential.  The complexification of the potential can be understood in
terms of the loss of probability of the elastic channel. 
In order to have an idea of the size of the inelastic hole $a_{\rm inel}$, we show in
Fig.~\ref{Fig:sig-inel} the inelastic $\pi\pi$ total cross sections in
different isospin channels, defined as 
\begin{eqnarray}
  \sigma^{\rm inel}_{I}(s)= \frac{4 \pi}{k^2}\sum_{J} (2 J+1) \left[ 1- \eta_{IJ}^2(s)  \right] \, . 
\label{eq:inel-sig}
\end{eqnarray}
If we take $\sigma^{\rm inel}_I = 4 \pi a_{\rm inel}^2$, 
the largest inelastic cross section in Fig.~\ref{Fig:sig-inel} is compatible with an inelastic hole of less than half a
fm, i.e.  $a_{\rm inel} \lesssim 0.5$ fm. Of course, this corresponds according
to~\eqref{eq:inel-sig} to the contribution of all partial waves.

\begin{figure}
\centering
\includegraphics[scale=0.7]{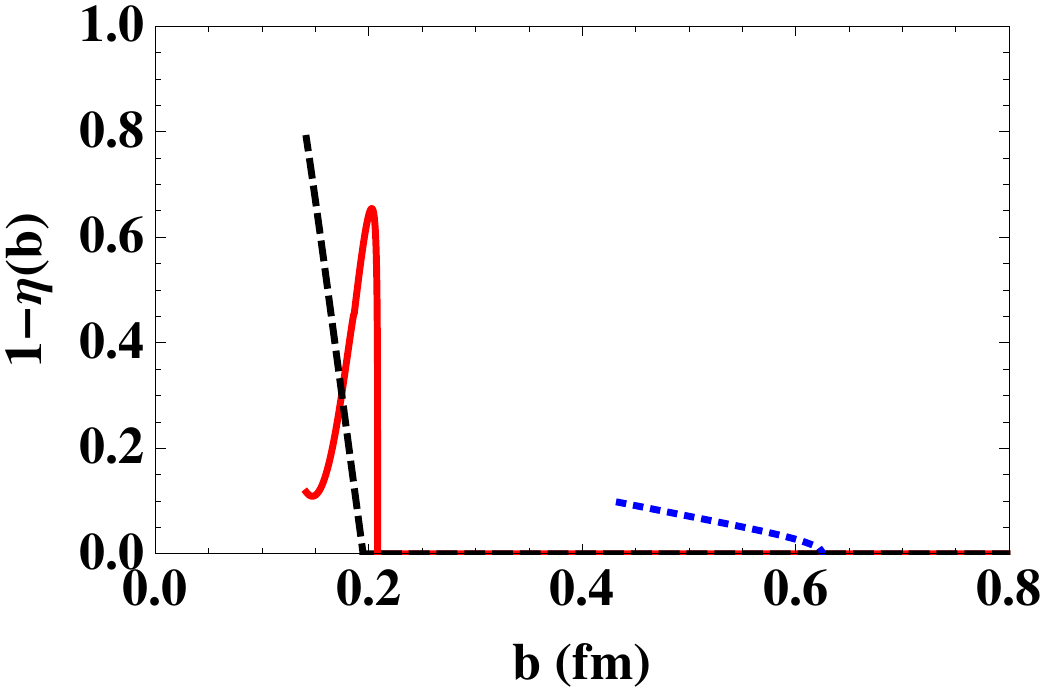} 
\caption{Color online: inelastic profile as a function of the impact
  parameter $b= (l+1/2)/p $ for the S0 (full, red) , P (dotted, blue)
  and S2 (dashed, black) waves when the maximum CM energy becomes
  $\sqrt{s}=1.42$~GeV.}\label{Fig:etab}
\end{figure}

As we have already seen, this loss of probability at the partial-wave 
level is parametrized in terms of a momentum-dependent inelasticity $\eta_I^J(s)$, see~\eqref{eq:pw-def}.
The imaginary part of the potential plays exactly the same role.
At this point, we will assume that this complexification of the
potential can be implemented just in the most inner layer of the
potential. This assumption can be justified by analyzing the
phenomenological inelasticities as a function of the impact parameter
given by the relation $b = (l+1/2)/p$, with $l$ the angular momentum
quantum number. In order to provide the range of the inelasticity, we
show this dependence in Fig.~\ref{Fig:etab}.  For the S waves it is
found that only for the {\it smallest} value, $b_{\rm min} = 1/(2
p_{\rm max}) \lesssim \Delta r $, the inelasticity is $\eta \ll 1$. As
we see, if we take $\Delta r \sim 0.3 $ we may parameterize the
inelasticity by {\it one} single energy dependent and complex
parameter~\footnote{The fixed-$r$ dispersion relation in~\eqref{eq:dr} suggests
  that in fact this parameter is an analytical function of the energy.
  (see also the discussion below)}.

\begin{figure*}
\epsfig{figure=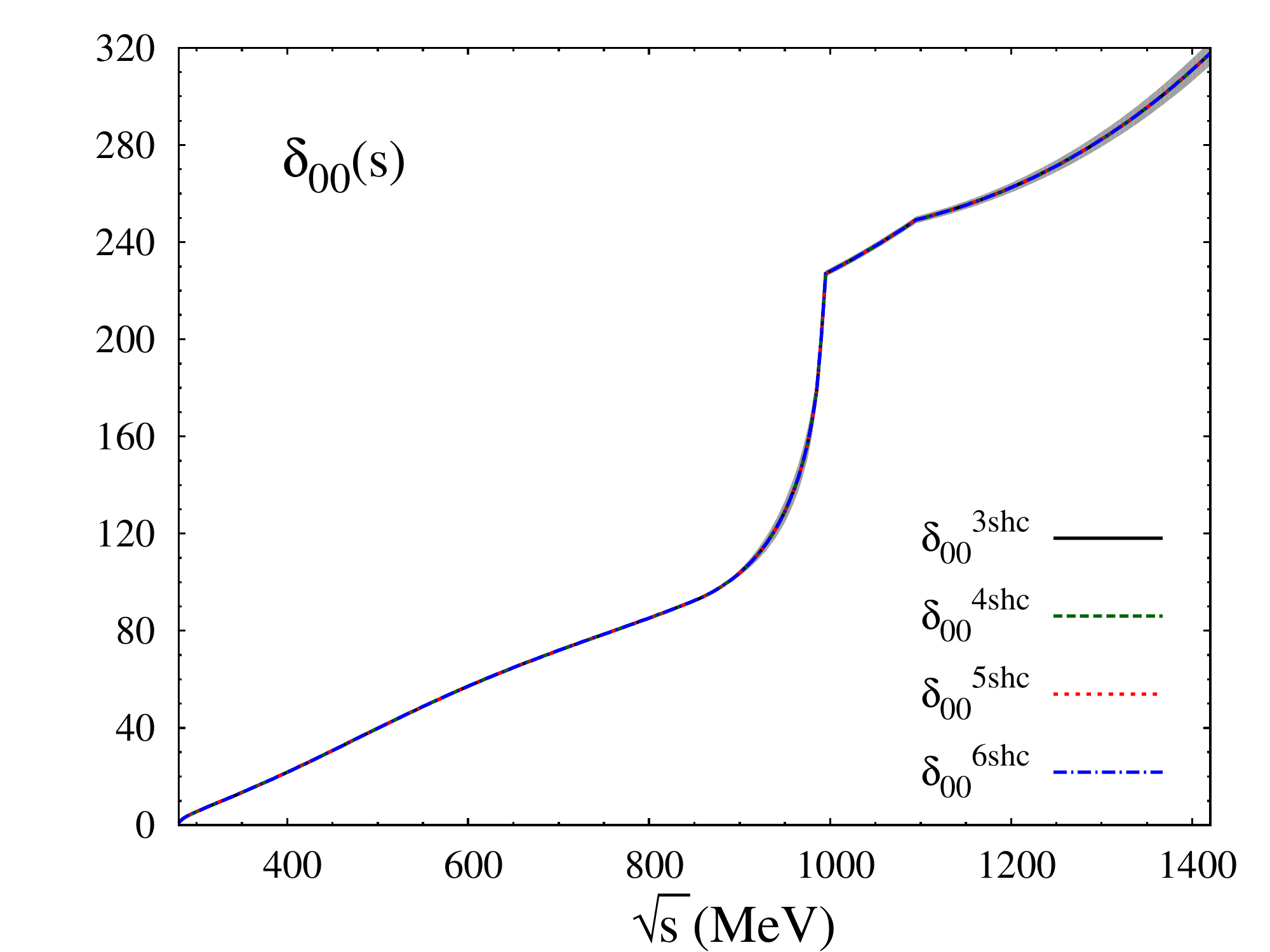,width=5.5cm,angle=-90}\epsfig{figure=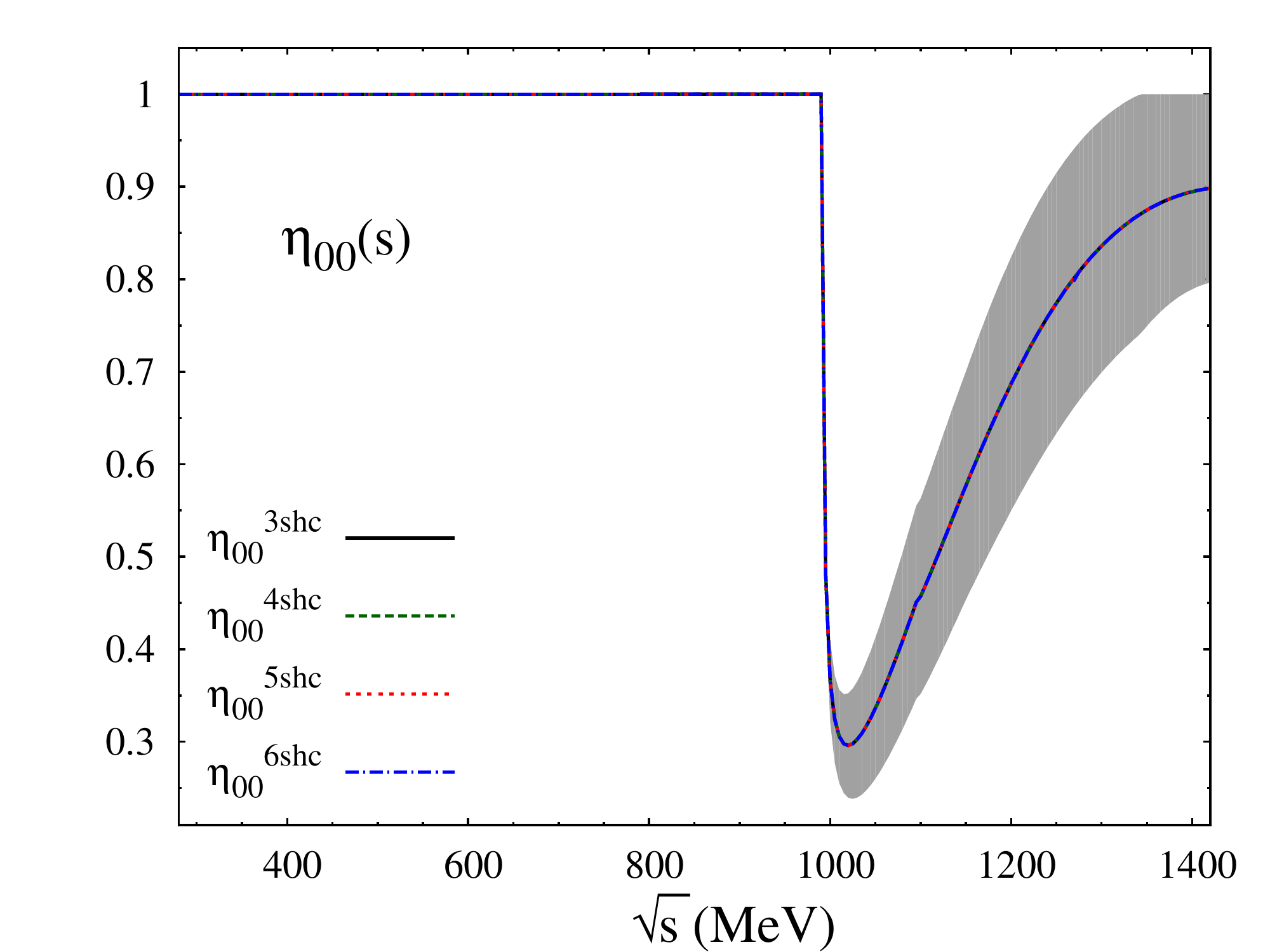,width=5.5cm,angle=-90}\\
\epsfig{figure=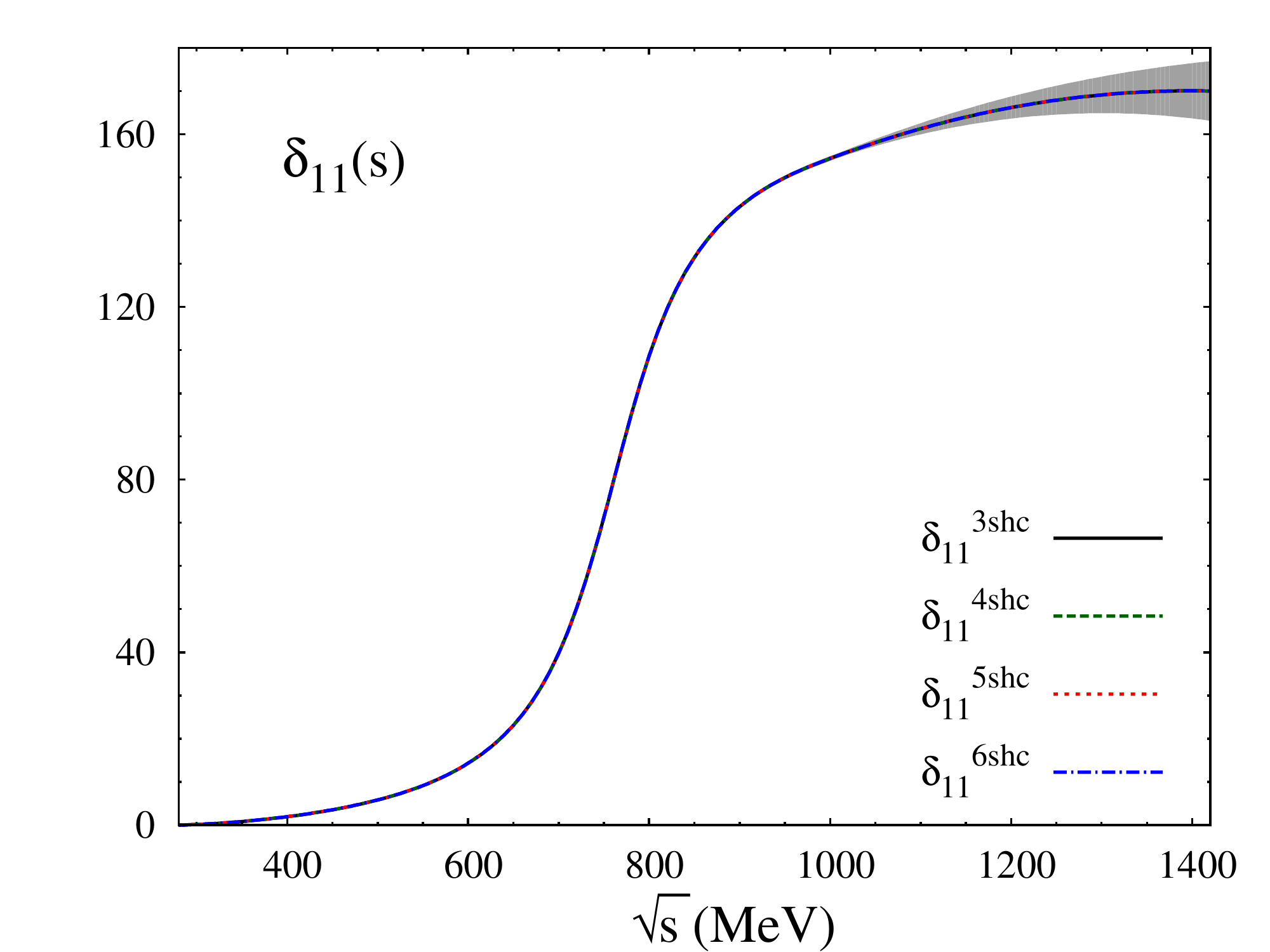,width=5.5cm,angle=-90}\epsfig{figure=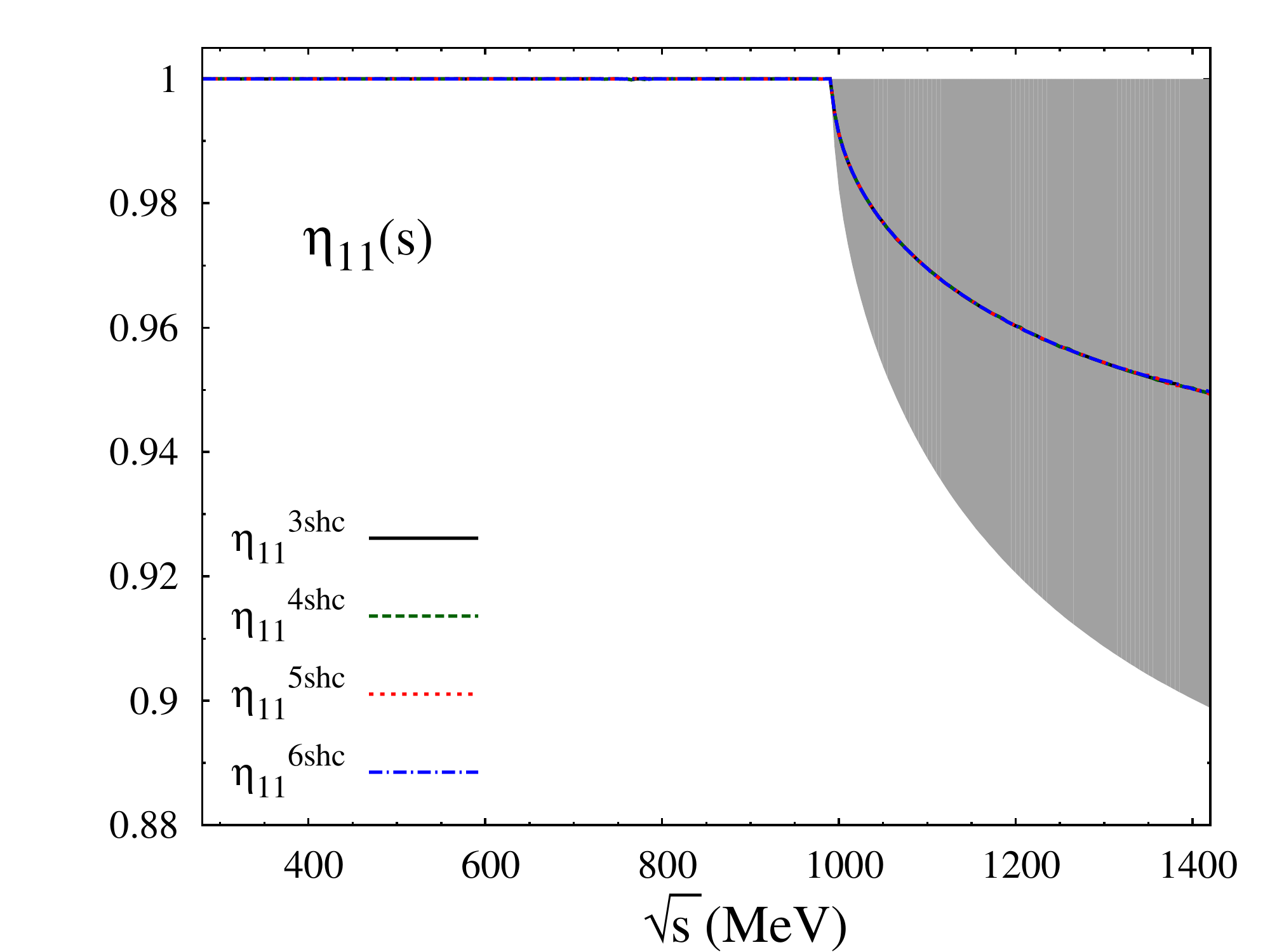,width=5.5cm,angle=-90}\\
\epsfig{figure=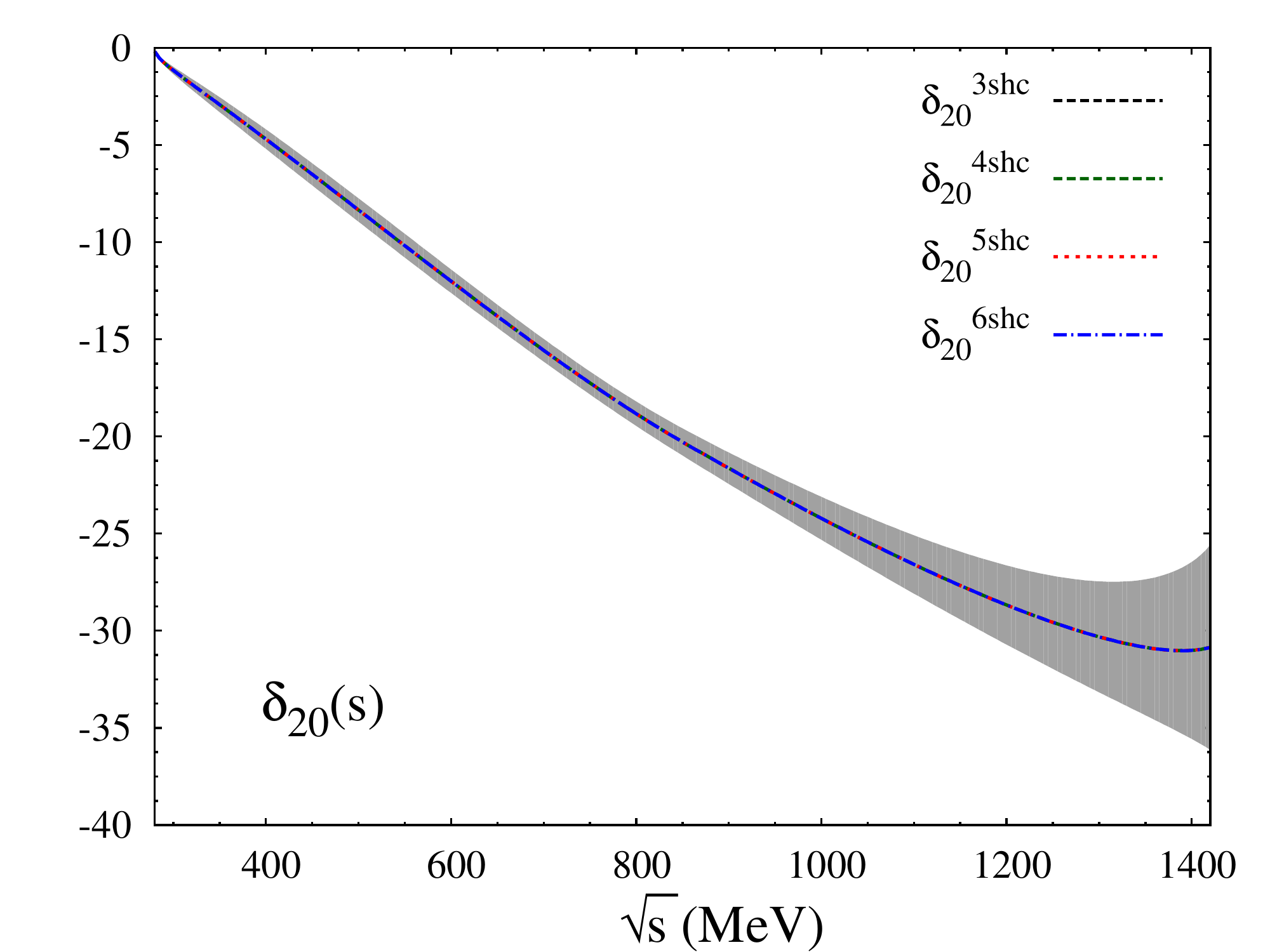,width=5.5cm,angle=-90}\epsfig{figure=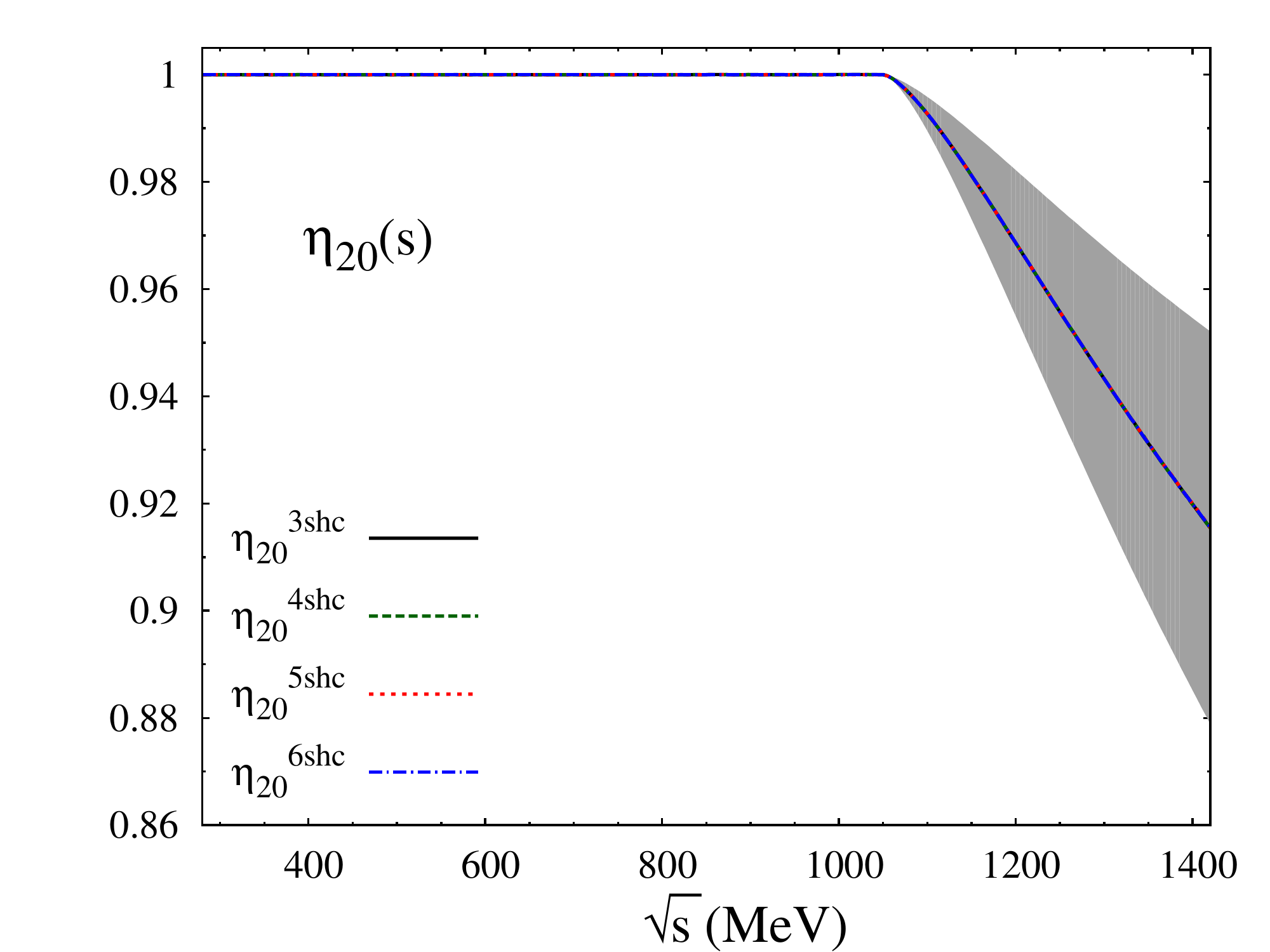,width=5.5cm,angle=-90}
\caption{S0-, P- and S2-wave phase shifts (left panels) and
  inelasticities (right panels).  The uncertainties are those quoted
  in~\cite{GarciaMartin:2011cn}, whereas solid-black, green-dashed,
  red-dotted and blue dot-dashed lines stand for the central results
  for $r_c$=0.9, 1.2, 1.5 and 1.8 fm, respectively.  Nevertheless, the
  procedure described in the main text allows one to describe the
  input exactly at each energy point, so the four lines coincide
  exactly.}\label{Fig:pipidata}
\end{figure*}

\begin{figure*}
\epsfig{figure=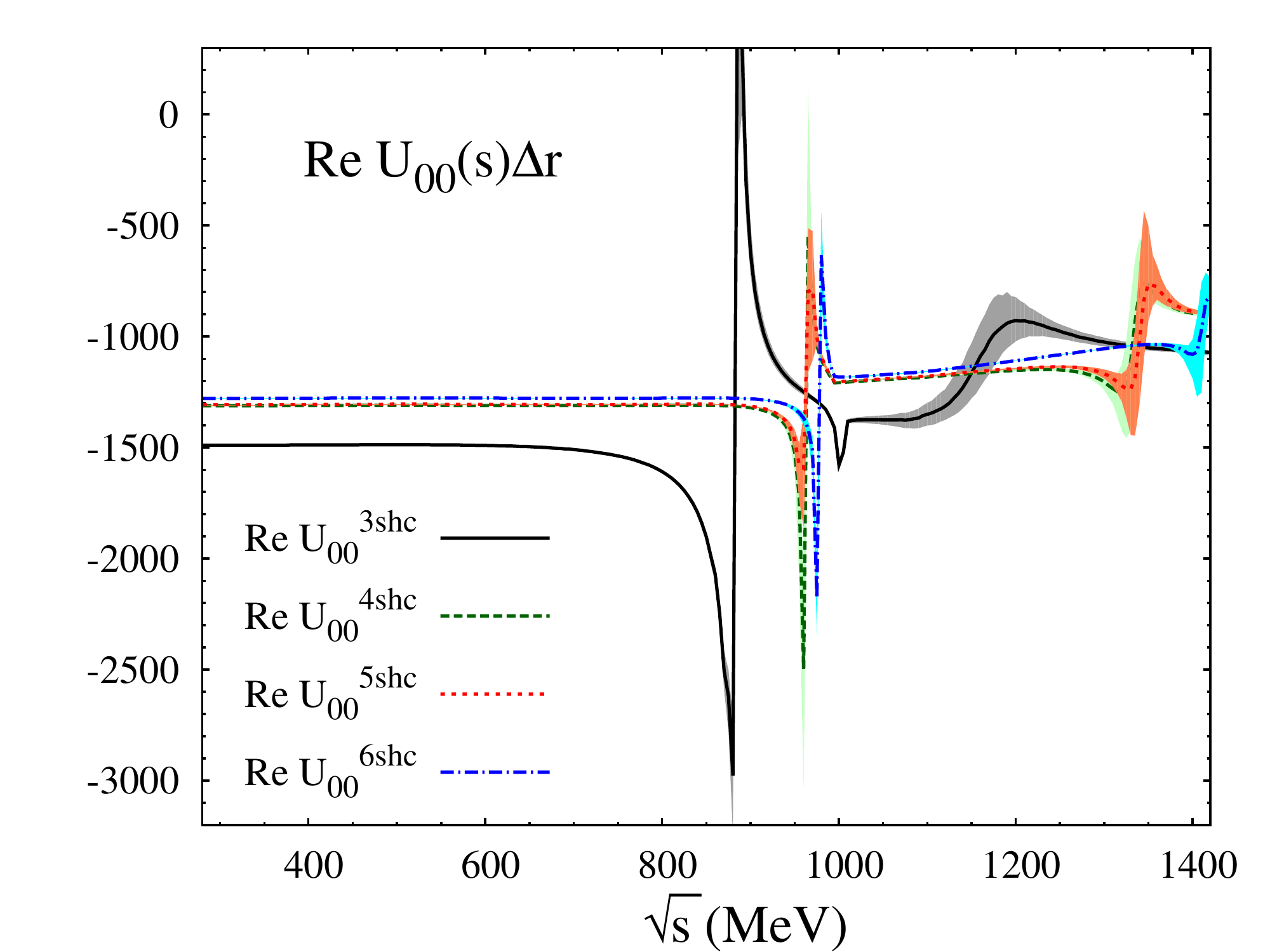,width=5.5cm,angle=-90}\epsfig{figure=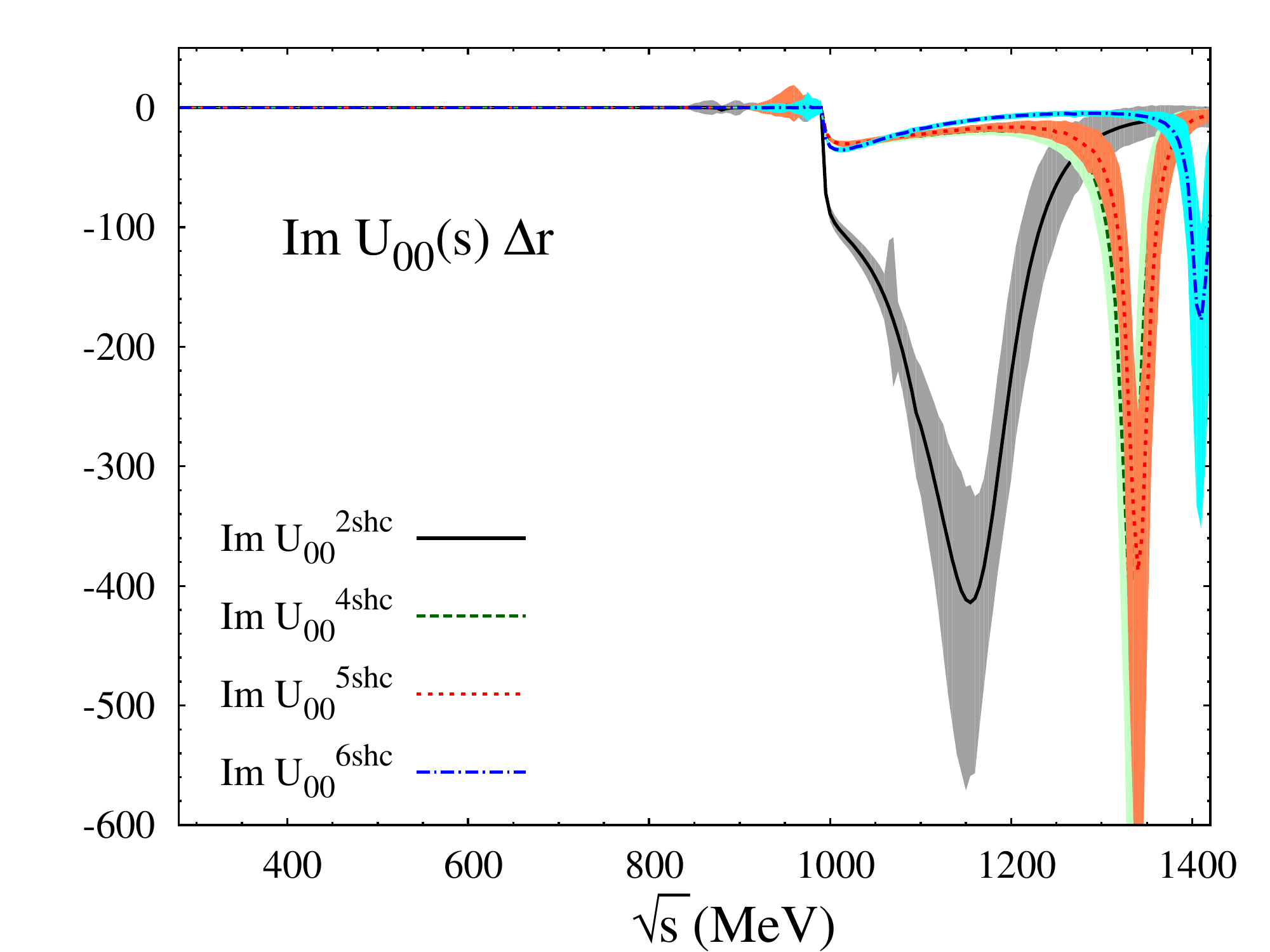,width=5.5cm,angle=-90}\\
\epsfig{figure=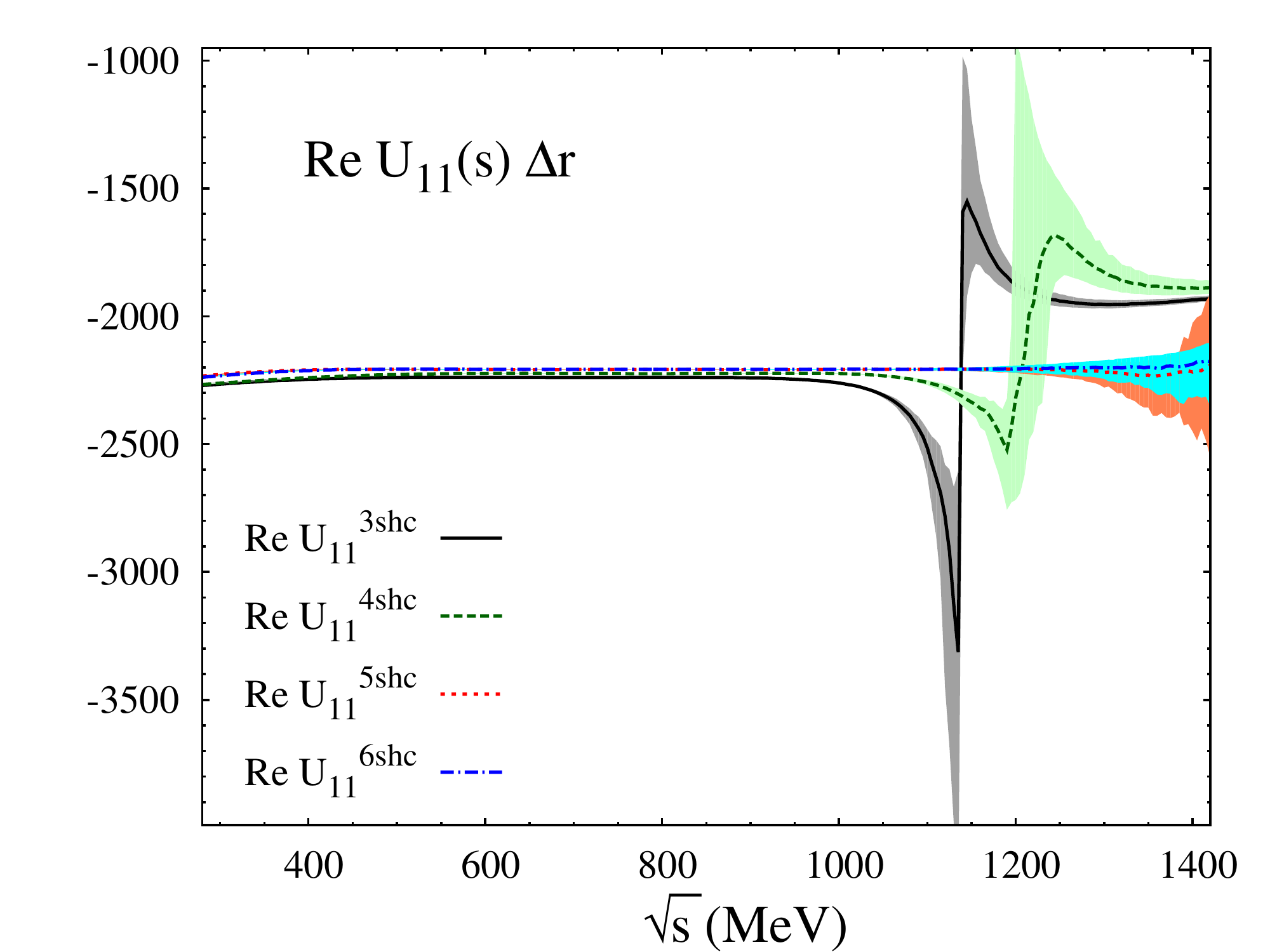,width=5.5cm,angle=-90}\epsfig{figure=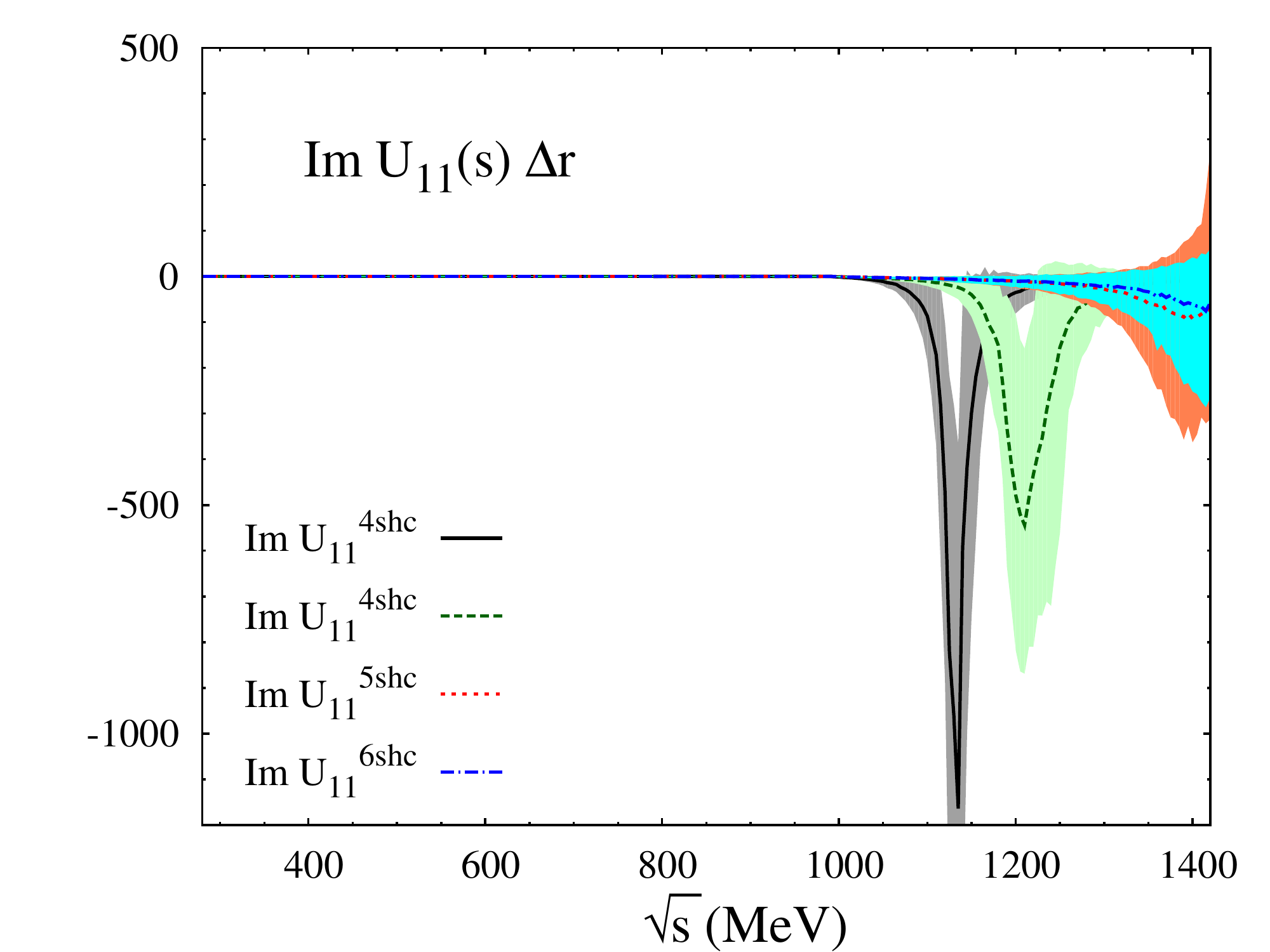,width=5.5cm,angle=-90}\\
\epsfig{figure=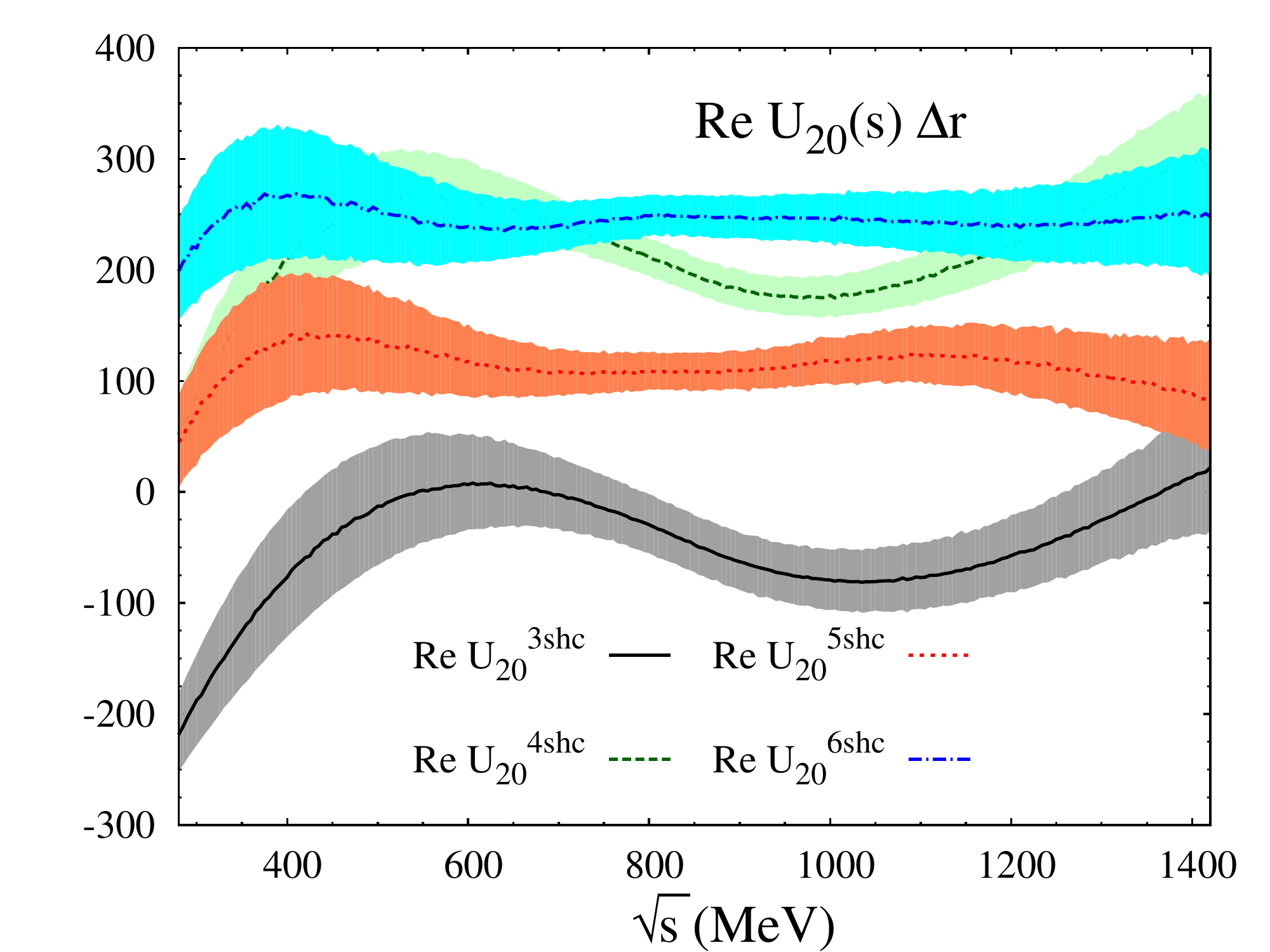,width=5.5cm,angle=-90}\epsfig{figure=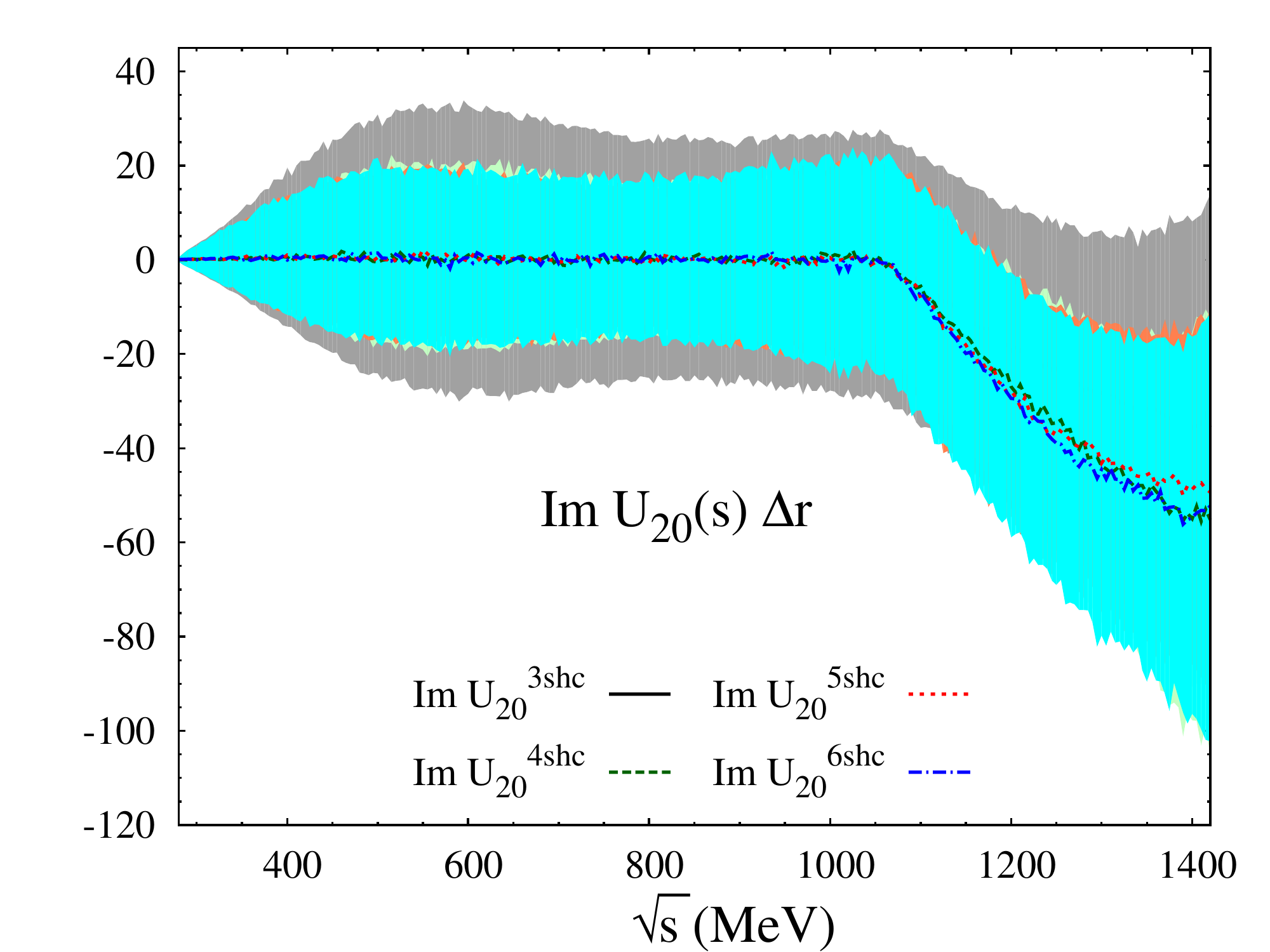,width=5.5cm,angle=-90}
\caption{Real and imaginary part of the inner delta-shell potential
  for the S0, P and S2 partial waves whereas solid-black, green-dashed, red-dotted and blue dot-dashed lines correspond to
  the results with $r_c=0.9$, $1.2$, $1.5$ and $1.8$ fm, respectively.  The error
  bands have been computed using at each energy point a bootstrap with
  a uniformly distributed sample of 1000 points and taking the 68\% of
  their distribution as the standard deviation.}\label{Fig:U0c}
\end{figure*}

Hence, starting from our elastic previous description, we
will assume $\lambda_0$ in~\eqref{eq:deltash} to be a complex unknown
function of the momentum and we will fit again the pseudo data given in~\cite{GarciaMartin:2011cn} for
the phase shift and inelasticity in~\cite{GarciaMartin:2011cn} up to
the maximum energy value provided $\sqrt{s_{\rm max}}=1.42$ GeV.
This procedure allows us to describe each partial wave (phase shift
and inelasticities) exactly at each energy point.  The results for the
four $r_c$-values considered are plotted in Fig.~\ref{Fig:pipidata},
whereas the value of the real and imaginary part of the inner
delta-shell layer is depicted in Fig.~\ref{Fig:U0c}.

\subsection{Traces of analyticity in the inelastic case}

An interesting feature which can be appreciated in Fig~\ref{Fig:U0c}
is the close resemblance of the real and imaginary parts of the inner
delta-shell coefficient with the expected behavior from dispersion
relations around a pole or a inelastic threhsold. These two effects reflect in the S0 and P channels
higher resonances or inelastic channels not explicitly included in the present optical
potential analysis. Interestingly, Cornwall and
Ruderman~\cite{cornwall1962mandelstam} found the fixed-r dispersion
relation for the optical potential $V(r,s)$ given in~\eqref{eq:dr}. The implementation of
such a dispersion relation in our analysis would reduce the number of
fitting parameters in the inelastic region
but it would also require a clear understanding of the high energy
behavior.  We leave this interesting investigation for future
research.

\begin{figure}
\centering
\includegraphics[scale=0.45,angle=-90]{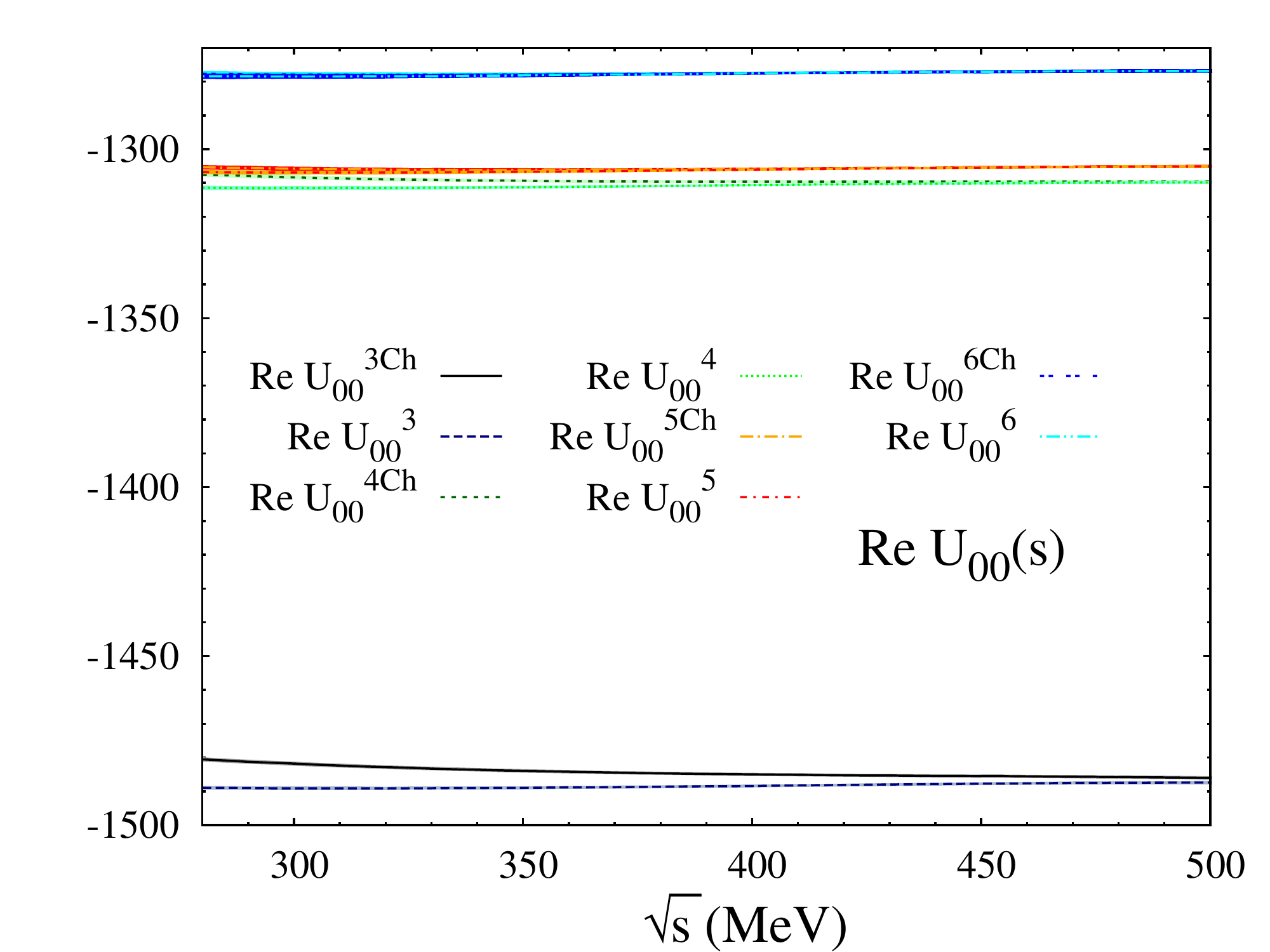}\\
\includegraphics[scale=0.45,angle=-90]{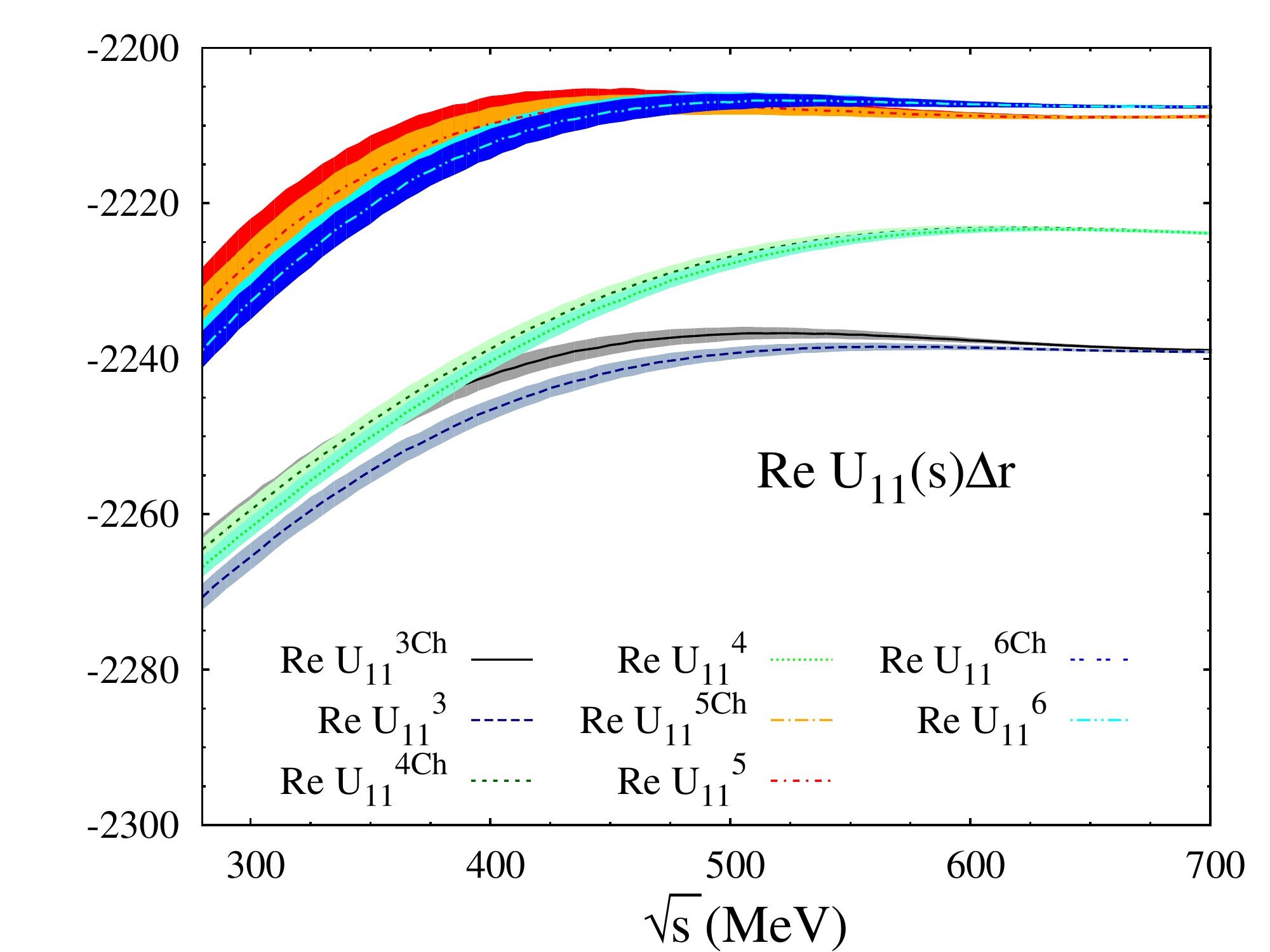}\\
\includegraphics[scale=0.45,angle=-90]{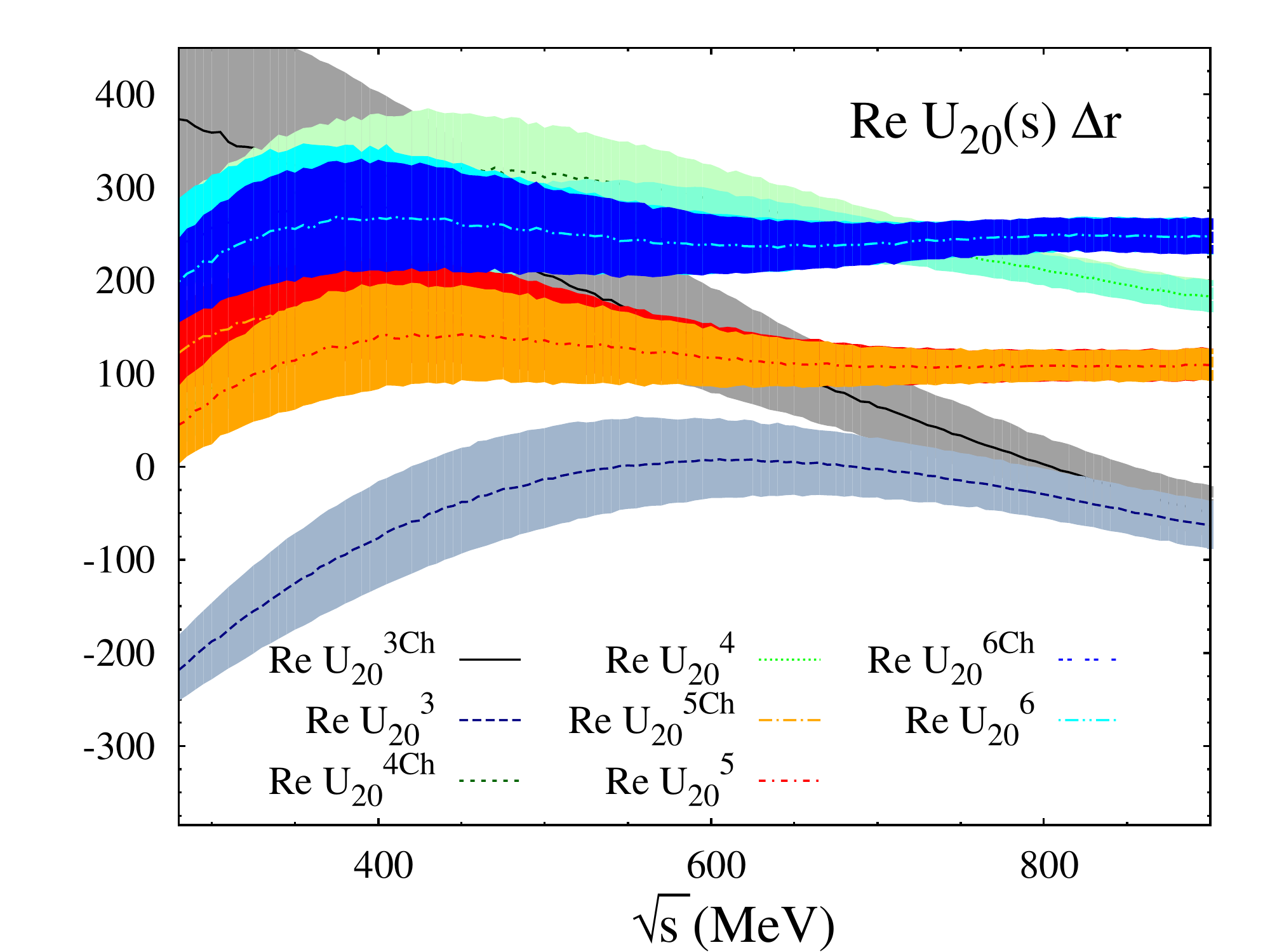}
\caption{Real part of the inner delta-shell potential for the S0, P
  and S2 partial waves when the chiral tail is included or
  excluded. The only differ at low energies. Gray (light-gray), green
  (light-green), red (orange) and blue (cyan) bands correspond to the
  results with (without) the chiral tail for $r_c=0.9$, $1.2$, $1.5$
  and $1.8$ fm, respectively.}\label{Fig:Uchi}
\end{figure}

The results for the real and imaginary part of the energy-dependet inner delta-shell
when the chiral tail is included differ from those without,  plotted in
Fig.~\ref{Fig:U0c}, only at low energies. The comparison is depicted in
Fig.~\ref{Fig:Uchi} showing again the rather small effect introduced
by chiral corrections in the inelasticity parameter.

Another interesting possibility which deserves some further
investigation is the generalization to the coupled channel case, to
account explicitly for the opening of the $K \bar K$ and $\eta \eta$
thresholds, while keeping multi-pion channels in the
inelasticity factor.

\section{Subtractions, low energy constants and 
the number of parameters}
\label{sec:LECs}

In our construction of the chiral $\pi\pi$ potential the low energy
constants (LECs) have been discarded as they do not contribute for $r
\neq 0 $. In fact, the spectral representation does provide the same
coordinate dependent potential regardless on the number of
subtractions.  

This raises the problem of where is this counterterm information gone
within the present approach. In this last section we want to address
the relation among subtraction constants in dispersion relations, low
energy constants and the number of independent parameters within our
coarse graining approach. We warn the reader that we have not succeeded
in finding a unambiguous relation, which may ultimately be traced to
two aspects. Firstly, it has to do with known ambiguities in mapping
different regularization methods, namely the one used in $\chi$PT and the
coordinate space regularization used here. Secondly, there is a
difficulty in separating the short distance parameters in perturbation
theory, particularly if we use a non-perturbative resummation scheme
to fit the parameters. 

In order to elaborate on this issue and appreciate the difficulties,
we proceed in perturbation theory and re-write the problem in terms of
a Fredholm integral equation,
\begin{eqnarray}
u_{k,l}(r) = {\hat j}_{l} (kr) + \int\limits_0^ \infty \diff r'\, G_{k,l}(r,r') U_{l}(r')\, u_{k,l}(r') \, , 
\label{eq:coupled_int}  
\end{eqnarray}
where ${\hat j}_{l}(x)=x j_{l}(x)$ is a reduced spherical Bessel function
of the first kind and $G_{k,l}(r,r')$ is the Green's function satisfying 
\begin{eqnarray}
\left[ \frac{\partial^2}{\partial r^2} - \left( 
\frac{l(l+1)}{r^2} - k^2 \right) \right] G_{k,l}(r,r') 
= \delta(r-r') \, . 
\label{eq:green-de}  
\end{eqnarray}
An analytic expression for the Green's function $G_{k,l}(r,r')$
 can be written in terms of two function ansatzs $u(r)$ and $v(r)$ as
\begin{eqnarray}
 G_{k, l}(r,r') = u(r) v(r') \theta(r-r') + u(r') v(r) \theta(r'-r) \, .
\label{eq:var}  
\end{eqnarray}
Inserting this in~\eqref{eq:green-de}, it follows that 
$u(r)$ and $v(r)$ are solutions of the homogeneous equation 
with a unit Wronskian, i.e.,  
\begin{eqnarray}
\left[ \frac{\partial^2}{\partial r^2} -\left( 
\frac{l(l+1)}{r^2} - k^2 \right) \right] u(r) &=&0 \nonumber\\ 
\left[ \frac{\partial^2}{\partial r^2} - \left(
 \frac{l(l+1)}{r^2} - k^2 \right) \right] v(r) &=&0 \nonumber\\
  u'(r) v(r) - u(r) v'(r) &=& 1 \, . 
\label{eq:sols}  
\end{eqnarray}
We choose one of the two solutions to be proportional to 
the regular one, ${\hat j}_{l} (kr)$. Then the
other linearly independent solution with the desired Wronskian 
has to be proportional to
 ${\hat  y}_{l}(kr)=kr\; y_{l}(kr)$, i.e. the
reduced spherical Bessel function of the second kind. Therefore
the Green's function of the ordinary differential equation with the
proper normalization can be written as
\begin{eqnarray}
 G_{k,l}(r,r') = \frac1{k}\,{\hat j}_{l} (kr_<)\, {\hat y}_{l} (kr_>) \, , 
\label{eq:green}  
\end{eqnarray}
where $r_<= {\rm min} \{ r,r'\}$ and $r_>= {\rm max} \{ r,r'\}$.

For the normalization of~\eqref{eq:coupled_int} the scattering amplitude  can be written as 
\begin{eqnarray}
t_{IJ} (s) = -\frac{\sqrt{s}}{p^2} \int\limits_0^{\infty} \, \dr \, \hat j_J
(pr) U^I (r) u_{k,l}(r) \,,
\end{eqnarray}
which using the perturbative expansion inferred from reiteration
of the integral equation gives 
\begin{eqnarray}
t_{IJ}^{(2)} (s) &=& -\frac{\sqrt{s}}{p^2} \int\limits_0^{\infty} \, \dr \,
[\hat j_J (pr)]^2 U^{(2)} (r) \,, \\ 
t_{IJ}^{(4)} (s) &=&
  -\frac{\sqrt{s}}{p^2} \int\limits_0^{\infty} \, \dr \, \dr' \, 
\hat j_J (pr)
  U^{(2)} (r) G_J (r,r') U^{(2)} (r') \hat j_J (pr') \nonumber \\
&-&\frac{\sqrt{s}}{p^2} \int\limits_0^{\infty} \,  \dr \, [\hat j_J (pr)]^2 U^{(4)} (r) 
\,.
\end{eqnarray}
Actually, the singularity structure in coordinate space suggests
introducing a short distance cut-off $r_c$ and hence a short distance
potential $U_{\rm Short}(r)$~\footnote{Here we assume for simplicity a
  local form.  More generally we may assume a nonlocal form of the
  type $ [U \psi] (r) =\int\limits_0^{r_c} U(r,r') \psi(r') \diff r'$ with
  $U(r,r') =0$ for $r>r_c$. The coarse graining interpretation below
  reduces effectively this non locality to a local form within a
  sampling distance $\Delta r$.}.  That means that in order to
identify numerically the counterterms in perturbation theory we may
split the integrals as 
\begin{eqnarray}
\int\limits_0^\infty = \int\limits_0^{r_c} + \int\limits_{r_c}^\infty 
\end{eqnarray}
so that we get 
\[
\int\limits_0^{r_c} r^2 \,  \dr \, [j_l (pr)]^2 U_{\rm Short} (r)
+ \int\limits_{r_c}^\infty r^2 \,  \dr \, [j_l (pr)]^2 U_{\rm Long} (r)\,
\]
and apparently the matching to the one-loop result could be undertaken
in a straightforward fashion. This is actually not so, since this
implies disentangling the fitting parameters in a chiral expansion,
namely  
\begin{eqnarray}
\lambda_n \equiv U(r_n) \Delta r = \lambda_n^{(2)} + \lambda_n^{(4)} +
\dots
\end{eqnarray}
where we can arbitrarily shift $\lambda_n^{(2)} $ and $
\lambda_n^{(4)} $ by equal but opposite constants keeping $\lambda_n$
constant, say the values of Tables~\ref{Tab:phasessh} or
\ref{Tab:phasessh-ch}. 

This situation is not exclusive to the present approach and in fact is
common to all unitarization schemes. For instance in the IAM method
the fitted LEC's are different than those of $\chi$PT or the
predicted unitarized amplitudes from $\chi$PT develop huge
uncertainties~\cite{Nieves:2001de}. 

The fact that this perturbative matching can ultimately provide a
successful description and still fulfilling the condition
$\lambda_n^{(2)} \gg \lambda_n^{(4)} $ is expected. We note, however,
that the small changes between the $\lambda_n$ parameters
corresponding to the case {\it without} TPE potential listed in
\ref{Tab:phasessh} and the case {\it with} TPE listed in
\ref{Tab:phasessh-ch} suggest this hypothesis.

\section{Conclusions}
\label{sec:conl}

The optical potential in $\pi\pi$ scattering is a meaningful object
under the most common and general assumption of the validity of
the Mandelstam double spectral representation. Therefore, it plays a
relevant role in the analysis of such an interaction within a
invariant mass formulation of the relativistic two-body problem.
Contrary to the more employed Bethe-Salpeter equation, such an
approach is free from well-documented spurious singularities, 
which are triggered by incomplete calculations embodying subsets of Feynman diagrams with
particle exchange. 
In addition, it is a much simpler approach in the CM frame, 
as it effectively reduces to a Schr\"odinger equation for equal mass particles. 

Within such a framework, in the present paper we have analyzed
$\pi\pi$ scattering from a coarse-grained point of view at distances
smaller than the elementarity radius of the pion.  This means sampling
the interaction in coordinate space at a resolution of the order of
the shortest de Broglie wavelength. In our case, where we choose a
maximal CM energy of $s_{\rm max}=2$ GeV$^2$, the resolution
turns out to be $\Delta r \sim 0.3$ fm.
As a results, we obtain successful fits in
the $S0$, $P$ and $S2$ partial waves with the expected number of
parameters. We have also analyzed the role of inelasticities by an
energy dependent coarse grained interaction. The implications from
chiral symmetry have also been analyzed in terms of a long-distance
potential featuring the two-pion-exchange mechanism.  
This potential has been determined for the first time.

A non-perturbative renormalization of the amplitude, based on boundary
conditions in coordinate space, is precluded by the energy dependence
of the chiral potential. For a finite short distance cut-off
about 1.2-1.5 fm this energy dependence becomes irrelevant.

A somewhat surprising result of our analysis is that explicit chiral
corrections play a minor role, since at the distances above 1.2-1.5 fm
chiral potentials are almost negligible. A rewarding consequence in
this regard concerns the extraction of phase-shifts from energy shifts
calculated in a finite box in the relative distance by means of the
Luscher formula~\cite{Luscher:1990ux}; its applicability requires the
interaction to vanish above a given size which provides a lower limit
to the size of the box, $L$. While the mere ${\cal O} (e^{-m_\pi L})$
nominal estimate suggests $L \gtrsim 2 {\rm fm}$, our analysis is
compatible with taking $L \sim 1-1.5 {\rm fm}$, a much smaller value.

We have also shown that the quantum mechanical re-interpretation of
the problem does not spoil the proper analytical properties of the
partial wave scattering amplitude, which are explicitly fulfilled by
the Feynman diagrams. Thus, the conventional dispersion relations
with both a left-hand cut due to particle exchange in the crossed channel 
and the right hand cut due to unitarity are fulfilled. 
The standard N/D decomposition of
the amplitude is explicitly realized, albeit without subtractions.
We expect subtractions to be encoded in the short distance components
of the potential, but the explicit determination and identification of
the subtraction constants in terms of the short-distance parameters
remains to be accomplished. 

An important advantage of the coarse graining
perspective is that the number of independent parameters is determined
{\it a priori} by the shortest wavelength and the available crossing
constraints.  These constraints become increasingly large as we increase the angular
momentum of the partial wave. This point deserves further
investigation and the determination of all partial waves with this
minimal number of parameters is left for future research.

A traditional objection to the successful data-driven unitarity
methods is based on their lack of a power counting scheme, reflecting
field dependence and off-shell ambiguities absent in the conventional
{\it bona fide} EFT framework. The present coarse graining approach to
$\pi\pi$ scattering makes no further assumptions than those usually
made.  Namely, it implements unitarity in the elastic regime and it
matches $\chi$PT in perturbation theory {\it above} a given separation
distance, which is estimated to be about 1.2-1.5 fm.  In contrast, it
does have the advantage that we can estimate the number of fitting
parameters {\it a priori}.  More demanding fits due to an increase in
the number of (consistent) data should not require more fitting
parameters, but rather determining short distance parameters more
accurately. Besides, the method is quite simple as it parameterizes
the unknown short distance behavior regardless of any power
counting. This also allows a discussion {\it a posteriori} of the
chiral contributions which turn out to be minor.

\vskip1cm

\section*{Acknowledgments}

We thank Jose Antonio Oller for a discussion and 
Jose Manuel Alarc\'on for reading the ms.  E.R.A is very
grateful to the AEC and the ITP in Bern for hospitality and finantial
support. Work partially supported by Spanish MINEICO and European
FEDER funds (grants FIS2014-59386-P, FIS2017-85053-C2-1-P and
FPA2015-64041-C2-1-P), Junta de Andaluc\'{\i}a (grant FQM-225) and the
Swiss National Science Foundation.

\newpage

\appendix

\section{Chiral $\pi\pi$ amplitudes and potentials}
\label{sec:pots-amps}

\subsection{$\pi\pi$ scattering at one loop in $\chi$PT}

At ${\cal O}(p^4)$ in $\chi$PT, the $\pi\pi$ elastic scattering amplitudes
can be written in the form \cite{Gasser:1983yg}:
\begin{widetext}
\begin{eqnarray}
A (s,t,u) &=& A_2 (s,t,u) + A_4 (s,t,u) \, ,\\
 A_2 (s,t,u) &=&
    \frac{s-m_\pi^2}{f^2}\, ,\\
A_4 (s,t,u) &=&
    \frac{1}{6\pi^2f^4}\Big \{ (2{\bar l}_1+{\bar l}_2-\frac72)s^2+
    ({\bar l}_2
-\frac56)(t-u)^2 + 4(3{\bar l}_4-2{\bar
    l}_1-\frac13) m_\pi^2 s-(3{\bar l}_3+12{\bar l}_4-8{\bar
    l}_1-\frac{13}{3})m^4_\pi \nonumber\\
  &+& 3(s^2-m_\pi^4){\bar
    J}(s)+ \left(t(t-u)-2m_\pi^2 t+4m_\pi^2 u-2m_\pi^4\right) {\bar
    J}(t)+ \left(u(u-t)-2m_\pi^2 u+4m_\pi^2
    t-2m_\pi^4\right) {\bar J}(u) \Big\} \, .\label{eq:a4chpt}
\end{eqnarray}
\end{widetext}
The lowest-order amplitude $A_2 (s,t,u)$ is identical to the
first term in~\eqref{eq:LOpot} (pion contribution) and only depends
on the pion mass and weak decay constant. The ${\cal O}(p^4)$
correction involves four SU(2) renormalization-scale-independent LECs:
${\bar l}_i$ ($i=1,2,3,4$).  In addition, $A_4 (s,t,u)$
includes one-loop chiral corrections, which are suppressed by one
power of $1/N_C$; they are parameterized through the loop function
\begin{equation}
{\bar J}(s)= \frac{1}{16\pi^2}\left(2+\sigma(s)\,\log \left[
\frac{\sigma(s)-1}{\sigma(s)+1} \right]\right)\, .
\end{equation}

We list here for completeness the non-polynomial contributions to the
$\pi \pi$ scattering amplitudes at one-loop order~\cite{Gasser:1983yg}
\begin{eqnarray}
A_4 (s,t,u) &=& \frac{1}{6 \pi ^2 f^4}  \left[\,3 J(s) \left(s^2-m^4\right) 
\right. \nonumber \\
&+&  J(t) \left(-2 m^4-2 m_\pi^2 t+4 m_\pi^2 u+t(t-u)\right)\\
 &+& \left. J(u) \left(-2 m^4+4 m_\pi^2 t-2 m_\pi^2 u+u (u-t)\right)\,\right] \nonumber
\end{eqnarray}
As said, only the piece containing the loop integral $J(t)$
contributes to the potential, according to~\eqref{pol-part}.


\end{document}